\documentclass[10pt]{article}
\pdfoutput=1
\usepackage{jheppub} 

\usepackage{slashed}
\usepackage{color, verbatim}
\usepackage{latexsym}
\usepackage{epsfig}
\usepackage{amsmath,accents}
\usepackage{mathrsfs}

\usepackage{amssymb}
\usepackage{graphicx}
\usepackage{bm}
\usepackage{stackrel}

\usepackage[font={small}]{caption}

\usepackage{hyperref}
\usepackage{epstopdf}
\usepackage{cancel}

\usepackage[skins]{tcolorbox}

\definecolor{myred}{rgb}{0.7, 0, 0}
\definecolor{myblue}{rgb}{0, 0, 0.7}
\definecolor{mygreen}{rgb}{0.04, 0.7, 0.5}
\hypersetup{colorlinks,citecolor=myred,linkcolor=myblue,urlcolor=myblue,linktocpage=true}

\makeatletter

\@addtoreset{equation}{section}
\makeatother

\newcommand{\be}{\begin{equation}}
\newcommand{\ee}{\end{equation}}
\newcommand{\bea}{\begin{eqnarray}}
\newcommand{\eea}{\end{eqnarray}}

\newcommand{\tr}{\operatorname{tr}}

\def\Mp {M_{\rm Pl}}

\def \bse {\begin{subequations} \begin{eqnarray}}
\def \ese {\end{eqnarray} \end{subequations}}

\begin{document}

\thispagestyle{empty}

%\begin{comment}
\begin{center}

\begin{center}

\vspace{.5cm}

{\Large\sc
Baryogenesis from combined \\ Higgs – scalar field inflation\\\vspace{1cm}
}

\end{center}

\vspace{1.cm}

\textsc{
Yann Cado, 
%$^{\,a}$
\ Mariano Quir\'os%$^{\,a}$
}\\

\vspace{.5cm}

%${}^a\!\!$ 
{\em {Institut de F\'{\i}sica d'Altes Energies (IFAE) and\\ The Barcelona Institute of  Science and Technology (BIST),\\ Campus UAB, 08193 Bellaterra, Barcelona, Spain
}}

\end{center}

\vspace{0.8cm}

\centerline{\bf Abstract}
\vspace{2 mm}

\begin{quote}\small
We study a modification of the Higgs inflation scenario where we introduce an extra scalar $\phi$, with mass $m$, coupled to the Ricci scalar as $g\phi^2 R$, and mixed with the Higgs field $h$ via the Lagrangian term $\mu \phi h^2$. Both fields participate in the inflation process in a unitary theory that predicts values of the cosmological observables in agreement with the results from the Planck/BICEP/Keck collaborations.
In addition, by means of a $\mathcal{CP}$-odd effective operator that couples $\phi$ to the Chern-Simons term of the hypercharge gauge group as $f_\phi^{-1}\phi \,Y_{\mu\nu}\tilde Y^{\mu\nu}$, maximally helical magnetic fields are produced during the last $e$-folds of inflation. We found a window in the coupling $ f_\phi$ where these fields survive all constraints until the electroweak phase transition, and source the baryon asymmetry of the Universe through the Standard Model chiral anomaly. 
%
%For values $m\gg m_h$ there is a one-loop SM-like naturalness problem. 
From a phenomenological perspective, the model can solve the Standard Model instability problem at the scale $\mathcal Q_I\simeq 10^{11}$ GeV, provided that $\mu\lesssim m \lesssim \mathcal Q_I$, and for $m\lesssim \mathcal{O}$(few)~TeV, the $\phi$-$h$ mixing becomes sizable while the theory turns natural.
The latter thus predicts modifications of the trilinear and quartic couplings that could be explored at the HE-LHC, as well as at future colliders, and allows for direct $\phi$ production at the LHC followed by decay into $hh$. Present results from ATLAS and CMS already put (mild) bounds on the mass of the heavy scalar as $m\gtrsim 0.55$~TeV at 95\% C.L.
\end{quote}

\vfill

\newpage

\tableofcontents

\newpage

\section{Introduction}
\label{sec:introduction}
Electroweak (EW) baryogenesis is an appealing mechanism to understand the baryon asymmetry of the Universe (BAU)~\cite{Sakharov:1967dj} (for reviews see Refs.~\cite{Cohen:1993nk,Quiros:1994dr,Rubakov:1996vz,Carena:1997ys,Quiros:1999jp,Morrissey:2012db}), which is testable at EW  energies. Although the Standard Model (SM) contains all necessary ingredients required by the three Sakharov conditions, it quantitatively fails as the amount of $\mathcal{CP}$-violation in the CKM phase is too small and moreover, given the experimental value of the Higgs mass, the electroweak phase transition (EWPT) is not strong enough first order, but a continuous crossover~\cite{DOnofrio:2015gop,Kajantie:1996qd}. This mechanism should then require beyond the SM physics. 

It was more recently realized, Refs.~\cite{Anber:2006xt,Bamba:2006km,Bamba:2007hf,Anber:2009ua,Anber:2015yca,Cado:2016kdp,Sfakianakis:2018lzf},  that maximally helical hypermagnetic fields can be produced at the end of (axial) inflation, and can
generate the observed BAU, via the $B + L$ anomaly, during the EWPT. In this kind of theories, $\mathcal{CP}$ is spontaneously violated by the effective dimension-five operator $a Y^{\mu\nu}\widetilde Y_{\mu\nu}$, where $a$ is the axial field, $Y^{\mu\nu}$ the strength of the hypercharge gauge field $Y^\mu$, and $\widetilde Y^{\mu\nu}$ its dual, whose generation requires an ultraviolet (UV) completion of the model. The generation of the observed BAU was further elaborated in a number of papers, see e.g.~Refs.~\cite{Kamada:2016eeb,Kamada:2016cnb,Jimenez:2017cdr,Domcke:2019mnd}.

In a recent paper~\cite{Cado:2021bia}, we proposed a mechanism where the helical hypermagnetic fields were produced after inflation by the Higgs doublet field $\mathcal H$ with a $\mathcal{CP}$-violating $|\mathcal H|^2 Y^{\mu\nu}\widetilde Y_{\mu\nu}$ dimension-six operator, thus entirely relying the nonperturbative production of gauge fields on SM physics. Of course generating the $\mathcal {CP}$-odd operator $|\mathcal H|^2 Y^{\mu\nu}\widetilde Y_{\mu\nu}$ requires a UV completion, which can be similar to that giving rise to the $\mathcal {CP}$-even operator $a Y^{\mu\nu}\widetilde Y_{\mu\nu}$, for which $\mathcal{CP}$ is (spontaneously) violated for background values of the axial field $a$.

Moreover there are theories, dubbed as Higgs inflation (HI) models~\cite{Bezrukov:2007ep,Bezrukov:2008ej,Bezrukov:2010jz} (for a review see~\cite{Rubio:2018ogq}), where the inflaton is identified with the SM Higgs boson, thus linking the cosmological observables during the inflationary period of inflation with SM quantities. These models are based on assuming, in the Jordan frame, a coupling between the Higgs doublet $\mathcal H$ and the Ricci scalar $R$ as $\mathcal L=-(\Mp^2/2) R-\xi_{\mathcal H}|\mathcal H|^2R+\cdots$, where the ellipses refers to the SM Lagrangian. This model has been shown to have a (dynamical) cutoff $\Mp/\xi_{\mathcal H}$, for values of the Higgs at the electroweak scale, i.e.~$h\sim v$~\cite{Han:2004wt,Burgess:2009ea,Barbon:2009ya,Lerner:2009na,Burgess:2010zq,Hertzberg:2010dc}, while at values of the Higgs where inflation happens, i.e.~$h\sim \Mp/\sqrt{\xi_{\mathcal H}}$, the cutoff has been proven to be $\sim \Mp/\sqrt{\xi_{\mathcal H}}$, at least for two-by-two tree level scattering amplitudes, avoiding thus unitarity violation~\cite{Antoniadis:2021axu,Ito:2021ssc,Karananas:2022byw}.  
Moreover, HI models have to face another challenge: for actual values of the Higgs boson and top-quark masses the SM potential becomes unstable at values of the Higgs field $h\sim \mathcal Q_I\sim 10^{11}$ GeV. This question has been tackled in Ref.~\cite{Bezrukov:2014ipa}, where the case of an unstable potential was considered, taking into account radiative corrections. Because of the Higgs-Ricci coupling the theory becomes nonrenormalizable in the Einstein frame and requires the addition of an infinite number of counterterms. By assuming a scale invariant UV completion it is found that there are threshold effects at scales $\sim \Mp/\xi_{\mathcal H}$ which generate jumps of the SM quartic coupling to positive values (although one cannot determine their amplitude from the theory) and therefore HI can proceed in the usual way. Still the potential has two minima: the EW minimum and a much deeper (unphysical) minimum associated to the instability of the original SM potential. The evolution of the Higgs field after inflation will depend on the reheating process, and in particular on the reheating temperature. If the reheating temperature is high enough such that the unphysical minimum is dominated by the thermal corrections, then the Higgs will relax to the symmetric phase, otherwise the Higgs would go to the unphysical vacuum and it would stay there forever.

Motivated by HI, we will propose a model where the SM potential is simply stabilized by a scalar field $\phi$ coupled to the Higgs (this coupling was already pursued in Refs.~\cite{Giudice:2010ka,Barbon:2015fla}) and with a mass $m\lesssim \mathcal Q_I$, opening up the possibility of direct or indirect detection at present (LHC) and future accelerators. Moreover if the stabilizing field has a weak enough self-coupling $\phi^4$ and is coupled to the Ricci tensor as $\sim g\phi^2 R$, it can trigger cosmological inflation, as the potential becomes flat in the Einstein frame, while the COBE normalization does not impose strong constraints on the $g$ coupling. In this theory the inflaton can couple to the Chern-Simons component of the SM hypercharge and trigger baryogenesis via the production of helical magnetic fields. Finally through the coupling of the inflaton and the Higgs field, the latter will also be a component of the inflaton sector, although we will work out a model where the parameters are such that cosmological inflation will be mainly driven by the stabilizing field $\phi$. The model thus combine HI, baryogenesis via production of helical magnetic fields and stabilization of the SM potential by modifying the renormalization group running, to provide a successful history of the Universe.

In the present paper we will follow the above guideline in order to build such a model of inflation, which consists in a modification of the HI model by the introduction of a scalar field $\phi$, with 
%
% in 
 a two-field potential $V(h,\phi)$ in which analytical relations between both fields are enforced by its shape. One major difference with respect to a previous attempt, Ref.~\cite{Barbon:2015fla}, is that $\phi$ is coupled to the Ricci scalar as $(g/2) \phi^2 R$, with $\phi\lesssim \Lambda_\phi\equiv \Mp/g$~\footnote{\label{foot}It has been proved, in Refs.~\cite{Hertzberg:2010dc,Lerner:2009na}, that there is no tree-level unitarity problem for the amplitude $\mathcal A(\phi\phi\to\phi\phi)$ as, in the Einstein frame, see Eq.~(\ref{eq:gen-kin-term}), there appears the effective operator $\phi^2 (\partial_\mu\phi)^2/\Lambda_\phi^2$ that—upon integration by parts gives, on-shell, the correction $m^2\phi^4/\Lambda_\phi^2$—leads to a four-point function that does not grow with the energy, and thus does not violate unitarity. A similar result is obtained in the Jordan frame, where the amplitude $\mathcal A(\phi\phi\to\phi\phi)$ grows, in the $s$-channel, with the energy, and behaves as $s/\Lambda_\phi^2$. However, considering the cross channels, there is a cancellation, and the four-point amplitude behaves as $(s+t+u)/\Lambda_\phi^2\propto m^2/\Lambda_\phi^2$. However, the quick conclusion that unitarity is not violated at the scale $\Lambda_\phi$ has been challenged in Refs.~\cite{Hertzberg:2010dc,Burgess:2010zq}, where it was pointed out that, in the Jordan frame, the above cancellation is very unlikely to appear in loop-induced corrections to the same process $\phi\phi\to\phi\phi$, leading to a cutoff at the value $\sim 4\pi\Lambda_\phi$, where a loop factor has been included. The observation is similar for higher order processes, since e.g.~$\phi\phi\to\phi\phi+n\phi$ has a cross section that scales as $\lambda_\phi^2 s^{n/2-1}g^n/\Mp^n$, where $\lambda_\phi$ is the $\phi$ quartic coupling. This indicates that the perturbative description breaks down for energies $\sqrt{s}\gtrsim \lambda_\phi^{-2/n}\Lambda_\phi$, which goes to $\Lambda_\phi$ for large values of $n$. Similarly, in the Einstein frame, on top of the nonproblematic effective operator $\phi^2(\partial_\mu\phi)^2$, other higher order operators, as e.g.~$\phi^2(\partial_\mu\phi)^4$, are expected to be generated by loop effects, and so are expected to trigger violations of unitarity beyond the scale $\Lambda_\phi$. In view of these arguments we will conservatively consider in this paper $\Lambda_\phi$ as the scale at which unitarity is violated.}, where $\Lambda_\phi$ is the theory cutoff, while $\xi_\mathcal{H} \ll 1$, thus satisfying the most naive unitarity requirement.
Therefore, in our model the Higgs field is not the only inflaton, but a component of the inflaton system, as inflation is really driven along a particular path in the two-field space, while its orthogonal direction has a strong curvature around its minimum, where the field system is anchored.

This paper is organized as follows. In Sec.~\ref{sec:model} we introduce the potential in the Jordan frame, as a function of the fields $\phi$ and $h$, which (for Planckian values of the field $\phi$) can be approximated by the most general renormalizable polynomial satisfying the $\mathbb Z_2$ symmetry $\phi\to -\phi$. As the Ricci term is quadratic in $\phi$, $g\phi^2 R$, the beginning of inflation will be controlled by the quartic term $\lambda_\phi\phi^4$, and the size of the amplitude of density perturbations is provided by the smallness of $\lambda_\phi$, for values of $\phi\leq \Lambda_\phi$, consistent with the naive unitarity of the theory. The smallness of the coupling $\lambda_\phi$ is stable under radiative corrections, and so is technically natural, but the Higgs potential is unstable for values of the renormalization scale $\mathcal Q_I\simeq 10^{11}$~GeV. If the mass $m$ of the $\phi$ field is $m<\mathcal Q_I$, the field $\phi$ decouples for values of $\mathcal Q<\mathcal Q_I$. Then, in the presence of a potential term softly breaking the $\mathbb Z_2$ symmetry, $-\mu \phi |\mathcal H|^2$, there is a threshold correction in the one-loop $\beta$ function of the Higgs quartic coupling that can stabilize the EW vacuum. This mechanism was introduced in Ref.~\cite{Barbon:2015fla} and we will use it to constrain our parameter space.\footnote{Should we have, instead, considered a linear Ricci term, $g\phi R$, and a quadratic, $m^2\phi^2$, inflationary potential, one could also have achieved the amount of flatness required by the slow roll conditions during the inflationary period, but the size of the amplitude of density perturbations, now controlled by $m$, would have yielded a value $m>\mathcal Q_I$, which is too large to stabilize the EW vacuum.}

The properties of the inflationary model are presented in Sec.~\ref{sec:inflation}. There, we will prove that all observational constraints from the Planck and BICEP/Keck collaborations on the slow roll parameters, or equivalently on the spectral index, the spectral index running and the tensor-to-scalar ratio, can be satisfied for a range of the parameter $g$ such that $g\ll 1$, thus easily satisfying the unitarity condition for the model. 

The nonperturbative production of gauge fields at the end of inflation is presented in Sec.~\ref{sec:field-production}. In particular we will consider the $\mathcal{CP}$-violating dimension-five operator $1/(4f_\phi)\phi Y_{\mu\nu}\widetilde Y^{\mu\nu}$, provided by some UV completion, to trigger baryogenesis at the EWPT. We postpone to App.~\ref{sec:UV} the details of a particularly simple UV completion giving rise to such an operator. Similar UV completions were proposed in Ref.~\cite{Dine:1990fj}, and recently in Refs.~\cite{Carena:2018cjh,Carena:2019xrr}, to generate the BAU using various mechanisms, so we can be agnostic about its origin. We have found a critical value of the parameter $f_\phi$, such that for $f_\phi\gtrsim f_\phi^c$ the backreaction of the produced gauge fields on the inflationary dynamics can be neglected, and so we have explicitly considered this region in the numerical analysis. Moreover, in the presence of magnetic fields, as those produced in this work, there appear fermionic currents, a phenomenon called the Schwinger effect, and, for sufficiently strong magnetic fields, their backreaction on the gauge fields cannot be neglected. As exactly solving the equations of motion of gauge fields, in the presence of the Schwinger fermionic currents, is beyond the scope of the present paper, we have followed recent proposals in the literature for gauge field estimates~\cite{Domcke:2018eki,Domcke:2019mnd}, and have worked out two simple approximations: the \textit{maximal estimate}, obtained upon maximizing the value of the helicity, and the \textit{equilibrium estimate}. A detailed recent analysis~\cite{Gorbar:2021zlr} shows that the exact solutions lie in between both estimates, so we can reliably corner the final allowed region in the relevant parameter space.

The reheating mechanism has been studied in Sec.~\ref{sec:reheating}, and we have consistently considered the region $f_\phi>f_\phi^c$, where the reheating takes place perturbatively by the leaading inflaton decay process $\Gamma(\phi\to hh)$. The inflaton width, as well as the reheating temperature $T_{\rm rh}$, are then functions of the inflaton mass parameter $m$. In order to stabilize the EW vacuum, the latter must lie in the interval $m\in[1\, \textrm{TeV}, \mathcal Q_I]$ which implies, for reheating temperatures, the interval $T_{\rm rh}/T_{\rm rh}^{\rm ins}\in[10^{-2}\, ,10^{-6}]$, where  $T_{\rm rh}^{\rm ins}\simeq 2\cdot 10^{15}$~GeV would be the instant reheating temperature, i.e.~the reheating temperature in the hypothetical case where $\Gamma(\phi\to hh)$ equals the Hubble parameter at the end of inflation.

We show in Sec.~\ref{sec:baryogenesis} how the baryon asymmetry is generated when helicity transforms into baryon number at the EW crossover. In particular we show that this mechanism works for $f_\phi>f_\phi^c$, and provides an upper bound on $f_\phi$ which depends on the value of the reheating temperature. 

In Sec.~\ref{sec:constraints} we consider all relevant conditions for the helical magnetic fields to survive from the end of inflation, when they are generated, to the EWPT, when they convert into the baryon asymmetry. In particular we have considered the constraints from magnetohydrodynamics (MHD) and Reynolds numbers, from chiral plasma instability, from the non-Gaussianity of the inflaton primordial fluctuations and from the baryon isocurvature perturbations. Some technical details about the latter are postponed to App.~\ref{sec:isocurvature}. Globally they constrain the region where the baryogenesis mechanism works, leaving an allowed range for the values of the parameter $f_\phi$, which depends on the ratio $T_{\rm rh}/T_{\rm rh}^{\rm ins}$. Readers not interested in the technical details of the analysis can straightforwardly go to Sec.~\ref{sec:summaryresult}, and in particular to Fig.~\ref{fig:parameter-space}, which summarizes the combined results.

In Sec.~\ref{sec:pheno} some phenomenological considerations, from the point of view of particle physics, are presented. First of all we study the naturalness problem generated by the mass hierarchy $m\gg m_h$, where $m_h$ is the Higgs mass, which leads either to a fine-tuning or considering $m=\mathcal O$(TeV). The latter case is phenomenologically appealing as the $\mathbb Z_2$-breaking term in the potential generates a mixing between the singlet $\phi$ and the Higgs field $h$. This mixing, which is negligible in the case of very large values of the parameter $m$, can be sizable, and with relevant phenomenological applications, for the case of $m=\mathcal O$(TeV) and, furthermore, is already bounded by the present LHC measurements of Higgs signal strengths. As the mixing angle is inversely proportional to $m$, the latter already provide mild lower bounds on $m$, as $m\gtrsim 0.4$~TeV, a region where electroweak observables are shown to be in agreement with their experimental values. Moreover, the mixing introduces modifications on the SM parameters $\lambda_3$ and $\lambda_4$, which could lead to constraints at the HE-LHC at $\sqrt{s}=27$~TeV, or even in future colliders with center of mass energies of 100 TeV. Finally the singlet state can be produced at LHC by means of its mixing with the Higgs field. Present upper bounds, from the ATLAS and CMS collaborations, on the production cross section lead to upper bounds on the mixing parameter and, consequently, to lower bounds on $m$ as $m \gtrsim 0.7$~TeV at 95\% C.L., while future runs are expected to provide stronger bounds on it. 

Finally we summarize the results and present our conclusions in Sec.~\ref{sec:conclusions}.

\section{The model}
\label{sec:model}
As stated in the previous section we consider, on top of the Higgs field $h$, the scalar field $\phi$ with the Lagrangian $\mathcal L_J$ as~\footnote{In our notation the Lagrangian $\mathcal L$ will not contain the factor $\sqrt{-g}$, so that the action $S$ is given by $S=\int d^4x \sqrt{-g}\mathcal L$.}
\be
\mathcal L_J=-\frac{\Mp^2}{2}R-\frac{g}{2}\phi^2R+\frac{1}{2}(\partial_\mu h)^2+\frac{1}{2}(\partial_\mu\phi)^2-U(\phi,h),
\label{eq:LagrangianJ}
\ee
which contains a coupling of the field $\phi$ to the Ricci scalar~\footnote{Notice that in our model we do not need to primarily introduce any $\xi_\mathcal{H} |\mathcal H|^2 R$ term. Although a small value of the parameter $\xi_\mathcal{H}$ will be generated anyway by radiative corrections~\cite{Espinosa:2015qea}, its effects on the inflation mechanism will always be negligible, even for values of $\xi_\mathcal{H}\simeq\mathcal O(1)$; so for simplicity we are assuming that $\xi_\mathcal{H}=0$.}, and the potential is given by
\be
\begin{aligned}
U(\phi,h)=&\;U_{\rm SM}(h)+\frac{1}{2}m^2\phi^2+\frac{1}{2}\lambda_{\phi h}\phi^2 h^2+\frac{1}{4}\lambda_\phi \phi^4
-\frac{1}{2}\mu \phi h^2\\
 U_{\rm SM}(h)=&-\frac{1}{2}\mu_h^2 h^2+\frac{1}{4}\lambda_0 h^4.
\label{eq:potencial}
\end{aligned}
\ee
The first four terms of the potential $U(\phi,h)$  in Eq.~(\ref{eq:potencial}) constitute the most general renormalizable potential invariant under the $\mathbb Z_2$ symmetry, $\phi\to -\phi$, while the last term is a soft breaking of such symmetry. Besides, for large Higgs field configurations we will be neglecting the mass term $\mu_h^2$, as compared to the $\lambda_0$ term, in $U_{\rm SM}(h)$.  

The parameters $\lambda_{\phi h}$ and $\lambda_\phi$ should be constrained by the slow roll conditions during inflation to very small values $\lambda_{\phi h},\lambda_\phi\ll 1$, as we will see later on. Their smallness is radiatively stable, as can easily be deduced from their one-loop $\beta$ functions~\footnote{We are defining conventionally here $\beta_X\equiv dX/dt$.}
\bse
\beta_{\lambda_{\phi h}}&=&\dfrac{\lambda_{\phi h}}{16\pi^2}\left[12\lambda_0+8\lambda_{\phi h}+6\lambda_\phi-\left( \dfrac{9}{2}g_2^2+\dfrac{9}{10}g_1^2-6 y_t^2  \right)   \right]\theta(t-t_0), 
\label{eq:RGE1}\\
\beta_{\lambda_\phi}&=&\dfrac{1}{16\pi^2}\,(8\lambda_{\phi h}^2+18 \lambda_\phi^2)\theta(t-t_0),
\label{eq:RGE2}
\ese
where $t-t_0=\log (\mathcal Q/m)$, and $\mathcal Q$ is the renormalization scale. In particular the choice $\lambda_{\phi h}=0$ is technically natural at one loop, as can be seen from Eq.~(\ref{eq:RGE1}). For simplicity we will adopt hereafter the value $\lambda_{\phi h}=0$. Moreover, from the amplitude of density perturbations, we will see that typically $\lambda_\phi\simeq 10^{-12}$, a value that is very mildly changed by radiative corrections.

\subsection{Jordan frame}
\label{sec:jordan-frame}

The previous Lagrangian is defined in the so-called Jordan frame, and it is a valid framework provided that the field $\phi$ satisfies the condition $\phi\ll\Mp/\sqrt{g}$. This region, as we will see, encompasses part of the inflationary period, and in particular the end of inflation. The trajectory of fields $\phi$ and $h$ will proceed along the submanifold given by the minimum of the two-dimensional potential surface, providing a relationship between both fields, as anticipated in the introduction of this paper.

To find the relationship between both fields $\phi$ and $h$, along the potential minimum direction, we will follow a general procedure summarized here. 
Given a potential $V(x,y)$ of two fields $x$ and $y$, the contour lines corresponding to constant values of the function $V(x,y)=\rm{constant}$, satisfy the relation $dV=0$, which reads
\be
\frac{\partial V}{\partial x}\, dx+\frac{\partial V}{\partial y} \, dy=0\quad \Rightarrow F(x,y)\equiv \frac{dy}{dx}=-\frac{\partial V/\partial x}{\partial V/\partial y},
\label{eq:dV0}
\ee
where, by definition, the function $F[x,y]$ is the slope along the contour lines at the point $(x,y)$. 
We wish to find the direction $y=f(x) $ that intersects orthogonally every contour line. The slope of this line is obviously $f'(x)$ and the slope of the orthogonal line is $-1/f'(x)$, so  the condition for the orthogonal intersection is
\be
F(x,f(x))=-1/f'(x).
\label{eq:ortogonal}
\ee 

The idea behind the regions is to divide the potential valley into segments such that $\phi = a h^n$. The regions are separated according to which term dominates in the potential.
Hence, we will find it useful to work with logarithmic variables
\be 
y=\log\phi,\qquad x=\log h,
\ee
where the $\phi$ and $h$ fields are considered in some arbitrary mass units, such that the relation between fields translate into straight lines $y=nx+ \log{a}$. Given the shape of our potential we find a unique solution to (\ref{eq:ortogonal}) in each region.

The direction $\phi=f(h)$ that intersects orthogonally every contour line in the plane $(h,\phi)$ is given by the solution to the equation
\be
\left.\frac{\partial V/\partial h}{\partial V/\partial \phi}\cdot\frac{h}{\phi}\right|_{\phi=f(h)}=\frac{1}{f'(h)}
\label{eq:orthogonal2}
\ee
where Eqs.~(\ref{eq:dV0}) and (\ref{eq:ortogonal}) have been used.

Therefore, the trajectory in the $(\phi,h)$ plane is given by relation (\ref{eq:orthogonal2}), which changes according to the different regions of the potential that we will now introduce. %In the following, for the different regions where we can approximate our potential in a simple way, we will find trivial solutions to Eq.~(\ref{eq:orthogonal2}) which will be
This is validated by the plot of the total potential exhibited in Fig.~\ref{fig:submanifold}.
In all cases, the valley acts as an attractor for the fields, as shown in Ref.~\cite{Barbon:2015fla}.

\subsubsection*{Region A}
\begin{figure}[htb]
\begin{center}
\includegraphics[width = 7.5cm]{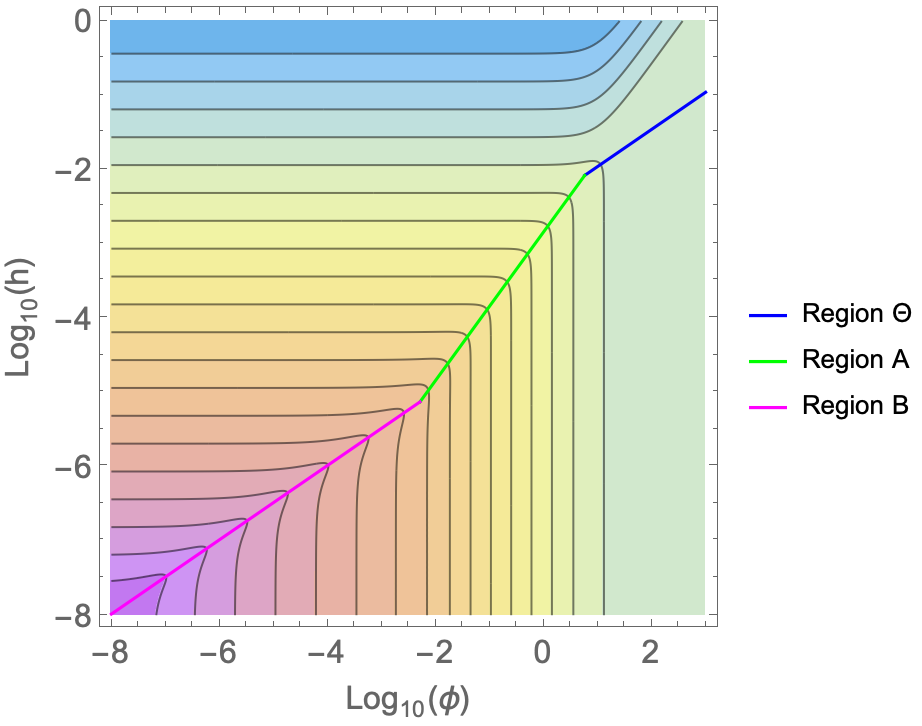}
\includegraphics[width = 7.9cm]{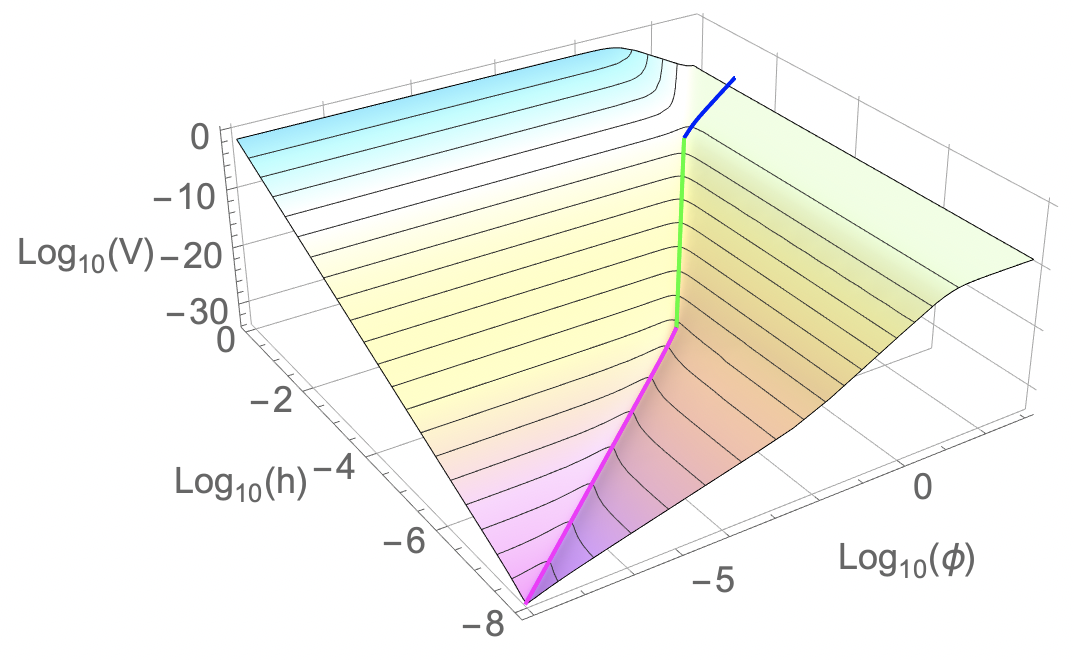}
\caption{ \it 
Left panel: Contour plot of the potential $V(\phi,h)$ in units where $\Mp=1$ with Regions $\Theta$, A and B and the corresponding minimum submanifolds. Right panel: 3D plot of the potential with the same color code. We use the following numerical values: $g=0.01$, $m=10^{10} $~GeV, $ \lambda_0 = 0.23$, $\delta_\lambda = 0.15$, $ \lambda_\phi =10^{-12}$.
\label{fig:submanifold}}
\end{center}
\end{figure}

In this region both fields take their maximum allowed values in the Jordan Frame, and the potential can be approximated by the quartic coupling terms
\be
U_{\rm A}\simeq \frac{1}{4}\lambda_0 h^4+\frac{1}{4}\lambda_\phi\phi^4\,.
\label{eq:UA}
\ee
The direction along the minimum can be found, after applying Eq.~(\ref{eq:orthogonal2}) to the potential (\ref{eq:UA}), with the function $f(h)=(\lambda_0/\lambda_\phi)^{-1/4}h$, i.e.
\be
h=\left(\frac{\lambda_\phi}{\lambda_0}\right)^{\frac{1}{4}}\phi\ .
\label{eq:minA}
\ee
We plot in Fig.~\ref{fig:submanifold} the complete inflationary potential in the Einstein frame (see Sec.~\ref{sec:einstein-frame}) and show the direction from Eq.~(\ref{eq:minA}) with a solid (green) line as specified in the figure caption.

Its region of validity is then given by
\be
\frac{\Mp}{\sqrt{g}}\gtrsim \phi\gtrsim \frac{2\sqrt{2}m}{\sqrt{\lambda_\phi}}\left(\frac{\delta_\lambda}{\lambda_0}  \right)^{\frac{1}{2}},\hspace{2cm} h\gtrsim \frac{2\sqrt{2}m}{\left(\lambda_\phi \lambda_0\right)^{\frac{1}{4}}}\left(\frac{\delta_\lambda}{\lambda_0}  \right)^{\frac{1}{2}},
\label{eq:regionA}
\ee
where we have defined the constant $\delta_\lambda$ as
\be
\delta_\lambda\equiv\frac{\mu^2}{2m^2}\,.
\label{eq:defdelta}
\ee
%
%Using the minimum direction in Eq.~(\ref{eq:minA}), we can write the Higgs-Ricci coupling term, $\xi_\mathcal{H} |\mathcal{H}|^2R$, as $\xi_\mathcal{H}=g\sqrt{\lambda_0/\lambda_\phi}$. For typical values of the parameters (e.g.~$g\simeq 0.01,\, \lambda_0\simeq 0.1,\, \lambda_\phi\simeq 10^{-12}$), we get $\xi_\mathcal{H}\sim 10^4$. 

Along the minimum direction (\ref{eq:minA}) the potential can be written, as a function of $\phi$, as
\be
U_{\rm A}(\phi)\simeq \frac{1}{2}\lambda_\phi \phi^4\,,
\label{eq:potA}
\ee
which will be used in the next section to describe the end of inflation.

% COPIED FROM INTRODUCTION
To make contact with HI results, in this region we can also use the Higgs field as the explicit variable using the relation between the fields $h$ and $\phi$ given by Eq.~(\ref{eq:minA}). This implies that the Ricci term in Eq.~(\ref{eq:LagrangianJ}) can be equivalently written as $-\xi_{\rm A}/2 \ h^2R$, with $\xi_{\rm A}=g\sqrt{\lambda_0/\lambda_\phi}$. For typical values of the parameters (e.g.~$g\simeq 0.01,\, \lambda_0\simeq 0.2,\, \lambda_\phi\simeq 10^{-12}$, see Secs.~\ref{sec:stability} and \ref{sec:inflation}), we get $\xi_{\rm A}\approx 4\cdot \,10^4$, which is the value required by HI. Moreover the potential (\ref{eq:potA}) can be written, using again (\ref{eq:minA}) as
\be U_{\rm A}(h)\simeq \frac{1}{2}\lambda_0 h^4\,. \ee
This result shows how, in region A, the results of HI could be interpreted in our model with $g\ll 1$, being perfectly consistent with the unitarity condition $\phi\lesssim \Lambda_\phi$.

\subsubsection*{Region B}
In this region, where 
\be
\phi\lesssim \frac{2\sqrt{2}m}{\sqrt{\lambda_\phi}}\left(\frac{\delta_\lambda}{\lambda_0}  \right)^{\frac{1}{2}},\hspace{2cm} m\lesssim h\lesssim \frac{2\sqrt{2}m}{\left(\lambda_\phi \lambda_0\right)^{\frac{1}{4}}}\left(\frac{\delta_\lambda}{\lambda_0}  \right)^{\frac{1}{2}}
\label{eq:regionB}
\ee
the potential can be approximated by
\be
U_{\rm B}\simeq -\frac{1}{2}\mu\phi h^2+\frac{1}{2}m^2 \phi^2+\frac{1}{4}\lambda_0h^4,
\label{eq:potB}
\ee
which, using Eq.~(\ref{eq:orthogonal2}), has its minimum along the direction
\be
\phi=f(h)\equiv \left(-\frac{3\mu}{4 m^2}+\sqrt{\frac{9\mu^2}{16 m^4}+\frac{2\lambda_0}{m^2}}   \right)h^2\,.
\label{eq:regionB}
\ee
Direction (\ref{eq:regionB}) is shown in the potential plot, Fig.~\ref{fig:submanifold}, with a solid (magenta) line.
If we define the coupling $\lambda$ as
\be
\lambda\equiv\lambda_0-\delta_\lambda\,,
\label{eq:deflambda}
\ee
in the limit $\lambda\ll 1$ we can write the minimum condition as
\be
\phi\simeq \sqrt{\frac{\delta_\lambda}{2}} \, \frac{h^2}{m} \; [1+\mathcal O(\lambda)]
\label{eq:relB}
\ee
and the potential (\ref{eq:potB}) becomes
\be
U_{\rm B}\simeq \frac{1}{4}\lambda h^4+\mathcal O(\lambda^2),
\label{eq:potBfinal}
\ee
which shows that the effective quartic coupling in this region is given by $\lambda$, instead of $\lambda_0$ as in the original potential (\ref{eq:potencial}).

\subsubsection*{Region C}
In this region $v<\mathcal Q\equiv h<m$, where $v$ is the Higgs vacuum expectation value (VEV) and $\mathcal Q$, the renormalization scale, is here identified with the classical value of the Higgs field $h$. The field $\phi$ hence decouples and is integrated out as
\be
\phi=\frac{\mu}{2m^2}h^2+\mathcal O(h^6)\simeq\sqrt{\frac{\delta_\lambda}{2}} \, \frac{h^2}{m},
\label{eq:minC}
\ee
which yields a potential
\be
U_{\rm C}\simeq \frac{1}{4}\lambda h^4+\mathcal O(h^8).
\label{eq:potC}
\ee
Notice that, to leading order, the solution to the equation of motion of $\phi$, Eq.~(\ref{eq:minC}), agrees with the minimum condition in Region B, Eq.~(\ref{eq:relB}), which guarantees the continuity between both regions. Moreover the stability of the potential in both Regions B and C is provided by the same condition, $\lambda>0$.

\subsection{Stability of the potential}
\label{sec:stability}

In Region C, $h<m$, the inflaton field $\phi$ is integrated out and the potential, as a function of the Higgs $h$, is given by Eq.~(\ref{eq:potC}), so that the parameter $\lambda$ runs as the quartic coupling in the SM potential, according to the SM $\beta$ function, $\beta_{\lambda}^{\rm SM}$.
In Regions B and A, $h>m$, the inflaton $\phi$ propagates and thus there is an extra contribution to the running of the parameter $\lambda$ as~\cite{Barbon:2015fla}
\be
\beta_\lambda=\beta_\lambda^{\rm SM}+\frac{1}{2\pi^2}\delta_\lambda(3\lambda+\delta_\lambda)\,\theta(t-t_0),
\label{running-lambda}
\ee 
where $\theta(x)$ is the Heaviside function, equal to 1 (0) for $x\ge 0$ ($x< 0$), and $t-t_0=\log(h/m)$. The parameter $\delta_\lambda$ also runs with the renormalization scale as
\be
\beta_{\delta_{\lambda}}=\frac{1}{2\pi^2}\delta_\lambda(3\lambda+2\delta_\lambda)\,\theta(t-t_0).
\ee

The extra contribution to the running of $\lambda$ in Eq.~(\ref{running-lambda}) can solve the Higgs vacuum instability problem provided that:
\begin{itemize}
\item
The inflaton mass $m$ is smaller than the SM instability scale, $\mathcal Q_I\sim 10^{11}$~GeV.
\item
The value of $\delta_\lambda$ at the scale $\mathcal Q=m$, $\delta_\lambda(m)$, is large enough in order to significantly change the value of $\beta_\lambda^{\rm SM}$. 
\end{itemize}
Of course, smaller values of $m$ (i.e.~wider regions where $\phi$ propagates) allow smaller values of $\delta_\lambda(m)$ to satisfy the second criterion.
Conversely, for values of $m$ close to $\mathcal Q_I$ the minimum value of $\delta_\lambda(m)$ that solves the instability is a largish one.

As we have seen, the condition for the stability of the potential is that the coupling $\lambda$ defined in Eq.~(\ref{eq:deflambda}) is positive definite, $\lambda\geq 0$.
We have solved at two-loop the RGE's of the theory for the following set of values of the input parameters~\cite{Khoury:2021zao} at the pole top mass $M_t=172.76$~GeV, 
\be\begin{aligned}
g_Y(M_t)&=0.358545,\quad g_2(M_t)=0.64765,\quad g_3(M_t)=1.1618, \\
\lambda(M_t)&=0.12607,\hspace{6.5mm} h_t(M_t)=0.9312.
\end{aligned}\ee

\begin{figure}[htb]
\centering
\includegraphics[width=7.5cm]{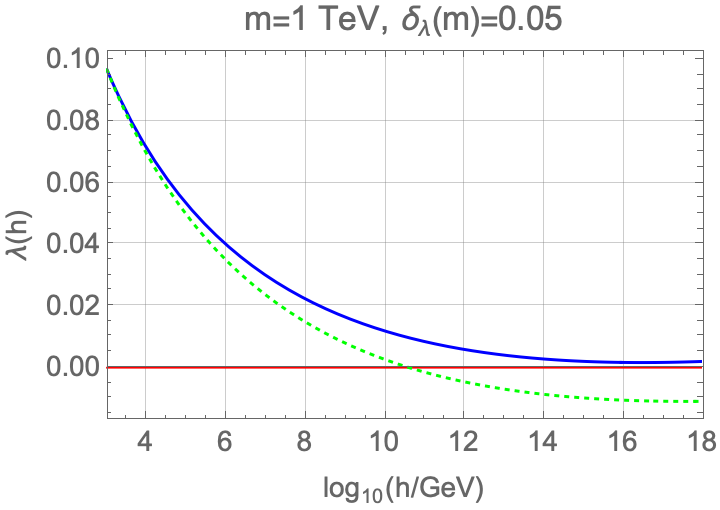}
\includegraphics[width=7.5cm]{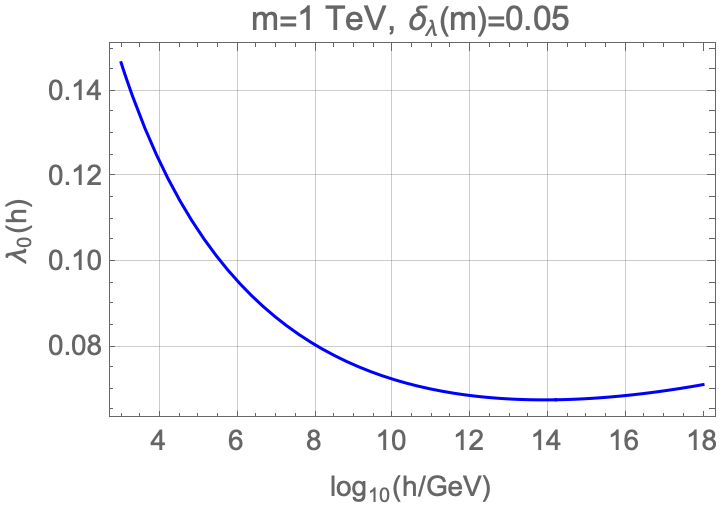} 
\includegraphics[width=7.5cm]{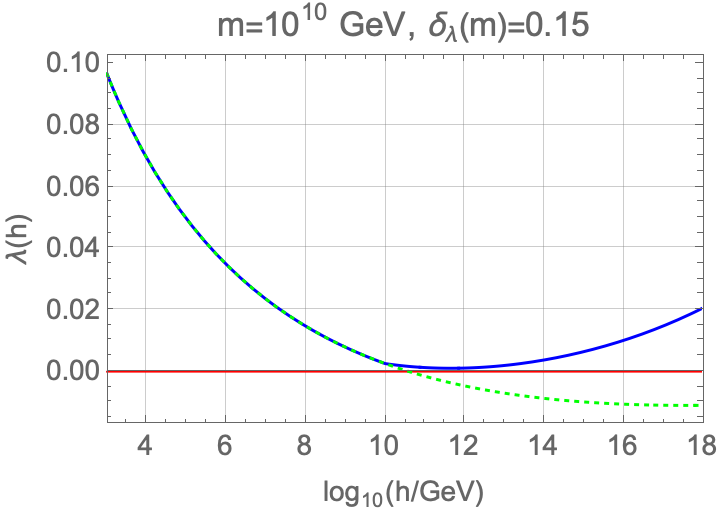}
\includegraphics[width=7.5cm]{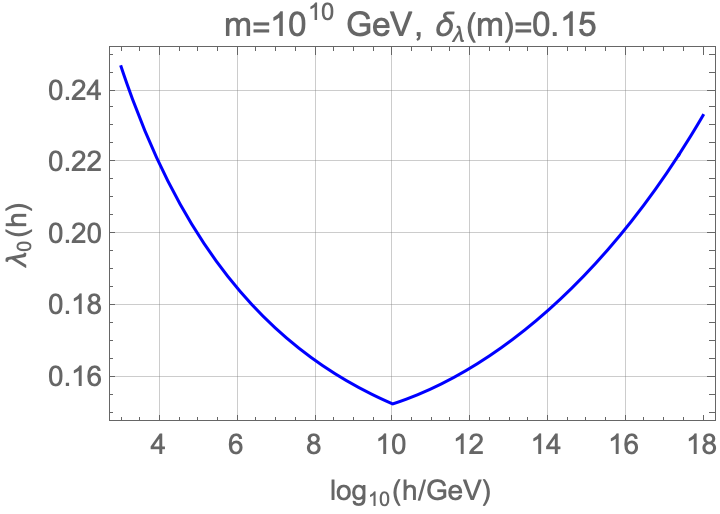} 
\caption{\it In blue, two-loop running of $\lambda$ (left panels) and $\lambda_0$ (right panels) for two cases. Top panels: with $m=1$~TeV, $\delta_\lambda(m)=0.05$. Bottom panels: with $m=10^{10}$~GeV, $\delta_\lambda(m)=0.15$. The green dashed line is the SM running. In both cases one has $\lambda_0 \simeq \delta_\lambda$ for $\mathcal Q \sim \Mp$. }
\label{fig:RGE}
\end{figure} 
In Fig.~\ref{fig:RGE} we show the two-loop running of the parameters $\lambda$ and $\lambda_0$ for two extreme cases, with a light ($m=1$~TeV, upper panels) and a heavy ($m=10^{10}$~GeV, lower panels) inflaton. As we can see typical values of $\delta_\lambda$ are smaller for smaller values of $m$. We have chosen $\delta_\lambda=0.05$ for $m=1$ TeV, and $\delta_\lambda=0.15$ for $m=10^{10}$ GeV. In both cases the value of $\delta_\lambda(m)$ can be tuned to smaller values, such that the corresponding values of $\lambda$ at high scales are smaller. On the other hand, larger values of $\delta_\lambda$ are bound by imposing that the theory remains in the perturbative regime up to the high scale. In particular we find, for large values of $m$, $m\simeq \mathcal Q_I$, $\delta_\lambda(m)\lesssim 0.35$, while for $m$ in the TeV region, $\delta_\lambda(m)\lesssim 0.2$. The dashed lines in the left panels are the SM running, shown for comparison.
On both left panels, we can see that the condition $0<\lambda\ll1$ is satisfied while $\delta_\lambda\gg\lambda$ at $\mathcal{Q}\sim\Mp$. 

\subsection{Einstein frame}
\label{sec:einstein-frame}

For values of the $\phi$ field such that $\phi >\Mp/\sqrt{g}$ we must redefine the metric and go to the so-called Einstein frame to recover the Einstein-Hilbert action for the Ricci scalar. To do so, we perform a Weyl redefinition of the metric:
\be g_{\mu\nu}\to \Theta \, g_{\mu\nu}, \hspace{3cm}\sqrt{-g}\to \Theta^2 \sqrt{-g}. \ee
For the Ricci scalar this implies
\be R\to \frac{ R}{\Theta} - \bar{R}, \quad \bar{R}= \dfrac{6}{\Theta^{3/2}} \dfrac{\partial_\mu\left(g^{\mu\nu}\sqrt{-g}\, \partial_\nu \sqrt{\Theta}\right)}{\sqrt{-g}}.
 \ee
Note that $R$ is absent in the correction term $\bar{R}$, hence we will define $\Theta$ by demanding that the explicit coupling between $\phi$ and $R$ disappears from the Lagrangian in the Einstein frame. The Ricci part of the action transforms then as
\be S_R \to S_R^E= - \int d^4x\sqrt{-g}\left(\dfrac{\Mp^2}{2} + \dfrac{g \phi^2}{2}\right) \left( R \Theta  - \bar{R}\Theta^2\right) \label{Ricci-transform} 
\ee
and so the definition
\be \Theta(\phi) =\left(1+\dfrac{g\phi^2}{\Mp^2}\right)^{-1} \label{def-theta-mirage}\ee
leads to 
\be  S_R^E=   \int d^4x \sqrt{-g}\; \left[ -\dfrac{\Mp^2}{2} R  + 3  \Theta^2  \dfrac{g^2\phi^2}{\Mp^2} \partial_\mu \phi \partial^\mu \phi  \right]. \label{eq:SRE}
\ee
We can see from the second (dimension-six effective operator) term in Eq.~(\ref{eq:SRE}) that the cutoff of the theory $\Lambda_\phi$ is identified as $\Lambda_\phi\equiv \Mp/g$ (see however comments in footnote~\ref{foot}).

In the meantime the kinetic terms of $\phi$ and $h$ get transformed to
$  \frac{\Theta}{2}(\partial_\mu \phi\partial^\mu\phi+\partial_\mu h\partial^\mu h)
$
so that the (noncanonical) kinetic terms are given by
\be
\mathcal L_{\rm kin}^E
=\frac{\Theta}{2} \left( 1+ \frac{6g^2\phi^2}{\Mp^2}\;\Theta \right) \; \partial_\mu \phi\partial^\mu\phi+
\frac{\Theta}{2} \partial_\mu h\partial^\mu h
\label{eq:gen-kin-term}
\ee
leading to the action in the Einstein frame
\be 
S_E= \int d^4x \sqrt{-g} \left(-\dfrac{\Mp^2}{2}R +\mathcal L_{\rm kin}^{\rm E}- V(\Theta,h) \right) , 
\ee
where the Einstein frame potential $V(\phi,h)$ is given by
\be  V(\phi,h)= \Theta^2(\phi) \, U(\phi,h),
\label{EF-potential} 
\ee
and $U(\phi,h)$ is given by Eq.~(\ref{eq:potencial}).
The potential region where the values of the field $\phi$ satisfy the condition $\phi>\Mp/\sqrt{g}$ is denoted as Region $\Theta$ and is explored hereafter.

\subsubsection*{Region $\Theta$}
As just stated, the Region $\Theta$ is characterized by the potential $V(\phi,h)$ in the Einstein frame, 
i.e.~Eq.~(\ref{EF-potential}) for $g\phi^2>\Mp^2$, and a straightforward application of Eq.~(\ref{eq:ortogonal}) shows that, using Eq.~(\ref{eq:orthogonal2}), the direction along the minimum in the two-dimensional potential is given by
\be
h^2= \Mp\,\left( \frac{\lambda_\phi}{3g \lambda_0}\right)^{\frac{1}{2}}\phi ,
\label{eq:minA0}
\ee
from where the function $f(h)$ in Eq.~(\ref{eq:orthogonal2}) can easily be read out.
Along this direction the potential is
\be V_{\Theta}(\phi)= \Theta^2(\phi) \;\frac{\lambda_\phi}{4}\,\phi^2 \left( \frac{\Mp^2}{3g} + \phi^2 \right) \simeq \Theta^2(\phi)\,\frac{\lambda_\phi}{4}\,\phi ^4, \label{potentialA0}  
\ee
where again the last equality comes from the very definition of the $\Theta$ region. Notice that the values of the field $\phi$ at the beginning of inflation, and in particular its value $\phi_\ast$ at horizon crossing of the present Universe, belong to Region $\Theta$.

In Fig.~\ref{fig:submanifold} we plot the potential in the Einstein frame $V(\phi,h)$ for a chosen set of the parameters values, and we superimpose the lines of minimum submanifolds given by Eqs.~(\ref{eq:minA0}), (\ref{eq:regionA}) and (\ref{eq:regionB}), for Regions $\Theta$, A and B, respectively. As we can see they intersect orthogonally, by construction, the contour lines of the potential. In the left panel we plot the contour lines of the potential and in the right panel the three-dimensional plot with the same color codes. 

We can try to make contact with HI in Region $\Theta$, as we did in Region A, using the Higgs field $h$ as the explicit variable, by means of the relation between the fields $\phi$ and $h$ given in Eq.~(\ref{eq:minA0}), which we can write as
\be g \phi^2=\xi_{\Theta} \frac{h^4}{\Mp^2},\hspace{1.5cm} \textrm{with}\quad\xi_{\Theta}\equiv \frac{3 g^2\lambda_0}{\lambda_\phi} \,. \ee
The Ricci coupling can then be written as $-\xi_{\Theta} h^4 R / \Mp^2$, where $\xi_\Theta\simeq 3\cdot 10^7$ by using the typical values of the parameters, $g\simeq 0.01,\, \lambda_0\simeq 0.2,\, \lambda_\phi\simeq 10^{-12}$ (see Secs.~\ref{sec:stability} and \ref{sec:inflation}).
Similarly, we can also write the potential as
\be V_\Theta(h)\simeq \left(1+\xi_{\Theta}\frac{h^4}{\Mp^4} \right)^{-2}\  \frac{\lambda_\phi}{4g^2}\;\xi_\Theta^2\,\frac{h^8}{\Mp^4}\,.%+\frac{\lambda_0}{4}h^4\right) 
\ee
These two expressions show that our model, written in terms of the Higgs field, departs from the conventional HI as it requires an effective dimension-eight operator for the potential which could only appear when the Standard Model is completed by some UV theory, giving rise, after decoupling, to higher dimensional operators.

\section{Inflation}
\label{sec:inflation}

Inflation takes place only in Regions  $\Theta$ (for $\sqrt{g}\phi>\Mp$), and A (for $\sqrt{g}\phi<\Mp$), thus we will choose conditions (\ref{eq:minA0}) and (\ref{eq:minA}), respectively, to relate $h$ and $\phi$. In this case the kinetic term (\ref{eq:gen-kin-term}) along the minimum direction can be written in both Regions $\Theta$ and A, as
\be
\mathcal L_{\rm kin}^{\rm R}=\frac{\Theta}{2}\left[  1+6\frac{g^2\phi^2}{\Mp^2} \Theta +\Delta_{\rm R}  \right]\partial_\mu\phi \partial^\mu\phi,\quad {\rm (R=\Theta,A)}
\label{eq:kinetic}
\ee
where $\Delta_{\rm R}$ corresponds to the (tiny) contribution of the Higgs kinetic term 
\be
\Delta_{\rm A}=\left(\frac{\lambda_\phi}{\lambda_0}  \right)^{\frac{1}{2}},\hspace{2cm} \Delta_{\Theta}=\frac{\Mp}{4\phi}\left(\frac{\lambda_\phi}{3g\lambda_0} \right)^{\frac{1}{2}}<\left(  \frac{\lambda_\phi}{48\lambda_0} \right)^{\frac{1}{2}}
\ee
and the last inequality comes from the condition $\sqrt{g}\phi>\Mp$. Putting numbers we obtain that $\Delta_{\rm A}\sim 10^{-6}$ and $\Delta_{\Theta}\lesssim 10^{-7}$, so that $\Delta_{\rm R}$ can be safely neglected for numerical calculations in Eq.~(\ref{eq:kinetic}).

As for the potential in both inflationary regions, $\Theta$ and A, using the previous results we can write it as
\be
V_{\rm R}(\phi)\simeq c_{\rm R}\,V(\phi),\quad V(\phi)= \Theta^2(\phi)\,\frac{1}{4}\,\lambda_\phi\, \phi^4\,,\quad c_{\rm A}=2,\quad c_{\Theta}=1\,,
\label{eq:infpotential}
\ee
so that, in both regions, they only differ by a global factor. As the slow roll parameters do depend on ratios of the potential and its derivatives, they  will not depend on the global factor $c_{\rm R}$ and can thus be given a universal expression. So for the computation of the slow roll parameters we will just remove the global factor $c_{\rm R}$ and use $V(\phi)$ as the inflationary potential.

We can now define the inflaton $\chi$ as a field with canonical kinetic term as
\be
\mathcal L_{\rm kin}=\frac{1}{2} \partial_\mu\chi\partial^\mu\chi ,
\ee
where the change of variable $\phi \to \chi$ is done by
\be \frac{d\chi}{d\phi} \simeq  \; \left[\Theta(\phi) \left(1+ \frac{6g^2\phi^2}{\Mp^2}\;\Theta(\phi)\right) \right]^{\frac{1}{2}}\equiv f(\phi), \label{derivative-correction} 
\ee
the last equality simply being the definition of the function $f(\phi)$ for later use.
Solving the above differential equation gives the approximation
\be \chi \simeq  \Mp \sqrt{\frac{1+6g}{g}} \; \text{arcsinh}{\sqrt{g (1 + 6 g)} \,\frac {\phi}{\Mp} }, 
\ee
which, for $\sqrt{g}\phi\gtrsim \Mp$, can be inverted to get
\be
\phi\simeq \frac{\Mp}{2\sqrt{g(1+6g)}} \exp\left(\sqrt{\frac{g}{1+6g} }\, \frac{\chi}{\Mp}\right) , 
\ee
while $\phi\simeq \chi$, for $\sqrt{g}\phi\lesssim \Mp$, as in this limit the Jordan and Einstein frames should coincide. 

However, although the slow roll parameters must be computed from the inflaton potential $V(\chi)$, we will not need to use this explicit solution to obtain the inflationary parameters.
Instead, we can keep $\phi$ as the explicit variable, since performing the change of variables (\ref{derivative-correction}) in the slow roll parameters definition allows us to avoid making inevitable approximations stemming from the relationship between the fields $\phi$ and $\chi$. 
Hence, we can keep the description of the model in terms of the $\phi$ field and the potential $V(\phi)$ given in Eq.~(\ref{eq:infpotential}).\footnote{From here on we will use primes to denote derivatives of a function with respect to its functional dependence, e.g.,~$V'(\chi)=dV(\chi)/d\chi$ and $V'(\phi)=dV(\phi)/d\phi$.} In this framework the slow roll parameters can be written as~\cite{DeSimone:2008ei}
\bse \epsilon(\phi) &=& \frac{\Mp^2}{2} \left(\dfrac{V'(\chi)}{V(\chi)}\right)^2=\dfrac{\Mp^2}{2} \left(\dfrac{V'(\phi)}{V(\phi)}\right)^{2}f^{-2}(\phi) ,\\
\eta (\phi)&=& \Mp^2 \;\dfrac{V''(\chi)}{V(\chi)}=\Mp^2\;\left[ \dfrac{V''(\phi)}{V(\phi)} f^{-2}(\phi)   -\frac{V'(\phi)}{V(\phi)} \,f'(\phi)f^{-3}(\phi)\,  \right], \\
\xi^2(\phi) &=&\Mp^4\; 
\dfrac{V'(\chi)V'''(\chi)}{V^2(\chi)}=\Mp^4\;\frac{V'(\phi)}{V(\phi)}f^{-4}(\phi)\\
&&\cdot \left[ \frac{V'''(\phi)}{V(\phi)} -3 \frac{V''(\phi)}{V(\phi)} f'(\phi)f^{-1}(\phi)
+\frac{V'(\phi)}{V(\phi)} \left( 3 f^{\prime\, 2}(\phi) f^{-2}(\phi)-f''(\phi)f^{-1}(\phi)     \right)\right] ,\nonumber\ese
where the function $f(\phi)$ was defined in (\ref{derivative-correction}). Their current observational bounds are, from Ref.~\cite{Planck:2018jri}:
\be \begin{aligned}  \epsilon < 0.0044 \hspace{3.2cm} (95\% \text{C.L.}), \\
 \eta = -0.015 \pm 0.006 \hspace{1.8cm}(68\% \text{C.L.}),\\
 \xi^2 = 0.0029^{+0.0073}_{-0.0069} \hspace{2.cm}(95\% \text{C.L.}).\end{aligned}\label{slow-roll-constraint}\ee

\begin{figure}[t]
\begin{center}
\includegraphics[width = 7.8cm]{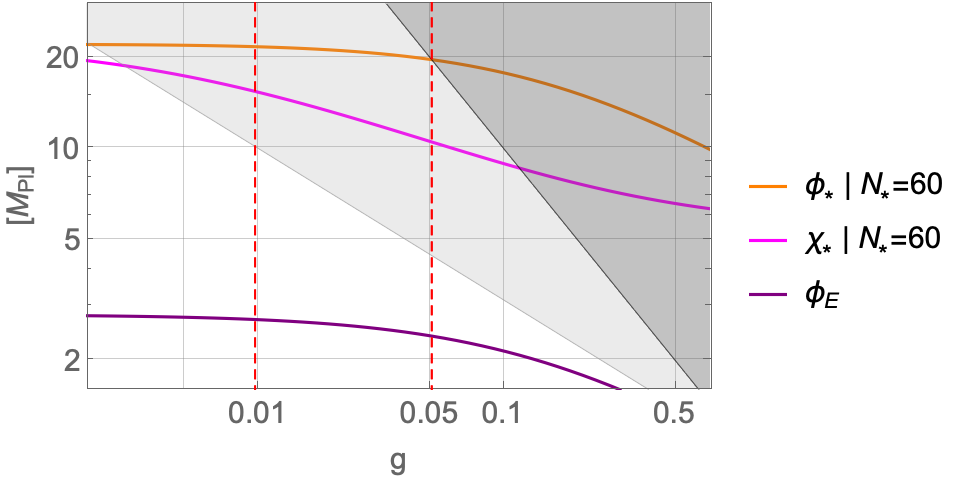}
\includegraphics[width = 7.6cm]{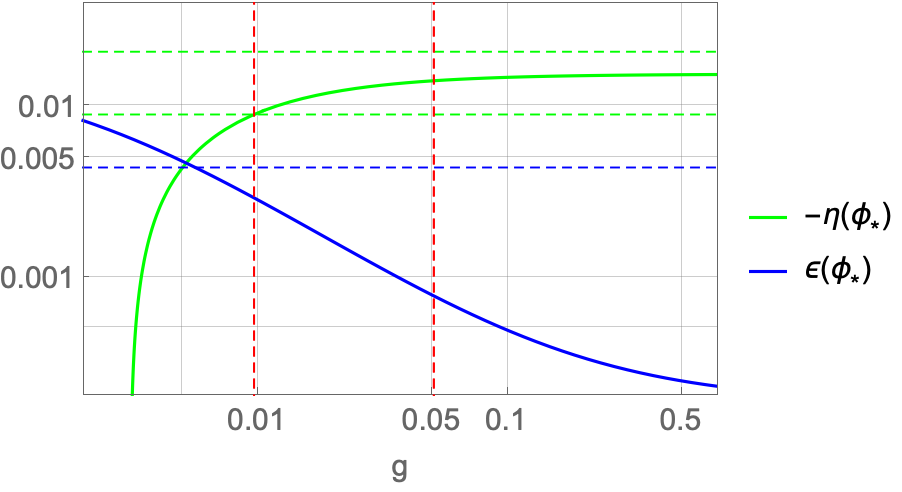}
\caption{\it Left panel: $\phi_E$, $\phi_\ast$ and $\chi_\ast$ in unit of $\Mp$, as functions of $g$. 
%We did not display $\chi_E$ as $\chi_E \simeq \phi_E$. 
The dark shading region violates the unitarity bound $\phi<\Mp/g$. The white area corresponds to Region A and the light shading one to Region $\Theta$. Right panel: slow roll parameters evaluated at the beginning of inflation and their corresponding observational bounds (dashed, matching color). The bound for $\epsilon(\phi_\ast)$ is an upper bound.}
  \label{fig:phiEStar}
\end{center}
\end{figure}

We should evaluate the slow-roll parameters at the field value $\phi_\ast = \phi(N_\ast)$ with
\be N_\ast =\frac{1}{\Mp^2}\int_{\chi_E}^{\chi_\ast} \dfrac{V(\chi)}{V'(\chi)}\; d \chi  =\frac{1}{\Mp^2}\int_{\phi_E}^{\phi_\ast} \dfrac{V(\phi)}{V'(\phi)} \;f^2(\phi)\, d \phi,
\label{eq:Nphi}
\ee 
 being $N_\ast$ the number of $e$-folds at which the reference scale exits the horizon. Here $\phi_E$, the value of $\phi$ at the end of inflation, is defined by the condition $\epsilon(\phi_E)=1$ and can be computed analytically. A plot of its dependence on $g$ is shown on the left panel of Fig.~\ref{fig:phiEStar}.
One can evaluate the integral (\ref{eq:Nphi}) to find
\be N_\ast = \dfrac{1}{4}\left[ \dfrac{(1+6g)(\phi_\ast^2-\phi_E^2)}{2\Mp^2}- 3\log\frac{\Mp^2+g\phi^2_\ast}{\Mp^2+g\phi^2_E}\right]
\label{eq:phistar}
 \ee
and then solve for $\phi_\ast\equiv\phi(N_\ast)$.\footnote{
We can solve Eq.~(\ref{eq:phistar}) for $\phi(N_\ast)$ recursively, first ignoring the logarithm for the first iteration and then inserting each solution into the next iteration (which this time contains all terms). The sequence converges quickly to the exact solution. After 3–4 iterations the relative error is already $\sim10^{-3}$ at most.}
A plot of $\phi_\ast$, for $N_\ast=60$, and its dependence on $g$ is shown in the left panel of  Fig.~\ref{fig:phiEStar}. The dark shading region is excluded as there $\phi_\ast>\Lambda_\phi\equiv\Mp/g$ and so there is a unitarity violation (see however footnote~\ref{foot}). This constraint provides an upper bound on the value of the parameter $g$ as $g\lesssim 0.0508$.\footnote{ For $g<1$ the cutoff $\Lambda_\phi$ is trans-Planckian, and from Fig.~\ref{fig:phiEStar} we can see that during inflation $\Delta \phi \simeq 20 \,\Mp$, which satisfies the so-called Lyth bound~\cite{Lyth:1996im}. Such behavior induces nonnegligible quantum gravity corrections to the potential. However, the terms induced by quantum gravity effects are suppressed, not by factors $\phi^n / \Mp^n$, but by factors $V / \Mp^4$ and $m^2 / \Mp^2$, see Sec.~2.4 of Ref.~\cite{Linde:1990flp}. Hence, as long as the inflationary potential takes sub-Planckian values and $m\ll\Mp$ (like in our model), quantum gravity effects are insignificant, regardless of the values of $g$ or $\Lambda_\phi$.}

We display, in the right panel of Fig.~\ref{fig:phiEStar}, the functions $\epsilon(\phi_\ast)$ and $\eta(\phi_\ast)$ as functions of $g$. The observational constraints (\ref{slow-roll-constraint}) provide a lower bound on the Ricci coupling as $g\gtrsim 0.0096$. When combined with the upper bound from unitarity, the allowed region in the $g$ parameter is given in the range~\footnote{One should worry about the stability, under radiative corrections, of such small values of the $g$ parameter. Contributions to the one-loop $\beta_{\xi_{\mathcal H}}$ function, in the Ricci coupling $(\xi_{\mathcal H}/2)\,h^2 R$, from the contribution of the SM fields (top quark, gauge and Higgs bosons), have been computed in Refs.~\cite{Herranen:2014cua,Espinosa:2015qea} where it is shown that the renormalization from the weak to the high scale of $\xi_{\mathcal H}$ is $\lesssim$ 20\%. In the case of  our coupling $(g/2)\, \phi^2 R$, as $\phi$ is not directly coupled to the SM fields, the $g$ running between $m$ and the high scales is suppressed by the mixing angle $\alpha$ between the fields $\phi$ and $h$ (see Sec.~\ref{sec:pheno}), so that $\beta_g\simeq 2\delta_\lambda (v^2/m^2) \beta_{\xi_{\mathcal H}}\ll \beta_{\xi_{\mathcal H}}$. In this way the running of the $g$ parameter can be safely neglected.}
\be 0.0096\, \lesssim \, g \, \lesssim\, 0.0508. \label{g-range}\ee
Finally we have found that, in the relevant region of the $g$ parameter, the parameter $\xi^2$ is $|\xi^2|\sim10^{-4}$, well in agreement with the experimental range in Eq.~(\ref{slow-roll-constraint}). 

For the allowed region  of the slow roll parameters in Fig.~\ref{fig:phiEStar}, the cosmological observables,  the scalar spectral index $n_s\simeq 1-6\epsilon(\phi_\ast)+2\eta(\phi_\ast)$, the spectral index running $n'_s\simeq 16\epsilon(\phi_\ast)\eta(\phi_\ast)-24\epsilon^2(\phi_\ast)-2\xi^2(\phi_\ast)$, and the tensor to scalar ratio $r=16\,\epsilon(\phi_\ast)$, fall inside the experimental range given by~\cite{Planck:2018jri,BICEPKeck:2021gln},
\be
n_s=0.9649\pm 0.0042 ,\hspace{1cm} n'_s=- 0.0045 \pm 0.0067,\hspace{1cm} r=0.014^{+0.010}_{-0.011}
\label{eq:nsr}
\ee  
where we have included, in the last $r$ determination, the most recent combined result from the BICEP/Keck collaboration~\cite{BICEPKeck:2021gln}. In particular, for the allowed range in the coupling $g$ (\ref{g-range}) the theory predicts
\begin{subequations}\label{eq:allowed} \begin{eqnarray}
0.96448 \;\lesssim& \; n_s \;&\lesssim \;0.96695\hspace{1.28cm} (0.96783) \label{eq:allowed-a}\\
-0.00063 \; \lesssim&\; n'_s \;&\lesssim \;-0.00019  \hspace{10mm} (-0.00005 ) \label{eq:allowed-b}\\
0.0467\; \gtrsim &\; r\;& \gtrsim \; 0.0124 \hspace{1.45cm} (0.00296)
\label{eq:allowed-c}
\ese
where the unbracketed right-hand side (RHS) bounds come from the unitarity bound, while the bracketed ones come from disregarding the latter in view of the comments in footnote~\ref{foot}~{\footnote{If we disregard the unitarity bound, see the comments in footnote~\ref{foot}, there is no upper bound on $g$ from observational constraints and the cosmological observables for larger values of $g$ asymptotically go to the RHS values in parenthesis.}. As we can see both predicted ranges (with/without considering the unitarity bound) in (\ref{eq:allowed}) nicely fit inside the allowed range in Eq.~(\ref{eq:nsr}). These results also agree with those of model $(n,p)=(2,4)$ in the recent work of Ref.~\cite{Kodama:2021yrm}, where general inflationary models with nonminimal inflaton couplings to gravity have been analyzed.

We now use the constraint on the amplitude of scalar fluctuations to find an analytical relation for the inflaton self-coupling $\lambda_\phi$, since this quantity is obtained from the potential as
\be A_s = \dfrac{1}{24\pi^2\Mp^4}\dfrac{V_\Theta(\phi_\ast)}{\epsilon(\phi_\ast)},\ee
where we are using as inflaton potential $V_{\Theta}$, as the $\phi_\ast$ line in the left panel of Fig.~\ref{fig:phiEStar} is inside the light shading region, where the inflaton potential corresponds to that in Region $\Theta$, Eq.~(\ref{eq:infpotential}). We can then compute $\lambda_\phi$ as 
\be \lambda_\phi=96\,\pi^2\, g^2 \, A_s \, \epsilon(\phi_\ast) \left(1+\dfrac{\Mp^2}{g\phi_\ast^2} \right)^2. \label{inflaton-self-coupling}\ee
Using the observed value of $A_s$ from Ref.~\cite{Planck:2018jri}, $A_s^{\rm obs}=2.2\cdot10^{-9}$, as well as the values of $\epsilon(\phi_\ast)$ and $\phi_\ast$ from Fig.~\ref{fig:phiEStar}, we plot, in the left panel of Fig.~\ref{fig:lambdaphi} the parameter $\lambda_\phi$ as a function of the Ricci coupling $g$. Notice that, inside the allowed region in Eq.~(\ref{g-range}), we obtain $\lambda_\phi \sim 10^{-12}$ as postulated earlier.
%the chosen value in Fig.~\ref{fig:submanifold}
%
\begin{figure}[t]
\begin{center}
\includegraphics[width = 7cm]{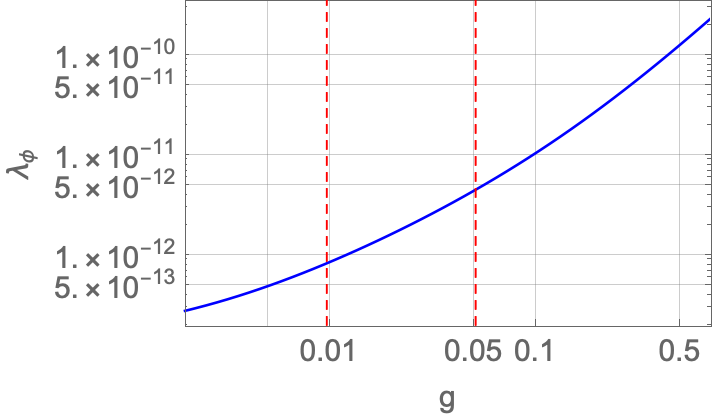}
\hspace{2mm}
\includegraphics[width = 8.1cm]{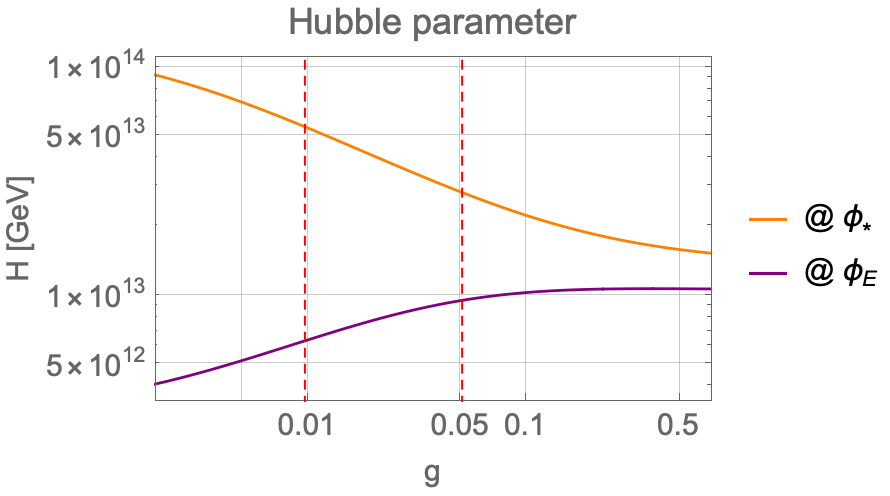}
 \caption{\it Left panel: The inflaton self-coupling $\lambda_\phi$ as a function of $g$. Right panel: The Hubble parameter $H$ at the end of inflation $H(\phi_E)$ and for the number of $e$-folds $N_\ast$, $H(\phi_\ast)$ as functions of $g$. In both plots the vertical red lines show the range for $g$ where the slow roll cosmological observables and unitarity constraints are met.} 
\label{fig:lambdaphi}
\end{center}
\end{figure}

Finally we can compute the Hubble parameter during inflation $H(\phi)$. From the Friedmann equation we have that the energy density of the inflaton reads as
$ \rho(\phi) = 3\Mp^2 H^2(\phi)$.
Since we are assuming a slow roll evolution of the inflaton, we can neglect the kinetic part in the energy density and consider $\rho(\phi)  \simeq V_{\rm R}(\phi)$. Therefore in Region~A, i.e.~around the end of inflation and, in particular, at $\phi_E$, $H_E\equiv H(\phi_E)$,
\be H(\phi)=  \dfrac{\Mp}{g} \sqrt{\dfrac{\lambda_\phi}{6}}\left(1+\dfrac{\Mp^2}{g\phi^2}\right)^{-1}.\label{Hubble-exact}\ee
For the value $\phi=\phi_\ast$, in Region~$\Theta$, this further simplifies to
\be H_\ast \equiv H(\phi_\ast)= 2^{3/2}\pi \Mp \sqrt{\epsilon(\phi_\ast)A_s^{\rm obs}}. \ee

We plot in the right panel of Fig.~\ref{fig:lambdaphi} the Hubble parameters at the end of inflation, i.e.~for $\phi=\phi_E$, $H_E$, and at the beginning of inflation, for a number of $e$-folds $N_\ast=60$, $H_\ast$. As we can see from the right panel of Fig.~\ref{fig:lambdaphi}, for the lower bound of $g$, $g\simeq 0.01$, the Hubble parameter changes, between $\phi_E$ and $\phi_\ast$, by one order of magnitude, from $H_\ast\simeq 5.5\cdot 10^{13}$~GeV, to $H_E\simeq  6.4\cdot 10^{12}$~GeV. On the other hand,  for the upper bound of $g$, $g\simeq 0.05$, the Hubble parameter changes by a few, from $H_\ast\simeq 2.8\cdot 10^{13}$~GeV  to $H_E\simeq  10^{13}$~GeV. As we can see the absolute upper bound on the Hubble parameter in our model is $H_\ast\lesssim 5.5 \cdot 10^{13}$~GeV, or equivalently an inflation scale $V^{1/4}_{\Theta}(\phi_\ast)\lesssim 1.5 \cdot10^{16}$~GeV, in agreement with the observational upper bounds from the Planck collaboration, Ref.~\cite{Planck:2018jri}, given by
\be
H_\ast^{\rm obs}< 6 \cdot 10^{13}\ \textrm{ GeV},\hspace{1.5cm} (V_\ast^{\rm obs})^{1/4}<1.6\cdot 10^{16}\ \textrm{GeV}\quad (95\%\ \textrm{C.L.}).
\ee
Consequently our model, independently of the value of $m$, is a high scale inflation model, where the Hubble parameter does depend on the value of $g$ and is maximized for its lower bound.

\section{Gauge field production}
\label{sec:field-production}

In this section we will consider the generation of fully helical hypermagnetic fields that will be transformed into baryon asymmetry at the EWPT. Of course all modes produced during inflation, except the last modes that exit the horizon at the end of inflation reentering the horizon at the onset of reheating, get diluted~\cite{Anber:2015yca}. For that reason we will be concerned by the last $e$-folds of inflation, corresponding to the inflaton value $\phi\simeq\phi_E$, well inside Region A with a potential given by Eq.~(\ref{eq:infpotential}).

 We need a source of $\mathcal{CP}$-violation and we will assume the $\mathcal{CP}$-odd dimension-five operator given by
\be
   \sqrt{-g}\, \mathcal L_{\cancel{CP}}=-\frac{1}{4}\frac{\phi}{\tilde f_\phi}Y_{\mu\nu}\widetilde Y^{\mu\nu},
\label{eq:CPV}
\ee
where $Y^{\mu\nu}$ is the field strength of the hypercharge gauge field $Y^\mu$, and $\widetilde Y^{\mu\nu}=\frac{1}{2}\epsilon^{\mu\nu\rho\sigma}Y_{\rho\sigma}$ its dual tensor. This Lagrangian term is scale invariant (it does not change when going from the Jordan to the Einstein frame), and should appear in the effective theory after integrating out some UV physics, heavier than the inflaton field. A possible and simple UV completion, with a heavy vectorlike fermion coupled to the field $\phi$ by a $\mathcal{CP}$-violating Yukawa coupling, and giving rise to Eq.~(\ref{eq:CPV}) is presented in App.~\ref{sec:UV}. However, in the rest of this paper, we will be agnostic about the origin of such a term as it may arise from a great variety of models.

In addition to this, by virtue of the minimum condition in Region A, Eq.~(\ref{eq:minA}), the Higgs background value is nonzero (it is anchored to the value of the field $\phi$), and so the electroweak symmetry is broken, meaning we are producing ordinary $U(1)_{\rm EM}$ magnetic fields, as the $Z$ fields are very massive for those values of the background field $h$, and hence much harder to produce. 
In this way the $\mathcal{CP}$-violating term in the broken phase, at the end of inflation, will look like
\be
\sqrt{-g}\,\mathcal L_{\cancel{CP}}=-\frac{1}{4}\frac{\phi}{f_\phi}F_{\mu\nu}\widetilde F^{\mu\nu},
\label{eq:phiFFtilde}
\ee
where $F_{\mu\nu}$ is the electromagnetic field strength, corresponding to the photon field $A_\mu$, and we have rescaled the constant $f_\phi$ as
\be
f_\phi=\frac{ \tilde f_\phi}{\cos^2\theta_W},
\ee
where $\theta_W$ is the EW angle.

At reheating, $h$ will drop at its potential minimum at zero, because of the sudden dominance of the thermal correction terms, and we will recover the symmetric phase. This is a necessary requirement for a successful baryogenesis as the helical fields participating in the chiral anomaly must belong to the unbroken electroweak sector. This is because, in the symmetric phase of the electroweak plasma, the chiral anomaly induces the phenomenon where variations in baryon, $N_B$, and lepton, $N_L$, number can be induced by changes in the $SU(2)_L$ Chern-Simons number $N_{\rm cs}$ and/or $U(1)_Y$ hypermagnetic helicity $\mathscr H_Y$ as
\be \Delta N_B = \Delta N_L = N_g \left(  \Delta N_{\rm cs} - \dfrac{g^2_Y}{16\pi^2}\; \Delta \mathscr H_Y\right), \label{anomaly}\ee
where $N_g=3$ is the number of fermion generations and $g_Y$ the $U(1)_Y$ coupling. This equation tells us that any change in the $U(1)_Y$ helicity leads to a fermion asymmetry, in particular when projecting ordinary magnetic fields into hypermagnetic fields at the end of inflation. However, as long as $T\gtrsim 160$~GeV, the electroweak sphalerons are in equilibrium in the plasma, hence any fermion asymmetry gets washed out in less than a Hubble time, and only the $U(1)_Y$ helical fields remain.

As the $U(1)_Y$ helical magnetic fields participate in the baryogenesis process~\cite{Kamada:2016cnb}, while $U(1)_{\rm EM}$ helical magnetic fields are produced at the end of inflation, the projection of the latter on the former must be taken into account with a factor~\footnote{Bold characters stands for 3D vectors in space.}
\be \bm{A}_Y = \cos{\theta_W}\, \bm{A}, \hspace{1cm} \mathscr H_Y = \cos^2{\theta_W}\, \mathscr H\,. \label{Weinberg} \ee
The $Z$ fields can also project onto $U(1)_{\rm Y}$ fields but, as stated before, we will ignore this contribution as they were too heavy to be produced.

Moreover after inflation, in Region B, the Higgs will start relaxing to its minimum and, if some conditions are satisfied, the Higgs could source extra helical magnetic fields, as was studied in Ref.~\cite{Cado:2021bia}, and eventually overproduce the BAU from the induced coupling
\bse
\sqrt{-g}\,\mathcal L_{\cancel{CP}}&=&-\frac{1}{4}\sqrt{2\delta_\lambda}\frac{|\mathcal H|^2}{m f_\phi} \cos^2\theta_WY_{\mu\nu}\widetilde Y^{\mu\nu}\equiv 
-\frac{1}{4}\frac{|\mathcal H|^2}{\Lambda_\mathcal H^2} Y_{\mu\nu}\widetilde Y^{\mu\nu}\, , \\
 \Lambda_{\mathcal H}&\simeq& 8.7\cdot 10^{13}\ \textrm{GeV} \left(\dfrac{m}{10^{10} \text{ GeV}} \right)^{\frac{1}{2}}\left(\dfrac{\delta_\lambda}{0.15} \right)^{-\frac{1}{4}}\left(\dfrac{f_\phi}{0.1\, \Mp} \right)^{\frac{1}{2}},
\ese
where we have used the minimum condition in Eq.~(\ref{eq:relB}). Nevertheless the required relaxation mechanism found in Ref.~\cite{Cado:2021bia} should not work under the present conditions, because one necessary condition for the Higgs relaxing into the hypermagnetic fields is not fulfilled here, namely that $h \gtrsim 3 \cdot 10^{15}$~GeV at the end of inflation. In fact in our model, as a consequence of the definition (\ref{eq:regionB}) of Region B, $$h\lesssim \frac{2\sqrt{2}m}{\left(\lambda_\phi \lambda_0\right)^{\frac{1}{4}}}\left(\frac{\delta_\lambda}{\lambda_0}  \right)^{\frac{1}{2}} \lesssim \; 4 \cdot 10^{13}\ \textrm{GeV},$$ 
where we chose the parameters configuration that maximizes the RHS bound.
Therefore a significant production of helical magnetic fields from the Higgs decay after inflation is unlikely, and we can consider $\mathcal L_{\cancel{CP}}$ as inactive in Region B.

\subsection{Helical magnetic fields}
\label{sec:backreactionless-helicity}

As stated earlier, we will mainly be interested in values of the inflaton field near the end of inflation, i.e.~$\phi\simeq \phi_E$, which means that $\phi$ and $h$ evolve in Region A of the potential, given by (\ref{eq:infpotential}).
For such values of $\phi$, the differential equation~(\ref{derivative-correction}) admits the simple solution $\phi\simeq \chi+\mathcal O(g)$. We recall that $\chi$ is the true inflaton field with canonical kinetic term and action
\begin{align}
S\simeq & \int d^4x \left[ \sqrt{-g} \left(-\dfrac{\Mp^2}{2}R +  \frac{1}{2}\partial_\mu \chi\partial^\mu \chi -\dfrac{1}{4} F_{\mu\nu}F^{\mu\nu} - V(\chi) \right)-\dfrac{\chi}{4f_\phi}\;F_{\mu\nu}\tilde{F}^{\mu\nu}  \right]\nonumber \\
&+\int d^4x \;\sqrt{-g}\; i\,\bar\psi \cancel{\mathcal D}  \psi ,
\label{eq:actionfinal}
\end{align}
where we have included the interaction of fermionic currents, corresponding to charge $Q$ fermions, with the electromagnetic fields (encoded in the covariant derivative $\mathcal D_\mu\equiv \partial_\mu-e Q A_\mu$), the $\mathcal{CP}$-violation term and the inflaton potential
\be V(\chi)= \dfrac{\lambda_\phi}{2}\; \left(1+\dfrac{g\chi^2}{\Mp^2}\right)^{-2}\chi^4 \label{potential-baryo}.\ee

Varying the action (\ref{eq:actionfinal}) with respect to $A_\mu= (A_0,\bm{A})$ leads to the gauge equations of motion in the radiation gauge, $A_0=0$ and $\bm{\nabla} \cdot \bm{A} =0$,
\be \left( \dfrac{\partial^2}{\partial \tau^2}-\nabla^2-\dfrac{ \chi'}{f_\phi}\; \bm{\nabla} \times \right) \bm{A}=0 ,\ee
where $\tau$ is the conformal time, defined by $g_{\mu\nu}=a^2(\tau)\,\eta_{\mu\nu}$, and we assume a homogeneous inflaton with only zero mode, $\chi(\tau,\bm{x})=\chi(\tau)$. 
Unless otherwise specified, all quantities and fields are \textit{comoving}. 
 
During the inflationary period one has~\footnote{As for fields, we denote the derivative with respect to $\tau$ with a prime and the derivative with respect to the cosmic time $t$ with a dot, e.g.~$\chi'=d\chi/d\tau$ and $\dot{\chi}=d\chi/dt$.}
\be \chi' = \dot{\chi}a\simeq-\dfrac{\dot{\chi}}{\tau H(\chi)}\,, 
\label{eq:chiprime}
\ee
and the field velocity $\dot\chi$ is computed from the equation of motion for the inflaton obtained from the action (\ref{eq:actionfinal})
\be \ddot{\chi} +3H(\chi)\,\dot{\chi} + V'(\chi) = \dfrac{\bm{E} \cdot \bm{B}}{a^4f_\phi} ,\label{eom-inflaton} \ee
where we have used that $F_{\mu\nu}\tilde{F}^{\mu\nu}=-4\, \bm{E} \cdot \bm{B}$.
From the slow roll conditions, we can neglect $\ddot\chi$, since 
\be \dfrac{\ddot\chi}{3 H \dot\chi} = \dfrac{\epsilon - \eta}{3}, \ee
where we are already neglecting the backreaction of the generated magnetic field on the inflaton (i.e.~we are neglecting the term $\bm{E} \cdot \bm{B}/ a^4f_\phi$ in Eq.~(\ref{eom-inflaton})), a hypothesis that will be self-consistently checked a posteriori (see Sec.~\ref{sec:backreaction}).
During the last $e$-folds of inflation, our model provides $|\epsilon - \eta | \,/\,3<0.1$. 
Hence, we obtain 
\be  \chi^\prime=\frac{\Mp^2}{\tau}\,\frac{V^\prime(\chi)}{V(\chi)},
\label{eq:chiprimeequal}
\ee
where we have made use of Eq.~(\ref{eq:chiprime}).

We now quantize the gauge field $\bm{A}$ in momentum space
\be \bm{A}(\tau , \bm{x}) \, = \, \sum_{\lambda = \pm} \int \frac{d^3 k}{(2\pi)^3} \, \left [\bm{\epsilon}_\lambda(\bm{k}) \, a_\lambda(\bm{k}) \, A_\lambda(\tau, \bm{k}) \, e^{i \bm{k} \cdot \bm{x}} + \, \text{h.c.} \right]  ,
 \ee
where $\lambda = \pm$ is the photon polarization and $a_\lambda(\bm{k})$ ($a_\lambda^\dagger(\bm{k})$) are annihilation (creation) operators that fulfill the canonical commutation relations
\be
[a_\lambda(\bm{k}),a_{\lambda'}^\dagger(\bm{k}')]=(2\pi)^3\delta_{\lambda\lambda'}\delta^{(3)}(\bm{k}-\bm{k'})\,.
\ee
The polarization vectors $\bm{\epsilon}_\lambda(\bm{k})$ satisfy the conditions~\footnote{A simple realization can be given in terms of a real basis with the orthonormal vectors $(\bm{k}/|\bm{k}|, \bm{e}_i)$, ($i=1,2$), such that $\bm{k}\cdot \bm{e}_i=\bm{e}_1\cdot \bm{e}_2=0$ and $\bm{e}_i\cdot \bm{e}_i=1$, with $\bm{\epsilon}_\lambda\equiv(\bm{e}_1+i\lambda \bm{e}_2)/\sqrt{2}$, from where identities (\ref{eq:identitiespol}) follow.}
\be\begin{aligned} \bm{k} \cdot \bm{\epsilon}_\lambda(\bm{k}) &= 0\, , \hspace{2cm} &   \bm{k} \times \bm{\epsilon}_\lambda(\bm{k})  &= - i \lambda k \, \bm{\epsilon}_\lambda(\bm{k})\, , \\ \bm{\epsilon}^*_{\lambda'}(\bm{k}) \cdot \bm{\epsilon}_\lambda(\bm{k}) &= \delta_{\lambda \lambda'}\, , & \bm{\epsilon}^*_{\lambda}(\bm{k}) &= \bm{\epsilon}_\lambda(-\bm{k}) \, ,\end{aligned}
\label{eq:identitiespol}
\ee
where $k \equiv |\bm{k}|$.
The equation of motion for the modes yields
\be \dfrac{\partial^2A_\lambda}{\partial \tau^2} +k \left(k+\lambda \, \dfrac{2\xi}{\tau}\right)A_\lambda =0 ,
\label{eq:Apm}
\ee
which is the Coulomb wave equation, with
\be \xi = \dfrac{\Mp^2}{2f_\phi}\,\dfrac{V'(\chi)}{V(\chi)}=\dfrac{\Mp}{f_\phi}\sqrt{\dfrac{\epsilon(\chi)}{2}}  > 0, \label{xi-parameter-def} \ee
where Eq.~(\ref{eq:chiprimeequal}) has been used and $\epsilon(\chi)$ is the slow-roll parameter. Let us mention that, even if the first equality in Eq.~(\ref{xi-parameter-def}) looks model dependent, as it depends on the potential and its derivative, in fact it is very model independent because the last relation only relies on the slow roll regime of the inflationary potential, and $\epsilon(\chi)\simeq 1$ at the end of inflation. We have done a self-consistency check by comparing the numerical results of both expressions and found no significative difference, see below.

As already emphasized, all modes produced during inflation will get diluted, except the last mode that exits the horizon right before the end of inflation. This mode reenters the horizon at the onset of reheating and is the source for the BAU.
Hence, it is only necessary to consider the mode produced at $\phi_E\simeq \chi_E$, for which $\epsilon(\chi_E)\simeq 1$, and hence, using the last equality in Eq.~(\ref{xi-parameter-def}) we obtain for $\xi$ the constant value
\be \xi  \simeq \dfrac{\Mp}{\sqrt{2}f_\phi}\,. \label{xi-parameter-approx} \ee
We numerically checked that this approximation coincides with the exact solution:
\be \xi =\left[\dfrac{\Mp^2}{2f_\phi}\,\dfrac{V'(\phi)}{V(\phi)} \;\; \dfrac{d\phi}{d\chi}\right]_{\phi=\phi_E} \label{xi-parameter-exact} ,\ee
where equations (\ref{derivative-correction}) and (\ref{potential-baryo}) should be used. A plot of $\xi$ as a function of $g$ is shown in the top left panel of Fig.~\ref{xi-parameter} (solid lines) where we compute the exact solution in Eq.~(\ref{xi-parameter-exact}). As we can see the values of $\xi$ are nearly constant with respect to $g$, a behavior that is well approximated by the expression of $\xi$ in Eq.~(\ref{xi-parameter-approx}). A plot of $\xi$ as a function of $f_\phi$ is displayed in the bottom left panel of Fig.~\ref{xi-parameter} (solid lines) for the range of values of $g$ allowed by the final inflationary analysis. As all results of the following sections are very sensitive to the precise value of the parameter $\xi$ we will use next the exact expression for $\xi$ in all numerical calculations.

Notice that in this section we are neglecting, in the RHS of Eq.~(\ref{eq:Apm}), the possible effect of the fermion currents $eQ\,\bm{J}_\psi$, appearing in the action Eq.~(\ref{eq:actionfinal}), and in particular their backreaction on the produced helical magnetic fields. This phenomenon, known as the Schwinger effect, will appear for sufficiently strong magnetic fields, hence for large (small) enough values of the $\xi$ ($f_\phi$) parameter. In this section we will consider the case of backreactionless fermion currents (i.e.~small values of $\xi$) and will devote Sec.~\ref{sec:schwinger} (where those values will be quantified) to the analysis of the Schwinger effect and its backreaction on the helical magnetic fields.

The general solution of (\ref{eq:Apm}) is
%\be A_\lambda = \dfrac{1}{\sqrt{2k}} \left[i F_0 (\lambda\xi, -k \tau)+G_0 (\lambda\xi, -k \tau) \right] \ee
\be A_\lambda = \dfrac{i F_0 (\lambda\xi, -k \tau)+G_0 (\lambda\xi, -k \tau)}{\sqrt{2k}} \ee
where $F_0$ and $G_0$ are, respectively, the regular and irregular Coulomb wave functions with index 0~\cite{Anber:2006xt}.\footnote{See also Sec.~14 of Ref.~\cite{book:Abramowitz}.}  At early times, the above solution has the asymptotic behavior that corresponds to the Bunch-Davies vacuum of the modes. In fact during inflation, where $\epsilon(\chi)\ll 1$, we obtain, using Eq.~(\ref{xi-parameter-def}), that $\xi\ll 1$  and therefore $|k\tau|\gg 2\xi $, so we can write $A_\lambda \propto e^{-i k \tau}$.
However, at the end of inflation $\epsilon(\chi_E)\simeq 1$ and so we can have $\xi\gtrsim 1$. Then, only one mode develops both parametric and tachyonic instabilities for $k\simeq k_c$ where 
\be
k_c=2\xi a_E H_E, \qquad a_E=a(\tau_E), \label{last-mode}
\ee
while the other one stays close to its vacuum. As in our model $\xi>0$, and during inflation $\tau<0$, the mode exhibiting the instability is the one with the $\lambda=+$ polarization.
For late times, $k\ll k_c$ (i.e.~$|k\tau|\ll 2\xi $), $F_0$ can be neglected and the growing mode solution can be approximated by~\cite{book:Abramowitz,Anber:2006xt,Anber:2015yca}

\be A_\lambda \simeq \dfrac{G_0}{\sqrt{2k}}\simeq \dfrac{1}{\sqrt{2k}}\left(\dfrac{k}{2\xi a_E H_E}\right)^\frac{1}{4} \exp\left\{\pi \xi-2\sqrt{\dfrac{ 2\xi k}{a_E H_E}}\right\} . \label{Amplified-mode}
\ee

Another assumption in this solution is that $H(\chi)\simeq H_E$ during the last $e$-folds of inflation. 
As we have seen that the $g$ dependence of $\xi$ is mild, the main $g$ dependence of all our predictions from here on, will arise only from the $g$ dependence of the Hubble parameter $H_E$, see the right panel of Fig.~\ref{fig:lambdaphi}.
Moreover, from the approximated value of the  $\xi$ parameter in Eq.~(\ref{xi-parameter-approx}) we can see that $\xi$ can be traded for the value of the parameter $f_\phi$ such that $\xi\gg 1$ corresponds to $f_\phi\ll \Mp$.
Moreover, as we see from the explicit solution in Eq.~(\ref{Amplified-mode}), there is an exponential magnification for large values of $\xi$. However, as shown later on in Sec.~\ref{sec:backreaction}, for very large values of $\xi$, the backreaction on the inflation dynamics from magnetic fields cannot be neglected, which will lead to upper bounds on the values of $\xi$, or correspondingly to lower bounds on the values of $f_\phi$.

%\subsection{Helicity, energy, correlation length}
Assuming homogeneity in momentum space, the comoving $U(1)_{\rm EM}$ helicity is by definition 
\be \mathscr{H} = \lim_{V\to\infty}\dfrac{1}{V}\int_Vd^3x \;\langle\bm{A} \cdot \bm{B}\rangle =  \int_0^{k_c}d k \, \frac{k^3 }{2\pi^2} \left(|A_+|^2-|A_-|^2\right) \label{ABdef}, \ee
where $\langle\cdot\rangle$ is the expectation value of quantum fields and the integral $V^{-1}\int_V d^3x$ is the spatial average, which is trivial for space independent quantities.
Since the magnetic fields are maximally helical, we can neglect one mode and set the other one to (\ref{Amplified-mode}). We cut off the integral at the last mode to exit the horizon given by (\ref{last-mode}). The resulting computation gives the amount of comoving helicity at the time of the end of inflation, as
\be \mathscr{H}\simeq \dfrac{45}{2^{15}}\; \dfrac{a_E^3H_E^3}{\pi^2\xi^4} \; e^{2\pi \xi}, \label{Helicity-analytical}\ee
where we have used the approximation for $\xi\gg 1$.\footnote{\label{foot:app}In fact, we have found that the approximation is valid up to $\mathcal O(e^{-8\xi})$ terms, so that it is good enough for $\xi\gtrsim 2$-3.}

On the right panels of Fig.~\ref{xi-parameter}, we display the magnetic helicity in function of $g$ (top panel) and $f_\phi$ (bottom panel). The exponential regime given by Eq.~(\ref{Helicity-analytical}) is shown by solid lines for values $f_\phi\gtrsim 0.15\, \Mp$ ($\xi\lesssim 4.7$). For solid lines with $f_\phi\lesssim 0.15\, \Mp$, and dotted lines, the backreaction of fermion currents on the magnetic fields (Schwinger effect) cannot be neglected, as we will see in the next section where we will continue our comments on these plots.

\begin{figure}[htb]
\begin{center}
\includegraphics[width = 8.5cm]{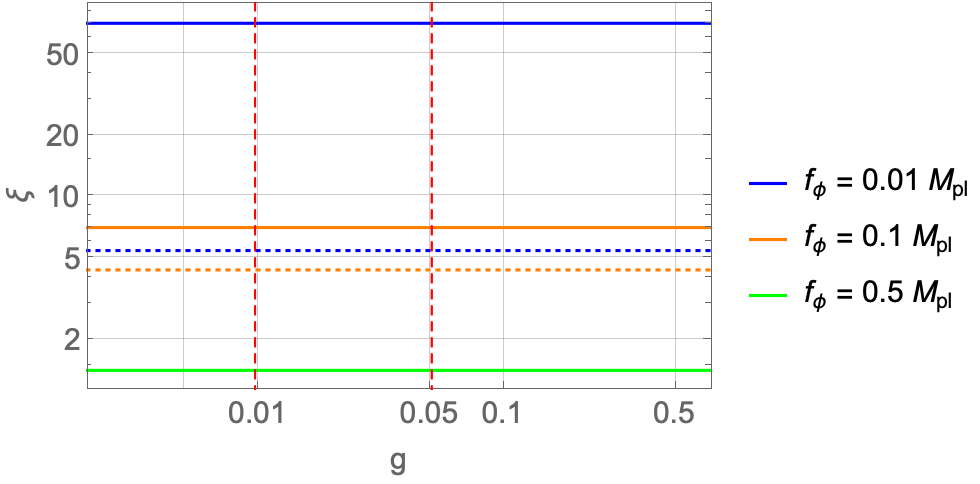}\hspace{3mm}
\includegraphics[width = 6.6cm]{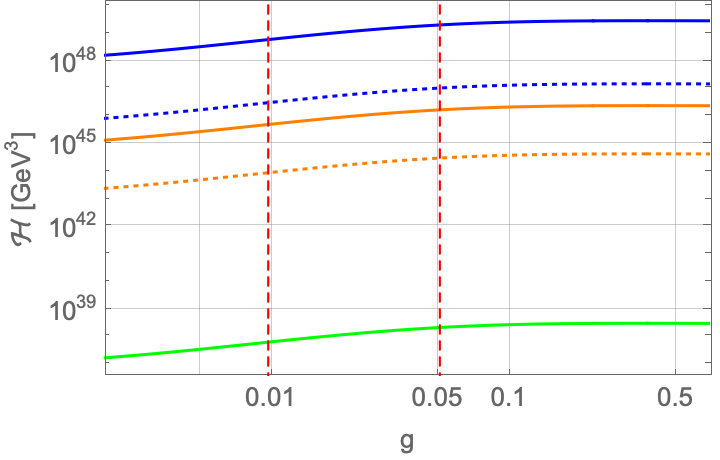}
\includegraphics[width = 6.4cm]{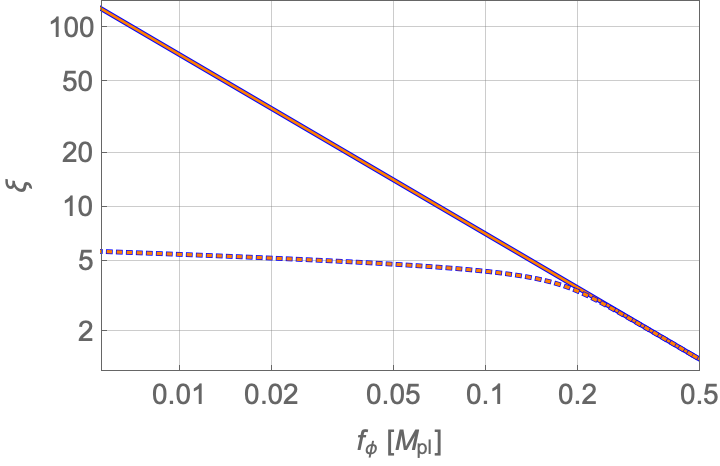}\hspace{2mm}
\includegraphics[width = 8.6cm]{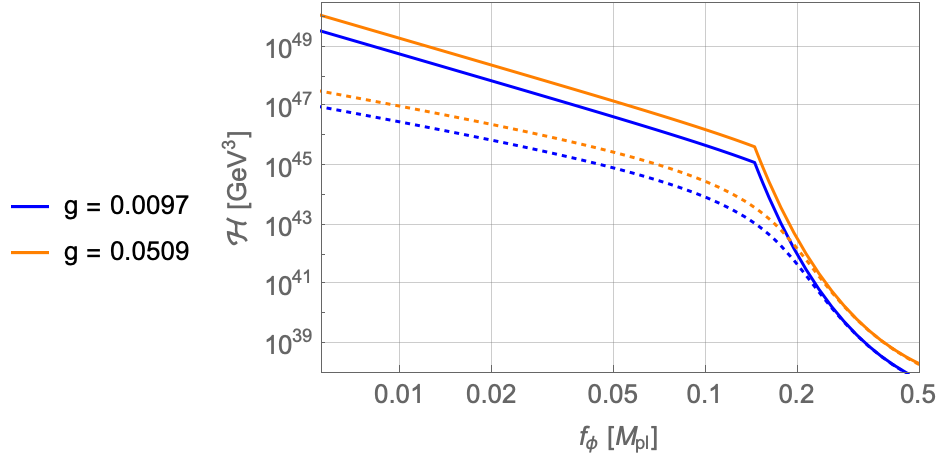}
 \caption{\it Top panels: the $\xi$ parameter (left panel) and produced helicity at the end of inflation (right panel) as a function of $g$ for various values of $f_\phi$. The vertical red lines display the range for $g$ where the inflation model is valid. Bottom panels: the same as top panels but as functions of $f_\phi$ in the same interval of values of $g$. In all panels solid lines correspond to the Schwinger effect maximal estimate while dashed lines are the equilibrium estimate. In the bottom left panel, blue and orange lines overlap since the result is insensitive to $g$.
\label{xi-parameter}}
\end{center}
\end{figure}

The comoving $U(1)_{\rm EM}$ energy density in the magnetic and electric fields are similarly computed as
\begin{subequations} \begin{eqnarray}
\rho_{B} &\equiv&  \lim_{V \to \infty} \frac{1}{2 \, V} \int_V d^3x \, \langle\bm{B}^2\rangle = \int_0^{k_c} d k \,  \frac{k^4}{4 \pi^2}\left(|A_+|^2 + |A_-|^2\right), \\
\rho_{E} &\equiv&  \lim_{V \to \infty} \frac{1}{2 \, V} \int_V d^3x \, \langle\bm{E}^2\rangle = \int_0^{k_c} d k \,  \frac{k^2}{4 \pi^2}\left(|\partial_\tau A_+|^2 + |\partial_\tau A_-|^2\right).\end{eqnarray} \end{subequations}
Using the (backreactionless) value (\ref{Amplified-mode}) for $A_+$, and neglecting the other mode, we similarly obtain the analytical solutions at the end of inflation, for $\xi\gg 1$ (see footnote~\ref{foot:app})
\begin{subequations} \label{EM-energy-densities} \begin{eqnarray}
\rho_{B} &\simeq&  \dfrac{315}{2^{18}}\; \dfrac{a_E^4H_E^4}{\pi^2\xi^5}\; e^{2\pi \xi} , \label{magnetic-energy}  \\
\rho_{E} &\simeq&  \dfrac{63}{2^{16}}\; \dfrac{a_E^4H_E^4}{\pi^2\xi^3}\; e^{2\pi \xi}  = \frac{4\xi^2}{5} \; \rho_B.  \label{electric-energy} \end{eqnarray} \end{subequations}
Hence the total comoving electromagnetic energy density is
\be \rho_{\rm EM}=\rho_E+\rho_B\simeq\rho_B \left( 1+ \frac{4\xi^2}{5}\right). \label{electromagnetic-energy}\ee

Finally, we will also need the correlation length of the magnetic field which can be estimated as~\cite{Durrer:2013pga}
\be \ell_B  =  \frac{2 \pi}{\rho_B}\int_0^{k_c} d k \,  \frac{k^3}{4 \pi^2}\left(|A_+|^2 + |A_-|^2\right) .\label{magnetic-correlation-length-def} \ee
Likewise its analytical solution at the end of inflation is given by
\be \ell_{B}\simeq \dfrac{8}{7} \dfrac{\pi\,\xi}{a_E H_E} , \label{magnetic-correlation-length} \ee

Note that the above three quantities are comoving and apply to ordinary electromagnetic field, while we will denote in subsequent sections their equivalents for the hypercharge $U(1)_Y$ in the symmetric phase with the index $Y$. The corresponding physical quantities are given by $\mathscr{H}^{\rm ph}=\mathscr{H} / a^3$, $\rho^{\rm ph}=\rho / a^4$, $\ell_B^{\rm ph}=a\ell_B$. Finally, we will conventionally set $a_E=1$.

To close this section, we would like to underline that maximally helical fields, $\bm{E}(\bm{k})$ and $\bm{B}(\bm{k})$, in (Fourier transformed) momentum space are collinear as, using the identities (\ref{eq:identitiespol}), one can easily check that both are proportional to $\bm{\epsilon}_\lambda(\bm{k})$. Besides, these fields in configuration space are, using the approximation (\ref{Amplified-mode}), (almost) collinear. %Nonetheless, in what follows for the backreactionless case, we have not assumed collinearity.
In fact, one can compute, using our approximated solution for the backreactionless solution, the angle $\theta$ measuring the collinearity of the electric and magnetic fields, as
\be
\cos\theta=\frac{\langle\bm{E}\cdot \bm{B}\rangle}{|E|\cdot|B|} ,
\ee
where we define~\footnote{Hereafter in this paper we are skipping the space average, as all background quantities are homogenous and so $\lim_{V\to\infty}\frac{1}{V}\int_Vd^3x=1$.}
\be
 |E|\equiv\sqrt{\langle \bm{E}^2  \rangle }\,,\hspace{2cm} |B|\equiv\sqrt{\langle \bm{B}^2  \rangle }\,. \label{abs-EB-def}
\ee
Using Eqs.~(\ref{EM-energy-densities}) and (\ref{EB}) we obtain, for $\xi\gg 1$
\be
\cos\theta\simeq\frac{3\sqrt{5}}{7}\simeq 0.958,
\ee
which corresponds to the angle $\theta\simeq 0.0016\pi$. As a result we have proved that the fields $\bm E$ and $\bm B$ are (almost) collinear, a property that will be used when applying the Schwinger effect in the next section. 

\subsection{Schwinger effect  \label{sec:schwinger}}
In the presence of strong gauge fields, i.e.~for $\xi\gg 1$, fermions charged under the gauge group are produced by the backreaction of gauge fields which source the fermion equations of motion. The corresponding currents can then, in turn, backreact on the produced gauge fields and change their (so-called backreactionless) solutions. This phenomenon is called the Schwinger effect and we will consider it in this section. Moreover, the fermions produced by this effect are at the origin of another phenomenon, called the \textit{chiral plasma instability}, that we will study in Sec.~\ref{sec:CPI}.

As the Higgs VEV is different from zero at the end of inflation, as we have already explained, the EW gauge bosons are massive and the system is in the broken phase. Massless gauge bosons are the photons, and we can consider the theory of electrically charged fermions in the presence of the $U(1)_{\rm EM}$ gauge group.
In the case of a Dirac fermion with mass $m$ and electric charge $Q$, the produced current satisfies the Ohm's law $\mathbf J=\sigma \mathbf E$, where $\sigma$ is the Schwinger conductivity given, for collinear $\bm{E}$ and $\bm{B}$ fields, by~\cite{Gorbar:2021zlr} 
\be
\sigma=\tr \frac{|eQ|^3}{6\pi^2}\frac{|B|}{a^2H}\;\coth\left(\frac{\pi|B|}{|E|} \right)\exp\left(-\frac{\pi m^2 a^2}{|eQ|\,|E|}\right),
\label{eq:sigma}
\ee
where the trace runs over all charged fermions $\psi_i$, with mass $m_i$ and charge $Q_i$, such that $\pi m_i^2\ll |eQ_i||E|$, and $e(\mathcal Q)=\sqrt{4\pi\alpha(\mathcal Q)} \simeq 0.33$ is the electromagnetic gauge coupling at the characteristic scale of Schwinger pair production $\mathcal Q\simeq \rho_{\rm EM}^{1/4}$~\cite{Gorbar:2021zlr}. 
However, as the Higgs VEV is suppressed with respect to the classical value of $\phi$, see Eq.~(\ref{eq:minA}), while Yukawa couplings for first and second generation fermions, are small, their corresponding square masses are much smaller than typical values of the produced electric field, and so we can make the simple reasonable approximation that only the first and second generation fermions are massless and contribute to the conductivity (\ref{eq:sigma}).

We have to stress here that, in spite of the fact that the Higgs VEV is large after inflation
$h_E\simeq 10^{15}$ GeV, which yields large masses $m_f$ for fermions, as the typical values of the produced electric fields are also large, typically $|E|\simeq h_E^2$, the small values of the Yukawa couplings for light fermions make that their pair production is not effectively blocked. Moreover, as their contribution to the conductivity is exponential in $m_f^2$ we can consider fermions that contribute to the conductivity as effectively massless. Here we have considered for simplicity that the two first generations of quarks and leptons contribute to the conductivity. Had we considered also the third generation would have amounted to a global factor in $\sigma$ of $3/2$, which would not change at all the qualitative results in this paper.

The backreaction of fermionic currents, the Schwinger effect, has been proven to roughly be encoded into a redefinition of the $\xi$ parameter, $\xi\to\xi_{\rm eff}$, as~\cite{Domcke:2018eki}
%
%\be \xi_{\rm eff} = \xi - \tr \frac{|eQ|^3}{12\pi^2}\coth\left( \frac{\pi|B|}{|E|}\right)\frac{|E|}{H_E^2}=\xi -  \frac{g_Y^3\cos^3\theta_W}{3\pi^2}\coth\left( \frac{\pi|B|}{|E|}\right)\frac{|E|}{H_E^2}
\be \xi_{\rm eff} = \xi - \Delta \xi, \qquad \Delta \xi= \frac{e^3}{3\pi^2}\coth\left( \frac{\pi|B|}{|E|}\right)\frac{|E|}{H_E^2}\,, \label{xi-effective}
%
%\Delta \xi, \hspace{1cm} \Delta \xi = \dfrac{41 g'^3}{144 \pi^2}\; {\rm coth}\left(\pi \sqrt{\dfrac{\rho_B}{\rho_E}} \right) \dfrac{\sqrt{\rho_E}}{H_E^2}, 
\ee
where only first and second generation fermions have been considered.
The correction becomes significant, $\Delta \xi / \xi \gtrsim 0.1$, for $\xi \gtrsim 3.7$, which corresponds to $f_\phi \lesssim 0.19\, \Mp$. Hence for $\xi \gtrsim 3.7$ the Schwinger effect must be taken into account, and the amplitudes of the gauge fields in equilibrium must satisfy the equation~\cite{Domcke:2018eki}
\be
2\xi_{\rm eff}H |E|\, |B|-2H(|E|^2+|B|^2)=\dot{\rho}_{\rm EM}=0
\label{eq:restriccion0}
\ee

Previous studies~\cite{Domcke:2018eki,Domcke:2019mnd,Gorbar:2021zlr} have considered two regimes: (i) Maximal estimate and (ii) Equilibrium estimate.
We will use them to compute the MHD quantities yielding the BAU, i.e.,~the helicity, its derivative, the electric and magnetic energies, as well as the magnetic correlation length. Both regimes follow different strategies: in the maximal estimate all quantities are capped by other relations that still depend on the parameter $\xi$, whereas in the equilibrium case the exponential relations from the previous section stay with the counterpart of the substitution (\ref{xi-effective}) on $\xi$.

\subsubsection{Maximal estimate}\label{sec:maximal}
In this case we assume the exponential behaviors of the backreactionless solutions to be valid until they saturate the maximal value that we will display hereafter. We numerically determine the value of crossing, which happens for $\xi \simeq 4.4$–$4.7$ depending on each quantity, corresponding to $f_\phi \simeq 0.15\, \Mp$. However, as we just saw, for such value the Schwinger effect can no longer be neglected, so there remains in this process a gray area of uncertainty as to the exact transition between the two regimes.

The maximum electric and magnetic energy density can be estimated as the solution of Eq.~(\ref{eq:restriccion0})~\cite{Domcke:2018eki}, i.e.
\be |E|^2+|B|^2 = \xi_{\rm eff} |E|\,|B| .
\label{eq:restriccion}
\ee
This replacement yields an equation relating the $|E|$ and $|B|$ fields that can be solved analytically. We then choose, as definition of our maximal estimate, the solution $(|E|,|B|)$ of (\ref{eq:restriccion}) that maximizes the product $|E|\cdot|B|$.\footnote{Notice that our definition of maximal estimate departs from that used in Ref.~\cite{Domcke:2018eki}, where separate maximal conditions to the configurations for the fields $E$ and $B$ (corresponding to absolute maximal values independently reached by the configurations $E$ and $B$) are imposed, so that their corresponding partners do not satisfy Eq.~(\ref{eq:restriccion}). Conversely, our criterium of maximizing the helicity guarantees that our solution satisfies Eq.~(\ref{eq:restriccion}).} For $\xi\gg 1$, the result approximates to:
\begin{subequations}\label{eq:EandBmax} \begin{eqnarray} |E|_{\rm max} &\simeq&  \frac{2\pi^2}{e^3}\; \xi H_E^2, \label{eq:Emax} \\
 |B|_{\rm max} &\simeq&  \frac{2\pi^2}{3e^3} \;\xi^2 H_E^2, \label{eq:Bmax}
 \ese
although, in the numerical calculations, we of course use the exact solutions.
Hence we obtain our maximal helicity estimate
%. 
\be \mathscr{H}_{\rm max} \simeq \frac{8\pi^4}{9e^6} \;\xi^3 H_E^3 \label{Helicity-Schw-max} \ee
as well as our maximal energy density estimate 
\be  \rho_{\rm EM}^{\rm max} \simeq  \frac{2\pi^4}{9e^6} \;\xi^4 H_E^4 .\ee
Finally, combining (\ref{ABdef}) and (\ref{magnetic-correlation-length-def}), and assuming maximally helical magnetic field, we get for the correlation length (still for large $\xi$)
\be \ell_B^{\rm max} = \pi\;\dfrac{ \mathscr{H}_{\rm max} }{\rho_B^{\rm max}}    \simeq \dfrac{4\,\pi}{\xi H_E}\,.
\label{eq:corrlength}
 \ee

In this case the upper labels ``max" on $\rho_{\rm EM}^{\rm max}$ and $\ell_B^{\rm max}$ mean that they are computed from maximal quantities, but do not necessarily mean upper bounds. In fact the estimate for $\ell_B^{\rm max}$ is a conservative one, as it matches the corresponding backreactionless quantity at a small value, $\xi\simeq 1.4$, so in principle we would expect higher values for $\ell_B^{\rm max}$, giving rise to bigger Reynolds numbers (see Sec.~\ref{sec:evolution-after-reheating}). Still we will use the estimate in Eq.~(\ref{eq:corrlength}) for our numerical calculations.

We finally recall that in this case the parameter $\xi$ remains as given by (\ref{xi-parameter-approx}), hence it corresponds to the solid lines displayed in the left panels of Fig.~\ref{xi-parameter}. For the solid lines of the right panels, however, the helicity has two regimes: it first obeys the exponential relation (\ref{Helicity-analytical}) until it reaches its maximal value, then follows (\ref{Helicity-Schw-max}).

\subsubsection{Equilibrium estimate}\label{sec:equilibrium}
In this case, we take into account the backreaction of the chiral fermions on the gauge fields by just replacing the parameter $\xi$ with the effective one given by (\ref{xi-effective})  in the backreactionless solutions. 
Using (\ref{EM-energy-densities}) and (\ref{abs-EB-def}) the latter becomes
\be  \,\dfrac{63}{2^{15}\pi^2}\; \dfrac{e^{2\pi \xi_{\rm eq}}}{\xi_{\rm eq}^3} =  \left( \frac{3\pi^2}{e^3}\right)^2 (\xi -\xi_{\rm eq})^2 \;\tanh^2\left(\sqrt{\frac{5}{4}} \,\frac{\pi}{\xi_{\rm eq}} \right). \label{Schw-Equ-Xieff} 
\ee
When the backreactionless solutions are used, and to make it explicit which case we are handling, we chose to label the effective parameter as $\xi_{\rm eq}$. 
The solution of Eq.~(\ref{Schw-Equ-Xieff}) provides the function $\xi_{\rm eq}=\xi_{\rm eq}(\xi)$ and, using (\ref{xi-parameter-approx}), we can obtain $\xi_{\rm eq}$ as a function of $f_\phi$ (and $g$) that we plot on the left panels of Fig.~\ref{xi-parameter} in dotted lines. 

Next, the MHD quantities are calculated in the same way as in the case without considering the Schwinger effect, but with the replacement $\xi \to \xi_{\rm eq}$, hence
\be \mathscr H_{\rm eq} = \mathscr H(\xi_{\rm eq}), \qquad  \rho_{B/E}^{\rm eq} = \rho_{B/E} \,(\xi_{\rm eq}),\qquad \ell_{B}^{\rm eq} = \ell_{B}\,(\xi_{\rm eq}), \ee
where (\ref{Helicity-analytical}), (\ref{EM-energy-densities}) and (\ref{magnetic-correlation-length}) should be used.
On the right panels of Fig.~\ref{xi-parameter}, we plot the equilibrium estimate for the helicity as a function of $f_\phi$ and $g$ in dotted lines. 

\subsubsection{Final comments}

Needless to say, neither the maximal nor the equilibrium estimates are true solutions to the gauge equations of motion in the presence of the Schwinger effect, which introduces highly nonlinear effects into them.
However, numerical solutions taking into account the backreaction from fermion currents have been recently considered in Refs.~\cite{Gorbar:2021rlt,Gorbar:2021zlr} which show that, for values of the $\xi$ parameter for which the 
Schwinger effect becomes relevant, the numerical solution for the different quantities, in particular for the helicity, lies between the maximal and equilibrium estimates. This feature remains if the Bunch-Davies vacuum is damped by the conducting medium, even for extreme cases of very large damping, leading to very suppressed vacua. Therefore, we expect that the solution to the complicated problem of taking into account all the backreaction from Schwinger fermions currents will be somewhere between the two considered estimates, and thus the allowed region by the BAU will be in between the allowed regions that we will exhibit in Secs.~\ref{sec:baryogenesis} and \ref{sec:constraints} for both estimates.

\subsection{Self-consistency condition \label{sec:backreaction}}

In previous subsections, we have computed
the helical gauge fields generated in the presence of the inflationary background, after estimating the backreaction of fermion currents on gauge fields, but we have neglected the backreaction of gauge fields on the inflaton dynamics.  We will now compute the conditions to have negligible backreaction of the generated gauge fields on the inflaton equations of motion, such that we can reliably trust the inflationary predictions, and therefore the actual generation of helical magnetic fields. Needless to say this condition is mainly a simplifying one, and allows to work out the inflationary model independently on the generated gauge fields. As we will see in Sec.~\ref{sec:reheating}, this condition is also related to the possibility of reheating the Universe after the inflationary period by the preheating mechanism, although this scenario deserves further studies.
  
Once we have obtained the helicity, we can compute the RHS of the inflaton equation of motion (\ref{eom-inflaton}), as in the radiation gauge they are simply related by
\be \langle\bm{E} \cdot \bm{B}\rangle = -\dfrac{1}{2} \dfrac{d}{d\tau} \langle\bm{A} \cdot \bm{B}\rangle. \label{ABtoEB}\ee
Ignoring for the moment the Schwinger effect on the produced gauge fields,
using (\ref{ABdef}), (\ref{Helicity-analytical}) and the relation in the de Sitter universe $aH=-\tau^{-1}$, one gets at the end of inflation, for $\xi\gg 1$
\be 
\left|\langle\bm{E} \cdot \bm{B}\rangle\right| \simeq \dfrac{135}{2^{16}}\dfrac{a_E^4 H_E^4}{\pi^2\xi^4} \; e^{2\pi \xi}. \label{EB} 
\ee
In the absence of backreaction of the gauge field on the inflaton equation of motion, the inflationary equation (\ref{eom-inflaton}) with slow roll conditions reduces to $3H\dot\chi\simeq -V'(\chi)$. Thus, in order to consistently neglect the backreaction on the inflaton, we must simply enforce that, in the inflaton equation of motion (\ref{eom-inflaton}), the RHS term is negligible compared to the potential term, i.e.
\be \dfrac{ \sqrt{2}\, \xi }{\Mp }  \left| \dfrac{\langle\bm{E} \cdot \bm{B}\rangle}{V'(\chi)}  \right|  \ll 1, \label{backreaction-condition} \ee
where we used (\ref{xi-parameter-approx}).
This condition is independent of the reheating temperature and should hold during the full magnetogenesis process, hence during the last few $e$-folds of inflation, so we can evaluate it 
using the above solutions for $\langle\bm{E} \cdot \bm{B}\rangle$ at the end of inflation.
Then, using the definition of the slow roll parameter $\epsilon(\chi)$, we can write $V'\simeq\sqrt{2}V / \Mp$ at the end of inflation, and hence, for $\chi \simeq \chi_E$, Eq.~(\ref{backreaction-condition})  becomes
\be \xi \left|\langle\bm{E} \cdot \bm{B}\rangle \right| \ll V(\chi_E). \ee

Moreover it is interesting to note that, if we ignore the Schwinger effect, combining $\rho_{\rm EM} \simeq \rho_{E} $ given by (\ref{electric-energy}), together with (\ref{EB}), we get 
\be  2\, \rho_{\rm EM}\simeq \xi \, \left|\langle\bm{E} \cdot \bm{B}\rangle\right| . \label{Energy-EB} \ee
Notice that, for collinear $\bm{E}$ and $\bm{B}$ with the substitution $\xi \to\xi_{\rm eff} $, this equation yields the starting point of the Schwinger maximal estimate, i.e.~Eq.~(\ref{eq:restriccion}).

Hence, the condition (\ref{backreaction-condition}) evaluated at the end of inflation is equivalent to imposing 
\be  2\, \rho_{\rm EM} \ll V(\chi_E). 
\label{eq:condition2}
\ee
Notice that the condition (\ref{eq:condition2}) is stronger than the condition for neglecting $\rho_{\rm EM}$ in the Friedman equation, i.e.,~$
\rho_{\rm EM}\ll 3H^2\Mp^2\simeq V(\chi_E)
$,
so that the latter does not need to be imposed. 

Now taking into account the Schwinger effect, the equilibrium estimate is obtained by the replacement $\xi \to \xi_{\rm eq}$ in the expression (\ref{EB}), as described in Sec.~\ref{sec:equilibrium}.
Hence, the consistency condition in the Schwinger equilibrium estimate is given by Eq.~(\ref{eq:condition2}) where $\xi \to \xi_{\rm eq}$, i.e.
\be  2\, \rho_{\rm EM}^{\rm eq} \ll V(\chi_E), 
\label{eq:condition3}
\ee
a stronger condition than the one coming from the Friedman equation $\rho_{\rm EM}^{\rm eq}\ll V(\chi_E)$, but much weaker than Eq.~(\ref{eq:condition2}) where we were ignoring the Schwinger effect, since $\rho_{\rm EM}^{\rm eq} \ll\rho_{\rm EM}$.

\begin{figure}[b]
\begin{center}
\includegraphics[width = 10.2cm]{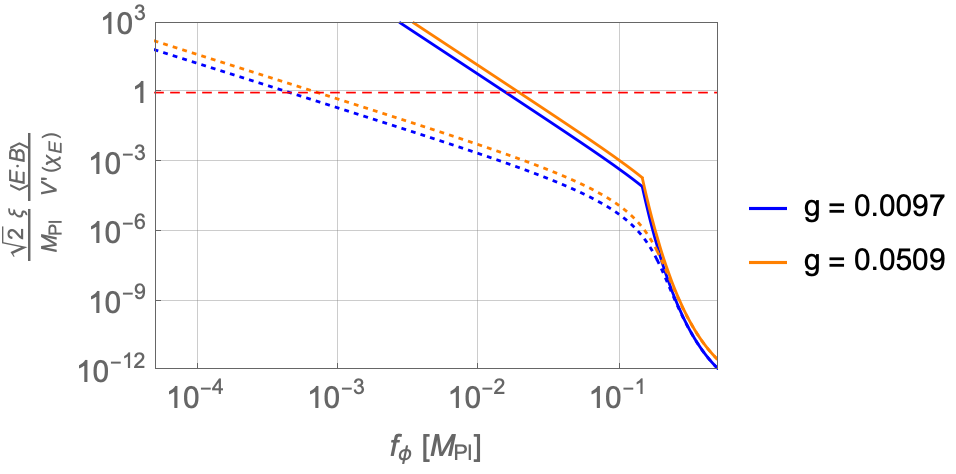}
\caption{\it Ratio between the potential term and the backreaction term in the inflaton equation of motion, see Eq.~(\ref{backreaction-condition}), for the range of values of $g$ allowed by inflation.
Solid lines are the maximal estimate while dashed lines are the equilibrium estimate, after taking into account the Schwinger effect.
\label{fig:back-reaction}}
\end{center}
\end{figure}

On the other hand, for the maximal estimate, using the results from Sec.~\ref{sec:maximal}, we can write 
\be
 \left|\langle\bm{E} \cdot \bm{B}\rangle\right|_{\rm max} \simeq |E|_{\rm max}\; |B|_{\rm max}\simeq \frac{1}{3}\left(\frac{2\pi^2}{e^3} \right)^2 \xi^3 H_E^4 \, ,
\ee
where in the first step we maximize the product by assuming a collinear configuration of $\bm{E}$ and $\bm{B}$~\cite{Domcke:2018eki}, while the second step is justified for large values of $\xi$.
A similar reasoning for the consistency condition does apply to this case, for which the total density is dominated by the energy stored in the magnetic field, $\rho_{\rm EM}\simeq \rho_B$, and such that the explicit maximal estimate found implies that, for large $\xi$,
\be
6\, \rho_{\rm EM}^{\rm max} \simeq  \xi \, \left|\langle\bm{E} \cdot \bm{B}\rangle\right|_{\rm max}.
\ee
Hence, in the maximal estimate, imposing condition (\ref{backreaction-condition}) is equivalent to requiring
\be  6\, \rho_{\rm EM}^{\rm max} \ll V(\chi_E),  \ee
which is again stronger than the condition for neglecting the total gauge energy density in the Friedman equation, $\rho_{\rm EM}^{\rm max}\ll V(\chi_E)$.

We display in Fig.~\ref{fig:back-reaction} the left-hand side of Eq.~(\ref{backreaction-condition}) as a function of $f_\phi$ for the range allowed on the parameter $g$ by the inflationary observables, using for the $|E|$ and $|B|$ fields both the maximal and the equilibrium estimates. 
In conclusion, condition (\ref{backreaction-condition}) is satisfied for:
\be\begin{aligned} f_\phi& \; \gtrsim \; 1.9 \cdot 10^{-2} \,\Mp\  \hspace{1.5 cm} \textrm{(Maximal estimate)}, \\
f_\phi& \;\gtrsim\; 7.2 \cdot 10^{-4} \,\Mp \hspace{1.63 cm} \textrm{(Equilibrium estimate)}.
\label{eq:nobackreaction}
\end{aligned}\ee

\section{Reheating}
\label{sec:reheating}
At the stage of inflation all the energy is concentrated in the slowly rolling inflaton field. Soon after, it begins to oscillate near the minimum of its effective potential and eventually perturbatively decays into SM particles that interact with each other, and come to a state of thermal equilibrium in a process called \textit{reheating}. However, the Universe can also be reheated nonperturbatively in a much quicker timescale through coherent fields effects while it oscillates in its potential, in a process known as \textit{preheating}~\cite{Kofman:1997yn}.

In inflationary models where the inflaton is coupled to the Chern-Simons term with coupling $1/f_\phi$, as in Eq.~(\ref{eq:phiFFtilde}), recent lattice simulations~\cite{Adshead:2015pva,Cuissa:2018oiw}, in the absence of fermionic currents, have shown that even for a negligible electromagnetic energy density at the end of inflation, the Universe can efficiently preheat provided that the coupling $1/f_\phi$ in Eq.~(\ref{eq:phiFFtilde}) is large enough. In particular preheating occurs when $f_\phi\lesssim f_\phi^c$, with $f_\phi^c\simeq 0.11 \Mp$, i.e.~for $\xi\gtrsim \xi_c$ with $\xi_c\simeq 6.7$. However, in this region the backreaction of the fermion currents on the helical gauge fields cannot be neglected, and we should adapt the previous results to our different estimates in the presence of the Schwinger effect. 

For the equilibrium estimate, we have seen in Sec.~\ref{sec:equilibrium} that the effect of the backreaction can be encoded into the redefinition of the parameter $\xi\to\xi_{\rm eq}$ while using the backreactionless solutions but as functions of $\xi_{\rm eq}$. Therefore a straightforward application of the results from Refs.~\cite{Adshead:2015pva,Cuissa:2018oiw} should provide the condition for efficient preheating as $\xi_{\rm eq}\gtrsim \xi_{\rm eq}^c$ with $\xi_{\rm eq}^c\simeq 6.7$ which translates, using Eq.~(\ref{Schw-Equ-Xieff}), to
\be
f_\phi\lesssim 2.4\cdot 10^{-4}\Mp.\hspace{1.5cm} \textrm{(Equilibrium estimate)}
\ee
This bound is outside the region where we can neglect the backreaction of the helical fields on the inflaton, Eq.~(\ref{eq:nobackreaction}).

In the case of the Schwinger maximal estimate of the electromagnetic fields, see Sec.~\ref{sec:maximal}, we can easily perform a similar translation on the same requirement.
To this end, we define a new effective parameter $\xi_{\rm max}$ that mimics the effect of the maximal estimate once plugged into the backreactionless solutions as
\be \left|\langle\bm{E} \cdot \bm{B}\rangle \right|_{\xi_{\rm max}} = |E_{\rm max}(\xi)| \; |B_{\rm max}(\xi)|, \label{xi-ximax-trick} \ee
where $|E_{\rm max}(\xi)|$ and $|B_{\rm max}(\xi)|$ are the maximal estimates of the electromagnetic fields from Eqs.~(\ref{eq:EandBmax}), and $\langle\bm{E} \cdot \bm{B}\rangle_{\xi_{\rm max}}$ is the corresponding backreactionless product given by (\ref{EB}) evaluated at $\xi=\xi_{\rm max}$.
From there, the condition for efficient preheating $\xi_{\rm max}\gtrsim \xi_{\rm max}^c$, with $\xi_{\rm max}^c\simeq 6.7$, translates into
\be
f_\phi\lesssim 5.6 \cdot 10^{-3}\Mp ,\hspace{1.5cm}  \textrm{(Maximal estimate)}
\ee
which again is inconsistent with the condition (\ref{eq:nobackreaction}) on no backreaction of helical fields on the inflaton dynamics. 

The previous results can be easily understood by considering that, in the presence of the backreaction of the Schwinger currents on the helical fields, the required coupling, between $\phi$ and the Chern-Simons term, for preheating must be much stronger than in the backreactionless case. This is because the produced helical gauge fields are much weaker, in the presence of backreaction, for a fixed value of the coupling $\Mp/f_\phi$.

We thus conclude that, after considering the backreaction of fermion currents on the helical fields in both estimates, the preheating mechanism is not consistent with the self-consistency condition obtained in Sec.~\ref{sec:backreaction}. Hence, in our model preheating does not occur and reheating should take place by perturbative decays of the inflaton into the SM matter only.

Reheating then takes place after inflation, during the inflaton oscillations around its minimum, by perturbative inflaton decays. In this period, between the end of inflation $t_E\sim 1/H_E$ and the reheating time $t_{\rm rh}\sim 1/\Gamma_\chi$, where $\Gamma_\chi$ is the inflaton decay width, the Universe temperature first grows from zero to a maximum temperature $T_0$ given by~\cite{Chung:1998rq,Giudice:2000ex}
\be
T_{\rm 0}\simeq 0.61 \sqrt{T_{\rm rh}T_{\rm rh}^{\rm ins}},
\ee
where, assuming thermalization,
\be
T_{\rm rh}= \left(\frac{90}{\pi^2 g_\ast}\right)^{\frac{1}{4}} \sqrt{\Gamma_\chi \Mp} 
\label{eq:Trh}
\ee
is the reheating temperature and $g_\ast=106.75$ is the number of relativistic degrees of freedom.
Also, in this work we define $T_{\rm rh}^{\rm ins}$ as a reference temperature given by the above equation with $\Gamma_\chi\simeq H_E$. It would correspond to the reheating temperature for instant reheating, and takes the value $T_{\rm rh}^{\rm ins}\simeq 2.13\; (2.61) \cdot 10^{15}$~GeV for $g\simeq 0.01 \; (0.05)$ in our model.

The temperature $T_{\rm 0}$ is attained at a time $t_0$ when the scale factor $a$ grows by an $\mathcal O(1)$ factor, i.e.~$a_0\simeq 1.5 \,a_E$, and, after that~\footnote{The energy density is dominated, after the end of inflation, by the inflaton energy density $\rho_\chi(t)$, which decays as $e^{-\Gamma_\chi t}$, so that at the reheating temperature the energy density is dominated by the radiation energy density $\rho_R(t)$.}, the Universe evolves toward the reheating temperature following the law $T\sim a^{-3/8}$~\cite{Espinosa:2015qea}, with a scale factor $a_{\rm rh}$ given by
%
%\be\frac{a_{E}}{a_{\rm rh}}=\frac{a_{E}}{a_0}\frac{a_0}{a_{\rm rh}}\simeq 2.5 \,\left( \frac{T_{\rm rh}}{T_{\rm rh}^{\rm ins}} \right)^{\frac{4}{3}}\simeq \left( \frac{2\,T_{\rm rh}}{T_{\rm rh}^{\rm ins}} \right)^{\frac{4}{3}}  .\ee
\be a_{\rm rh}\simeq 0.4 \; a_E  \left( \frac{T_{\rm rh}^{\rm ins}}{T_{\rm rh}} \right)^{\frac{4}{3}}.\label{a-reheating}\ee
At the reheating temperature, the inflaton energy density has completely decayed and the Universe is fully dominated by radiation, giving rise to a radiation dominated era where the temperature evolves as $T\sim 1/a$. Of course, the value of the inflaton decay width $\Gamma_\chi$, and the reheating temperature $T_{\rm rh}$, depend on the particular interactions between the inflaton and the Standard Model particles that we will now explore.

In the present model, the Lagrangian from Eq.~(\ref{eq:potencial}) contains the interaction term $\sqrt{\delta_\lambda/2}\,m\,\chi h^2$ which gives rise to the leading inflaton decay channel $\chi\to hh$, with a decay width given by~\cite{Kofman:1994rk}
\be
\Gamma(\chi\to hh)=\frac{\delta_\lambda\, m}{16\pi}\, \sqrt{1-\frac{4 m_h^2}{m^2}},
\label{eq:chitohh}
\ee
where $m_h=125.25$~GeV is the Higgs mass.
As the inflaton is stabilizing the EW vacuum (see Sec.~\ref{sec:stability}), which has an instability around $\mathcal Q_I\simeq 10^{11}$~GeV, we can reliably put the upper bound on $m$ as
$m\lesssim \mathcal Q_I$, and fix $m\simeq 5\cdot 10^{10}$~GeV while $\delta_\lambda\lesssim 0.35$ on perturbative grounds (see Sec.~\ref{sec:stability}). This gives for the decay width $\Gamma(\chi\to hh)\simeq 3.5\cdot 10^8$~GeV leading, using Eq.~(\ref{eq:Trh}), to a reheating temperature given by
$T_{\rm rh}\simeq  1.6\cdot 10^{13}$~GeV, which corresponds to $T_{\rm rh}/T_{\rm rh}^{\rm ins}\sim 10^{-2}$. On the other hand, the lowest bound on $m$, fixed by phenomenological considerations (see Sec.~\ref{sec:pheno}) to $m\simeq 10^3$~GeV, together with $\delta_\lambda\lesssim 0.2$ (see Sec.~\ref{sec:stability}) provide $\Gamma(\chi\to hh)$ of a few~GeV and correspondingly $T_{\rm rh}\simeq 10^{9}$~GeV, which corresponds to $T_{\rm rh}/T_{\rm rh}^{\rm ins}\sim 10^{-6}$. Hence from now on, we will consider the temperature ratio $T_{\rm rh}/T_{\rm rh}^{\rm ins}$ as a parameter of the model, which will become handy for the baryogenesis and constraints calculations. We also stress that this ratio mainly reflects the dependence of coming results on $m$, as just sketched above.

There are of course other channels that can contribute to $\Gamma_\chi$ but, as we will demonstrate hereafter, they are all subdominant. 
For instance, the coupling (\ref{eq:CPV}) gives rise to the decay channel $\chi\to AA$ into two gauge bosons with a decay width given by~\cite{Adshead:2015pva}
\be
\Gamma(\chi\to AA)\simeq\frac{m^3}{64\pi f^2_\phi},
\ee
which is subleading with respect to the channel $\chi\to hh$ for the relevant values of $m$ and $f_\phi$. In particular $\Gamma(\chi\to AA)\simeq 10^{-5}$  GeV for $m=5\cdot 10^{10}$ GeV, while $\Gamma(\chi\to AA)\simeq 10^{-28}$  GeV for $m=10^3$ GeV. Moreover, there is a mixing angle $\alpha$ between $\phi$ and $h$ (see Sec.~\ref{sec:pheno}), which is sizable for $m\sim \mathcal{O}$(few)~TeV, while of course is negligible for $m\gg 1$ TeV, given by $\sin\alpha\simeq \sqrt{2\delta_\lambda}v/m$. This mixing opens up the $\chi$ decays into the SM channels, with a total decay width into all SM channels given by $\Gamma(\chi\to\text{SM})=\sin^2\alpha\cdot \Gamma(h\to\text{SM})\simeq 4\sin^2\alpha$ MeV, in all cases subleading with respect to the decay width $\Gamma(\chi\to hh)$.

Note added: after this paper appeared on the arXivs, the possibility that the Lagrangian 
in Eq.~(\ref{eq:potencial}) could induce preheating by the explosive production of scalar fields after inflation was considered in Ref.~\cite{Cosme:2022htl}. In this case, we can identify the scalar fields of Ref.~\cite{Cosme:2022htl} with the Higgs field~\footnote{The relation between the parameters $q_3$ and $q_\chi$ in Ref.~\cite{Cosme:2022htl} and our parameters can be written as 
$$
q_3=\sqrt{2\delta_\lambda}\frac{\phi_E}{m},\quad q_\chi=\frac{\lambda_0 \phi_E^2}{m^2}.
$$
} and this mechanism, if implemented, would be more efficient than the perturbative production that has been considered so far in this section. First of all, as we wanted the inflaton to stabilize the Higgs potential we have imposed the condition on its mass $m< 10^{-7}\Mp$. This means that during preheating the inflaton potential term $\frac{1}{4} \lambda_\phi \phi^4$ will dominate the mass term $\frac{1}{2} m^2\phi^2$, as $\lambda_\phi\simeq 10^{-12}$ and $\phi_E\simeq \Mp$. In Ref.~\cite{Cosme:2022htl} it was proven that there is no runaway solutions provided that $\delta_\lambda<\lambda_0/4$, and preheating imposes the mild condition $2\delta_\lambda>\left( 100 \, m/\phi_E\right)^2$, always satisfied as $100 \, m/\phi_E<10^{-5}$. Still in the rest of the paper we will be agnostic about the (p)reheating mechanism and will consider the reheating temperature $T_{\rm rh}$ as a free parameter.

\section{Baryogenesis}
\label{sec:baryogenesis}

Using the helical gauge fields produced at the end of inflation, and assuming that their corresponding helicity remains after reheating (a hypothesis that will be self-consistently checked a posteriori, see Sec.~\ref{sec:evolution-after-reheating}), until the EWPT (which we assume to be the SM crossover), we can compute the conversion of the helicity into the $(B+L)$ asymmetry, and therefore the baryon asymmetry of the Universe (BAU).

At a temperature around the electroweak scale, $T_{\rm EW}\approx 160$~GeV, the Higgs VEV departs from zero and smoothly transitions to the SM VEV at $T=0$, $v=246$~GeV, making the off-diagonal elements of the gauge bosons mass matrix gradually compete with the thermal mass for $W^3_\mu$ on the diagonal, that decreases with decreasing temperature. This results in a phase transition controlled by the EW angle $\theta_W$ whose temperature dependence is subject to significant uncertainties \cite{Kajantie:1996qd, DOnofrio:2015gop}. Following Refs.~\cite{Domcke:2019mnd, Cado:2021bia} we define the parameter $f_{\theta_W}$, which encodes all the details of the EW transition and its uncertainties, as 
\be f_{\theta_W}  = -\sin (2 \theta_W) \, \dfrac{d\theta_W}{d\ln T}\bigg\rvert_{T=135\text{ GeV}}, \quad 5.6 \cdot 10^{-4} \lesssim f_{\theta_W}  \lesssim 0.32. 
\label{eq:ftheta}
\ee
This gives rise to a source term for the $(B+L)$ asymmetry, while the electroweak sphalerons are still in equilibrium for $T\gtrsim130$~GeV. In Ref.~\cite{Kamada:2016cnb}, it was shown in detail how the source and washout terms balance each other around an equilibrium value of the baryon asymmetry around $T=135$~GeV, which is finally given by
\be\eta_B  \simeq   4 \cdot 10^{-12} \, f_{\theta_W}  \frac{\mathscr{H}_Y}{H_E^3} \left( \frac{H_E}{10^{13} \, \text{GeV}} \right)^{\frac{3}{2}} \left(\frac{T_{\rm rh}}{T_{\rm rh}^{\rm ins}}  \right) \, \, \simeq \, 9 \cdot 10^{-11} ,\label{constraint-nB} \ee
where we have imposed the observed value~\cite{Zyla:2020zbs} in the right-hand side.

\begin{figure}[h]
\begin{center}
\includegraphics[width = 8.5cm]{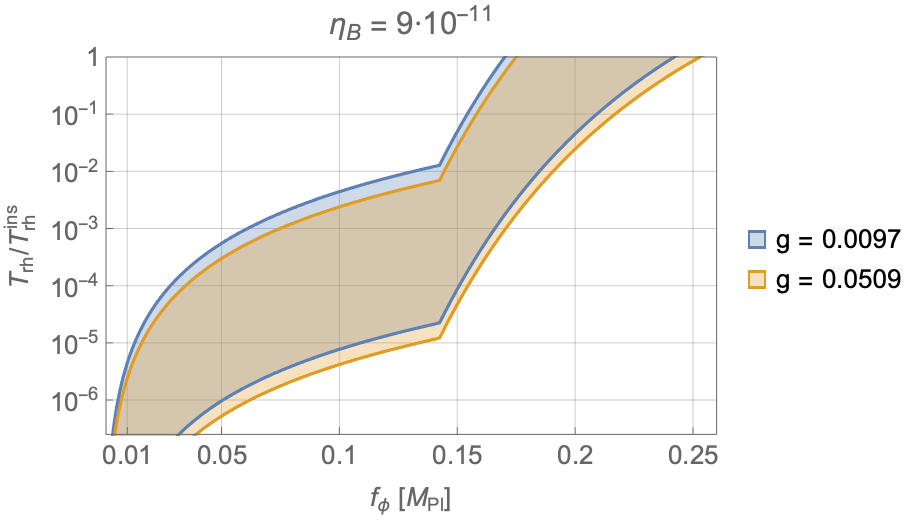}
\includegraphics[width = 6.8cm]{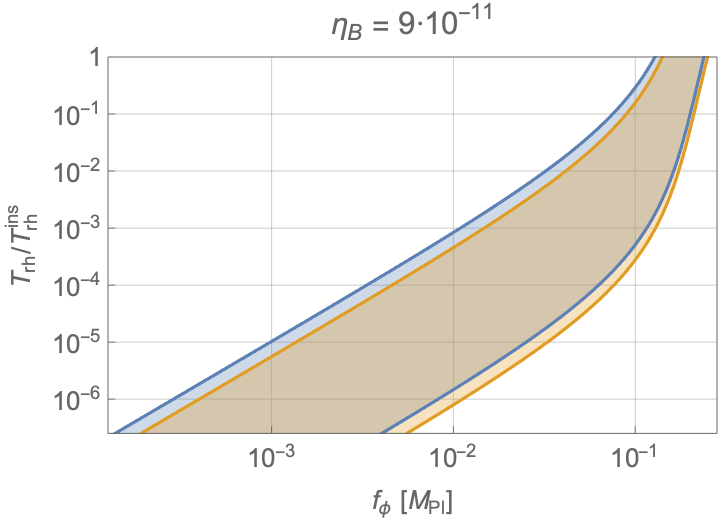}
\caption{\it The baryogenesis region. In the shading region the value of $\eta_B$ satisfies Eq.~(\ref{constraint-nB}). Left panel: Schwinger maximal estimate. Right panel: Schwinger equilibrium estimate.
\label{fig:nB-parameter-space}}. 
\end{center}
\end{figure}

In Fig.~\ref{fig:nB-parameter-space} we show—in the plane $(f_\phi,T_{\rm rh}/T_{\rm rh}^{\rm ins})$, for both bounds on the allowed $g$ range, Eq.~(\ref{g-range}), and for both Schwinger estimates for the magnetic fields, i.e., maximal (left panel) and equilibrium (right panel) estimates—the region where the value of $\eta_B$ satisfies Eq.~(\ref{constraint-nB}) taking into account the range in Eq.~(\ref{eq:ftheta}) for the quantity $f_{\theta_W}$. As we can see from both panels, together with the range (\ref{g-range}) on the $g$ parameter where inflationary conditions in Ref.~\cite{Planck:2018jri} are satisfied, there is an absolute upper bound on the parameter $f_\phi$ as $ f_\phi\lesssim 0.25\, \Mp$ for both Schwinger estimates, corresponding to the reference (instant) reheating temperature, where the baryogenesis conditions are met. Moreover, for the highest reheating temperature we can get from our model of inflation, $T_{\rm rh}\simeq 10^{-2}\;T_{\rm rh}^{\rm ins}$, the bound lowers to $ f_\phi\lesssim 0.19 \; (0.17)\, \Mp$ for the maximal (equilibrium) Schwinger estimate.  Putting together the lower bounds from (\ref{eq:nobackreaction}), and the requirement that $T_{\rm rh}/T_{\rm rh}^{\rm ins} \gtrsim 10^{-6}$, one gets the global ranges
\be\begin{aligned} &1.9 \cdot 10^{-2} \; \lesssim \; f_\phi \, / \,\Mp \; \lesssim \; 0.19  \hspace{1.5cm} \text{(Maximal estimate)}, \\
&7.2 \cdot 10^{-4} \; \lesssim \; f_\phi \, / \,\Mp \; \lesssim \; 0.17  \hspace{1.5cm} \text{(Equilibrium estimate)} , \label{window-baryogenesis}
\end{aligned}\ee
where baryon asymmetry can be generated consistently with the condition of no backreaction of the helical gauge fields on the inflationary dynamics.

However, as we will see next, the helical magnetic fields, produced at the end of inflation, interact after reheating with the thermal plasma and there are a number of constraints that have to be satisfied for the helicity to reach the temperatures where the EWPT takes place. As we will see, these constraints may reduce the allowed region in the parameters space.

\section{Constraints}
\label{sec:constraints}

We have computed, up to now, the baryon asymmetry generated by helical magnetic fields produced after inflation when their helicity decays into $(B+L)$ asymmetry at the electroweak crossover, and identified the region of the parameter space $(g,f_\phi,T_{\rm rh})$ where the observed value of the baryon asymmetry is reproduced. However, there are a number of constraints that can further narrow the region of the parameter space where the BAU can really be reproduced by our theory and will be analyzed in this section.

\subsection{Helicity evolution: Magnetohydrodynamics and Reynolds numbers}
\label{sec:evolution-after-reheating}

Helical magnetic fields are produced at the end of inflation, and %, since there is no preheating because the associated energy density is subdominant enough (see Sec.~\ref{sec:reheating}), 
we assume their comoving quantities stay constant until reheating, at temperature $T_{\rm rh}$. 
However, at reheating a thermal plasma is generated by the decay of the inflaton into the SM particles and consequently the electroweak symmetry is restored—by the appearance of thermal masses—until the EWPT.  Hence the helicity in photons $\mathscr{H}$ gets transformed into helicity in hypercharge gauge fields $\mathscr{H}_{Y}$, as sketched at the beginning of Sec.~\ref{sec:field-production}, see Eq.~(\ref{Weinberg}). 
The latter then interacts with the thermal plasma which, in turn, backreacts on the gauge fields.

This system can be described by the so-called MHD equations~\cite{Giovannini:1997eg,Durrer:2013pga,Vachaspati:2020blt}, and has been studied for the case at hand in Ref.~\cite{Cado:2021bia}. In a nutshell, the physical quantities of interest (amplitudes, energy densities, correlation length and helicity) do not scale adiabatically in such an environment, or equivalently their comoving quantities are not constant. Therefore there can be a magnetic \textit{diffusion} effect leading to the decay of the helicity.
If, on the other hand, the magnetic \textit{induction} is the leading effect, then the helicity can be conserved until the EWPT and the baryogenesis mechanism can take place. This effect is measured by the magnetic Reynolds number $\mathcal{R}_m$, and we will see that it is enough to require $\mathcal{R}_m>1$ at reheating for the helicity to be conserved until the EW crossover. 
Hence, in this section we will study how this constraint affects the region of the parameter space that yields the BAU. 

The magnetic Reynolds number is defined as the ratio of the magnetic induction term over the magnetic diffusion term of the corresponding MHD equation. It can be written as
\be
\mathcal R_m\equiv\sigma v \ell_{B_Y}, \label{Reynolds-m-def}
\ee
where $\sigma=c_\sigma T^c_{\rm pl}/(\alpha_Y\log(\alpha_Y^{-1}))$ is the conductivity of the thermal plasma, with $c_\sigma\simeq 4.5$, and
\be T^c_{\rm pl} \simeq 0.8\; T_{\rm rh}^{\rm ins} \left(\dfrac{T_{\rm rh}^{\rm ins}}{T_{\rm rh}} \right)^{\frac{1}{3}} \ee
is the typical (comoving) temperature of the plasma, where we have used Eq.~(\ref{a-reheating}) with $a_E=1$.

In addition, the typical bulk velocity of the plasma $v$ can be estimated from the MHD Navier-Stokes equation for the velocity field. The general solution should be computed numerically, but for asymptotic cases, when one term clearly dominates over the others in the equation, we can sketch some approximations.
To do so, like in the magnetic case, we can compute the electric Reynolds number $\mathcal R_e$, given by
\be
\mathcal R_e\equiv\frac{v\ell_{B_Y}}{\nu}, \label{Reynolds-e-def}
\ee
where $\nu=c_\nu/(\alpha_Y^2 \log(\alpha_Y^{-1})T^c_{\rm pl})$ is the kinematic viscosity, with $c_\nu\simeq 0.01$. If $\mathcal{R}_e > 1$, then there is an equipartition between the kinetic energy in the plasma and the magnetic energy. In the opposite case, where $\mathcal{R}_e <1$, the kinetic energy and velocity are smaller than the magnetic energy. 
Relying, in this way, on the value of $\mathcal{R}_e$, we can compute all quantities in these two separate cases:
\begin{itemize}
\item \textit{Viscous} regime: $\mathcal{R}_e<1<\mathcal{R}_m$\,,
\item \textit{Turbulent} regime: $1<\mathcal{R}_e<\mathcal{R}_m$\,.
\end{itemize}
We omit the other cases, where $\mathcal{R}_m<1$, since we will not be interested in them.

In summary, the evolution of these two scaling regimes with respect to conformal time $\tau$ behave  as~\cite{Domcke:2019mnd,Banerjee:2004df}
\begin{subequations} \label{eq:ScalingBoth} \begin{eqnarray}  \mathcal{R}_e < 1\;\text{:}& \qquad &B_Y \, \propto \, \tau^{-\frac{1}{2}}\, , \qquad\ell_{B_Y} \, \propto \, \tau\, , \qquad  v \, \sim\ell_{B_Y}B^2_Y/(\nu\rho)\propto \tau^0 \,, \label{eq:ScalingViscous} \\  \mathcal{R}_e > 1\;\text{:}& \qquad &B_Y \, \propto \, \tau^{-\frac{1}{3}}\, , \qquad \ell_{B_Y} \, \propto \,\tau^{\frac{2}{3}} \, , \qquad  v \, \sim B_Y/\sqrt{\rho}\propto \tau^{-\frac{1}{3}}\,,\label{eq:ScalingTurbulence}  \ese
where 
\be \rho \simeq 0.4\; \rho_\chi \left(\dfrac{T_{\rm rh}^{\rm ins}}{T_{\rm rh}} \right)^{\frac{4}{3}}, \qquad \rho_\chi \simeq 3 \Mp^2 H_E^2, \ee
is the plasma energy density. 
Inserting the latter relations for $v$ in (\ref{Reynolds-m-def}) for both cases, we can estimate the magnetic Reynolds number at reheating as~\cite{Cado:2021bia}
\bse  
\textrm{For}\quad \mathcal{R}_e^{\rm rh} < 1 \quad \Rightarrow\quad \mathcal{R}_m^{\rm rh}   &\approx &5.9 \cdot 10^{-6} \; \frac{\rho_{B_Y} \ell_{B_Y}^2}{H_E^2}  \left( \frac{H_E}{10^{13} \, \text{GeV}} \right) \left(\frac{T_{\rm rh}}{T_{\rm rh}^{\rm ins}}  \right)^{\frac{2}{3}},\label{constraint-Rm}\\ 
\textrm{For}\quad\mathcal{R}_e^{\rm rh} > 1 \quad \Rightarrow\quad \mathcal{R}_m^{\rm rh}   &\approx&  1.1 \cdot 10^{-1} \; \frac{\sqrt{\rho_{B_Y}} \ell_{B_Y}}{H_E} \left( \frac{H_E}{10^{13} \, \text{GeV}} \right)^{\frac{1}{2}} \left(\frac{T_{\rm rh}}{T_{\rm rh}^{\rm ins}}  \right)^{\frac{1}{3}} \label{constraint-Rmb}
\ese
where the magnetic energy density is roughly given by $\rho_{B_Y}\approx B_Y^2/2$. 
From (\ref{eq:ScalingBoth}), using (\ref{Reynolds-m-def}) and (\ref{Reynolds-e-def}), we see that in both regimes both Reynolds numbers grow with time according to the same scaling relations:
\bse 
 \mathcal{R}_e < 1\;\text{:}& \hspace{2cm} &\mathcal{R}_{m}  \, \propto \, \tau,\hspace{1.78cm}  \mathcal{R}_e  \, \propto \, \tau \, ,\\  \mathcal{R}_e > 1\;\text{:}& \hspace{2cm} &\mathcal{R}_{m}  \, \propto \, \tau^{\frac{1}{3}}\,, \hspace{1.5cm}  \mathcal{R}_e  \, \propto \, \tau^{\frac{1}{3}}\,. 
 \label{eq:scalings}
 \ese 

Hence, once the requirement $\mathcal{R}_m^{\rm rh}>1$ is reached, the magnetic Reynolds number remains greater than one, as long as there is a plasma filling the Universe. The conservation of helicity is due to an inverse cascade in which the helicity is transferred from smaller to larger scales, reflected in the growth of $\ell_{B_Y}$. Therefore, to guarantee the survival of the comoving helicity at the EWPT, 
it is enough to compute both Reynolds numbers at the end of inflation, allowing us to ignore the evolution of the plasma at later times.

Now, all we have to know is which regime (viscous or turbulent) does apply at the reheating temperature. This is given by the value of $\mathcal{R}_e$ at that time. 
Inserting the above expressions for $v$, Eqs.~(\ref{eq:ScalingViscous}) and (\ref{eq:ScalingTurbulence}), in the definition of $\mathcal R_e$,  Eq.~(\ref{Reynolds-e-def}), we obtain at reheating~\cite{Cado:2021bia}
\bse  \textrm{For}\quad \mathcal{R}_e^{\rm rh} < 1 \quad \Rightarrow\quad \mathcal{R}_e^{\rm rh} &   \approx & 2.5\cdot 10^{-9} \; \frac{\rho_{B_Y} \,\ell^2_{B_Y}}{H_E^2} \, \left( \frac{H_E}{10^{13} \, \text{GeV}} \right)\, \left(\frac{T_{\rm rh}}{T_{\rm rh}^{\rm ins}}  \right)^{\frac{2}{3}}, \\ 
\textrm{For}\quad\mathcal{R}_e^{\rm rh} > 1 \quad \Rightarrow\quad  \mathcal{R}_e^{\rm rh} &  \approx & 5.4\cdot 10^{-5} \; \frac{\sqrt{\rho_{B_Y}} \,\ell_{B_Y}}{H_E} \, \left( \frac{H_E}{10^{13} \, \text{GeV}} \right)^{\frac{1}{2}}\, \left(\frac{T_{\rm rh}}{T_{\rm rh}^{\rm ins}}  \right)^{\frac{1}{3}} \hspace{-2mm}.  \ese

\begin{figure}[htb]
\begin{center}
\includegraphics[width = 8.2cm]{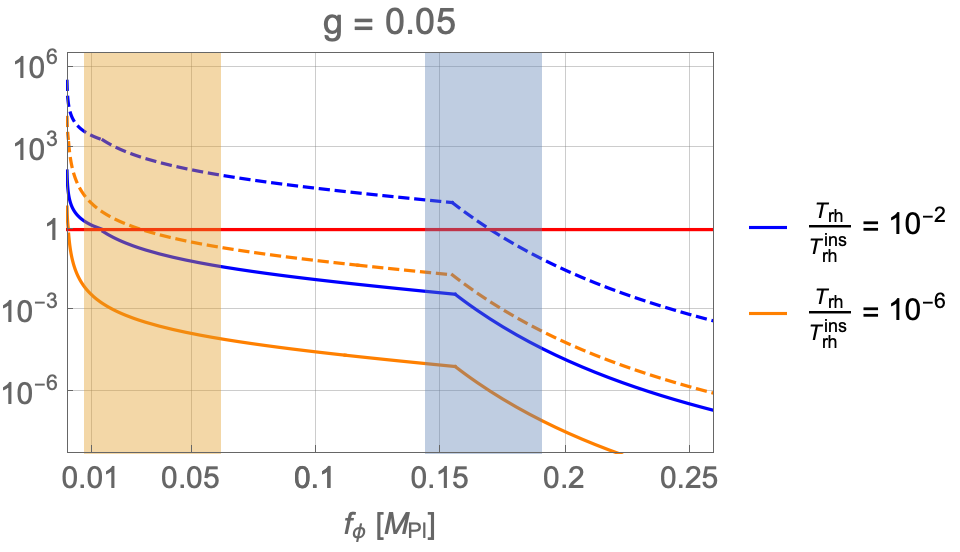}\hspace{1mm}
\includegraphics[width = 6.8cm]{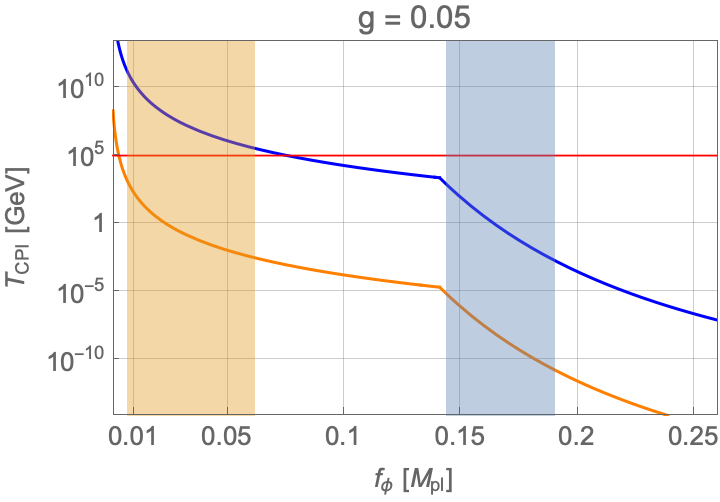}
\includegraphics[width = 6.7cm]{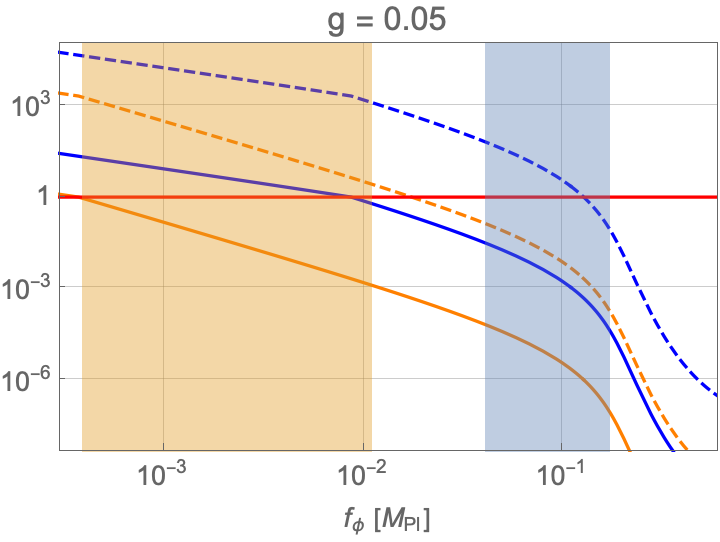}\hspace{13mm}
\includegraphics[width = 7.1cm]{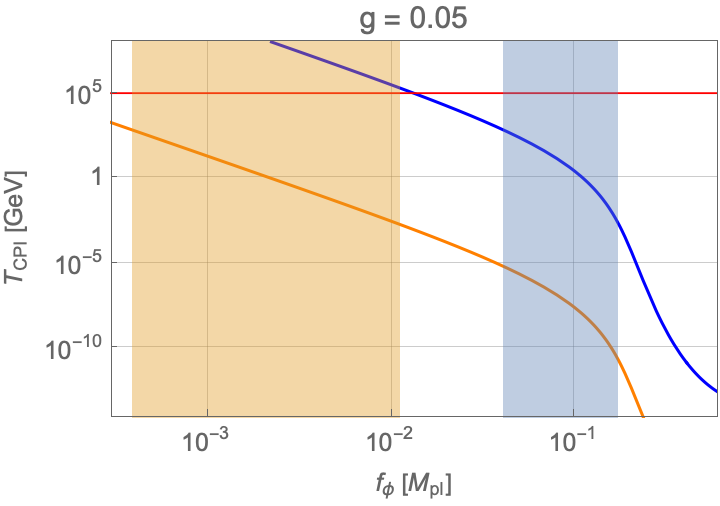}
 \caption{\it Left panels: Plot of the electric (solid lines) and magnetic (dashed lines) Reynolds number at reheating as a function of $f_\phi$ for different values of $T_{\rm rh}/T_{\rm rh}^{\rm ins}=$ $10^{-2}$ (blue color) and $10^{-6}$ (orange color). The ranges of successful baryogenesis for the different values of $T_{\rm rh}/T_{\rm rh}^{\rm ins}$ are displayed here by the vertical bands, whose colors match the corresponding lines color. We see that the production of the helical magnetic fields at reheating always occurs for $\mathcal{R}_e <1$ but not necessarily for $\mathcal R_m>1$, in the correct baryogenesis region. The latter condition must nevertheless be met for successful baryogenesis, which reduces the parameter window mainly (but not only) for high reheating temperatures.  Right panels: Plot of the $T_{\rm CPI}$ temperature. In the baryogenesis regions we always have $T_{\rm CPI}<10^5$~GeV. Top panels correspond to the Schwinger maximal estimate, and bottom panels to the equilibrium estimate.
\label{fig:baryoRe}}
\end{center}
\end{figure}
In our scenario it turns out that $\mathcal{R}_e^{\rm rh}<1$ for the range of parameters that provides a successful baryogenesis, as displayed in the left panels of  Fig.~\ref{fig:baryoRe} (solid lines) for the two extreme values of the parameter $T_{\rm rh}/T_{\rm rh}^{\rm ins}=$ $10^{-2}$ (blue color) and $10^{-6}$ (orange color). Thus, the plasma starts in the viscous regime and the magnetic Reynolds number should be computed using Eq.~(\ref{constraint-Rm}). Plots of $\mathcal R_m^{\rm rh}$, as a function of $f_\phi$, are shown in the left panels of Fig.~\ref{fig:baryoRe} (dashed lines) for the same values of the parameter $T_{\rm rh}/T_{\rm rh}^{\rm ins}$ and the same color codes. We consider the Schwinger maximal (top left panel) and equilibrium (bottom left panel) estimates for the gauge fields. In all cases we exhibit the regions allowed by the baryogenesis constraint, which depend on the corresponding values of the parameter $T_{\rm rh}/T_{\rm rh}^{\rm ins}$, using the same color code than for the different lines (both for $\mathcal{R}_e^{\rm rh}$ and $\mathcal{R}_m^{\rm rh}$) in the plot. Then even if  $\mathcal{R}_e^{\rm rh}<1$, at some later time $\tau$ the plasma will eventually fall into the turbulent regime where  $\mathcal{R}_e>1$, with evolution given by Eq.~(\ref{constraint-Rmb}).

As we can see from the dashed lines in the left panel plots of Fig.~\ref{fig:baryoRe} the condition $\mathcal{R}_m^{\rm rh}>1$ is not satisfied everywhere in the region allowed by baryogenesis. 
Therefore, as summarized in Fig.~\ref{fig:parameter-space}, the condition for magnetic induction dominance, $\mathcal{R}_m>1$, constrains the available region (\ref{window-baryogenesis}) from the baryogenesis window. Of course, once the condition $\mathcal{R}_m^{\rm rh}>1$ is satisfied (at the reheating temperature), its value increases with time, see Eq.~(\ref{eq:scalings}), which guarantees that the condition will be fulfilled until the EWPT.

\subsection{The chiral plasma instability \label{sec:CPI}}
When the symmetric phachiral plasma instabilityse is restored during reheating both, the asymmetries in the particle/antiparticle number densities, and the hypercharge helicity, are generated via the Schwinger effect, as described in Sec.~\ref{sec:schwinger}, and via the chiral anomaly, as stated at the beginning of Sec.~\ref{sec:field-production}, see Eq.~(\ref{anomaly}). In the absence of any other process, the newly generated asymmetry will relax into the same amount of the newly generated helicity but with opposite sign, as the gauge fields configuration has lower energy density than the fermion states configuration, resulting in a cancellation of the total helicity and hence no baryogenesis at the EWPT. This phenomenon is called the chiral plasma instability (CPI)~\cite{Kamada:2018tcs, Joyce:1997uy, Boyarsky:2011uy, Akamatsu:2013pjd, Hirono:2015rla, Yamamoto:2016xtu, Rogachevskii:2017uyc}, and has to be avoided for a successful baryogenesis. 

CPI can be avoided if we require that the CPI timescale is long enough to allow all fermionic states to come into chemical equilibrium (so that sphalerons can erase their corresponding asymmetries in particle number densities) before CPI can happen. The estimated temperature at which CPI takes place is~\cite{Cado:2021bia}
\be T_{\rm CPI}/\textrm{GeV}  \approx    4 \cdot 10^{-7} \,  \; \frac{\mathscr{H}_Y^2}{H_E^6} \, \left( \frac{H_E}{10^{13} \, \text{GeV}} \right)^3\left(\frac{T_{\rm rh}}{T_{\rm rh}^{\rm ins}}  \right)^2 \,.
\label{constraint-TCPI}\ee

The last fermion species to enter chemical equilibrium, through its Yukawa coupling with the left-handed electron $e_L$, is the right-handed electron, $e_R$, and it happens at the temperature $T\sim 10^5$~GeV. Indeed, when the fermionic states are in chemical equilibrium, their asymmetry is washed out through weak sphalerons and Yukawa couplings. Therefore the constraint $T_{\rm CPI}\lesssim 10^5$~GeV
guarantees that the CPI cannot occur before the smallest Yukawa coupling reaches equilibrium and all particle number density asymmetries are erased, preventing thus the cancellation of the helicity generated at the reheat temperature.

In the right panels of Fig.~\ref{fig:baryoRe} we show the plot of $T_{\rm CPI}$ as a function of $f_\phi$ for both, Schwinger maximal (top panel) and equilibrium (bottom panel), estimates and values of $T_{\rm rh}/T_{\rm rh}^{\rm ins}=$ $10^{-2}$ (blue color) and $10^{-6}$ (orange color). In each plot, the region between the vertical bands is that selected by the baryogenesis mechanism for the corresponding value of $T_{\rm rh}/T_{\rm rh}^{\rm ins}$ with the same color code. As we can see from Fig.~\ref{fig:baryoRe}, the range of values for $T_{\rm CPI}$ in the corresponding baryogenesis region is
\be
10^2\ \textrm{GeV} \;\gtrsim  \; T_{\rm CPI}\;\gtrsim \; 10^{-3}\  \textrm{GeV}
\ee
which then prevents the cancellation of any previously generated helicity. So, as we will explicitly exhibit in Fig.~\ref{fig:parameter-space}, this constraint is satisfied in all the region provided by the baryogenesis condition. 

\subsection{Primordial non-Gaussianity}
Inflation predicts that the statistical distribution of primordial fluctuations is nearly Gaussian. Measuring deviations from a Gaussian distribution, i.e.,~non-Gaussian correlations in primordial fluctuations, is a powerful test of inflation.
While the two-point function for $\delta\chi$ defines the power spectrum, the three-point correlation function encodes departures from Gaussianity~\cite{Komatsu:2001ysk,WMAP:2010qai}. Helical gauge fields yield a new source of cosmological perturbations for the inflaton field $\delta\chi$ as
\be \left( \dfrac{\partial^2 }{\partial t^2} + 3H \dfrac{\partial }{\partial t} - \dfrac{\nabla^2}{a^2} \right) \delta\chi=- \dfrac{4}{a^4 f_\phi}\;\bm{E} \cdot \bm{B}. \label{CMB-perturbations} 
\ee 

The magnitude of the three-point function is conventionally quantified using the parameters $f_{\rm NL}$. non-Gaussian effects from helical gauge fields are maximal when the three modes have comparable wavelength, the so-called equilateral form, which in the backreactionless case where gauge fields are given by Eq.~(\ref{Amplified-mode}) is given by~\cite{Barnaby:2010vf,Barnaby:2011qe}
\be
f_{\rm NL}^{\rm equil} \;\simeq \;4.7\, \cdot\, 10^{-16}\; \frac{e^{6 \pi \xi_{\rm CMB}}}{\xi^9_{\rm CMB}} ,\,
\label{eq:fNL}
\ee 
where $\xi_{\rm CMB}\equiv \xi(\chi_*)$. However, we have seen that the Schwinger effect significantly reduces the magnitude of the RHS of Eq.~(\ref{CMB-perturbations}), for a fixed value of $f_\phi$, and that we can mimic its effect by the replacement of the effective parameters $\xi_{\rm eq}$ and $\xi_{\rm max}$, for the equilibrium and maximal estimates respectively, in the backreactionless expression for $\langle\bm{E} \cdot \bm{B}\rangle$, Eq.~(\ref{EB}). Hence, in the same way as we did at the beginning of Sec.~\ref{sec:reheating}, we identify the primordial non-Gaussianity constraint with $\xi_{\rm eq}$ and $\xi_{\rm max}$ before translating them back to $f_\phi$ by the use of Eqs.~(\ref{Schw-Equ-Xieff}) and (\ref{xi-ximax-trick}).

Current observational bounds on non-Gaussianity of the cosmic microwave background (CMB) anisotropies lead to~\cite{Planck:2019kim}
\be
f_{\rm NL}^{\rm equil}=-26\pm 47
\ee 
which translate, from Eq.~(\ref{eq:fNL}) into $\xi_{\rm CMB}\lesssim 2.54$ (95\% C.L.). Using now the scaling relation
\be
\frac{\xi_{\rm eq/max}}{\xi_{\rm CMB}}=\sqrt{\frac{\epsilon(\chi_E)}{\epsilon(\chi_*)}}=\sqrt{\frac{1}{\epsilon(\chi_*)}}
\label{eq:scaling}
\ee
where $\xi_{\rm eq/max}\equiv \xi_{\rm eq/max}(\chi_E)$ is the value of the effective $\xi$ parameter at the end of inflation in the equilibrium/maximal Schwinger estimate, one can compute corresponding upper bounds on $\xi_{\rm eq/max}$, at the end of inflation. 
In fact, for the lower value of $g$ allowed by the cosmological observables in our inflation model, $g\simeq 0.01$, one gets $\xi_{\rm eq/max} \lesssim 47$ while for the upper bound, $g\simeq 0.05$, one gets $\xi_{\rm eq/max} \lesssim 91$.
 However those values of $\xi_{\rm eq/max}$ are never reached in our model, as they would correspond to negligibly small values of $f_\phi$ which are never met.

In conclusion, in the presence of the Schwinger effect the produced gauge fields are never strong enough to trigger non-Gaussianity in the distribution of the primordial inflaton fluctuations, in good agreement with present observations. In other words the model prediction in the presence of the fermionic Schwinger currents is $f_{\rm NL}^{\rm equil}\simeq 0$, and so we will not consider further this constraint.

\subsection{The baryon isocurvature perturbation}
Many models of baryogenesis using (hyper)magnetic fields try to simultaneously explain the origin of the large scale, intergalactic magnetic fields (IMF) measured today by the Fermi satellite~\cite{Neronov:2010gir, Tavecchio:2010mk, Ando:2010rb}. They all face a balance problem when addressing this issue.

While maximally helical fields can indeed generate the BAU without explaining the observed IMF, they would suffer from baryon overproduction should they try to accommodate IMF. 
In the case of a mixture of helical and nonhelical fields, the baryogenesis is less effective so that stronger hypermagnetic fields are needed to explain the present BAU, and hence they could meet the lower bound from the IMF observations.

However, it has been recently shown that such models are inconsistent with the baryon isocurvature perturbations, that are constrained by the observations of cosmic microwave background on large scales~\cite{Kamada:2020bmb}.
In particular, it was pointed out that the baryon isocurvature perturbations at a scale larger than the neutron diffusion scale at the Big Bang Nuclesynthesis (BBN) epoch is constrained by the deuterium overproduction due to the second-order effect \cite{Inomata:2018htm}. This translates into an upper bound on the volume average of the baryon isocurvature perturbation, as 
\be \overline{\mathcal S^2}_{\rm B,BBN} < 0.016 \quad (2\sigma)\label{eq:isocurvature}.\ee

It was shown that, regardless of their helicity properties, hypermagnetic fields with too large strength and coherence length are not allowed before the EWPT~\cite{Kamada:2020bmb}.
Still baryogenesis from the hypermagnetic helicity decay can be responsible for the present BAU, but additional magnetogenesis, or an unknown mechanism of the magnetic field amplification after the EWPT, is needed to fit the IMF observations. However the constraint becomes more severe for less helical hypermagnetic fields.
In our model the magnetic field produced at the end of inflation is maximally helical and we do not cope with the IMF observations. Hence we should be safe from this constraint. Nevertheless, we will deserve to App.~\ref{sec:isocurvature} the detailed calculation where it is proven that the bound (\ref{eq:isocurvature}) is indeed widely satisfied in our model, so that this constraint does not need to be taken into account any further.

\subsection{Summary of constraints}
\label{sec:summaryresult}

To close this section, we would like to compile all our results about baryogenesis into a single plot, see Fig.~\ref{fig:parameter-space}. Here, we have displayed in the plane $(f_\phi,T_{\rm rh}/T_{\rm rh}^{\rm ins})$ all the relevant constraints described in this section.
In particular:
\begin{itemize}
\item
The generated baryon asymmetry at the EW crossover, given by Eq.~(\ref{constraint-nB}), should be given by the observational value
\be
\eta_B\simeq 9\cdot 10^{-11},
\ee
where the broadness of the prediction band is associated to the uncertainty in the determination of the parameter $f_{\theta_W}$.
\item
The magnetic diffusion given by MHD, leading to the helicity decay should be smaller than the magnetic induction, to allow helicity to be conserved until the electroweak phase transition. This happens when the magnetic Reynolds number at reheating given by Eq.~(\ref{constraint-Rm}) is
\be
\mathcal R_m^{\rm rh}\gtrsim 1.
\ee
\item
As the symmetric phase is restored during reheating, an asymmetry via chiral anomaly is generated and decays into a helicity with opposite sign, resulting into a cancellation of the total helicity with no baryogenesis at the EWPT. This phenomenon, called chiral plasma instability, can be avoided if the temperature at which it is produced $T_{\rm CPI}$, given by Eq.~(\ref{constraint-TCPI}), is smaller than the temperature at which all fermion species enter chemical equilibrium through their Yukawa couplings, and in particular the last species to reach chemical equilibrium, $e_R$. This condition is satisfied provided that
\be
T_{\rm CPI}\lesssim 10^5 \textrm{ GeV}.
\ee
\end{itemize}

 We shall choose the overlapping region as that meeting all the constraints. We removed the dependence on $g$ as the results are not sensitive to it, 
\begin{figure}[h]
\begin{center}
\includegraphics[width = 8.8cm]{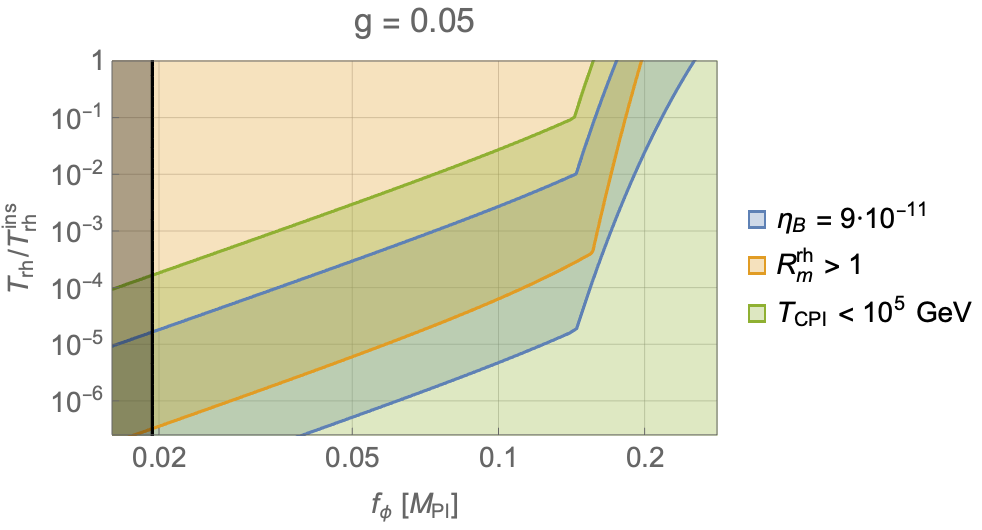}
\includegraphics[width = 6.5cm]{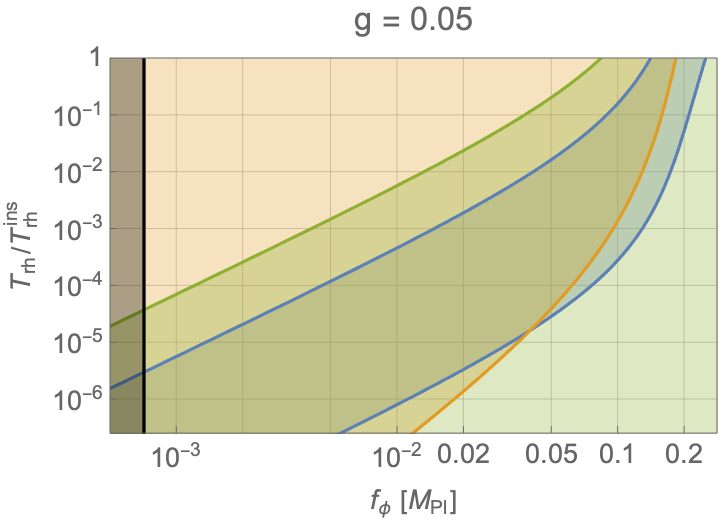}
\caption{\it Summary of constraints on baryogenesis for $g=0.05$ (the dependence on $g$ is tiny) in the plane $(f_\phi/\Mp,T_{\rm rh}/T_{\rm rh}^{\rm ins})$. The considered constraints are on $\eta_B$ (blue area), on the magnetic Reynolds number (orange area) and on chiral plasma instability (green area). We seek for the overlapping region. On the left side of each plot, the black band displays the region where the backreaction of gauge fields on the inflaton can no longer be neglected. Left panel: Schwinger maximal estimate. Right panel: Schwinger equilibrium estimate.
\label{fig:parameter-space}} 
\end{center}
\end{figure}
preferring to choose the value $g=0.05$ in the allowed range from the inflation model.

From Fig.~\ref{fig:parameter-space} we can conclude that the CPI constraint is satisfied in all the region where the constraint of having enough baryon asymmetry $\eta_B$ holds. On the other hand the constraint from the magnetic Reynolds number is effective for the case of the Schwinger maximal estimate, by cutting off the larger available values of the parameter $f_\phi$ for every value of $T_{\rm rh}$. However, for the Schwinger equilibrium estimate the magnetic Reynolds number constraint is effective for the larger values of $T_{\rm rh}/T_{\rm rh}^{\rm ins}$,  by cutting off the larger values of the parameter $f_\phi$, while for the smaller values of $T_{\rm rh}$, in particular for $T_{\rm rh} \lesssim 5\cdot 10^{-4}\;T_{\rm rh}^{\rm ins}$, it entirely covers the region satisfied by the constraint on $\eta_B$. Finally, given the range $m\in[10^3\, ,5\cdot 10^{10}]$~GeV, for the corresponding range on $T_{\rm rh}/T_{\rm rh}^{\rm ins}\in [10^{-2}\, ,10^{-6}]$, we get the available (approximated) regions, for $g\in [0.01\, ,0.05]$,

\be\begin{aligned} & \begin{aligned}
f_\phi/\Mp&\in[ 0.14\, ,0.17 ]\hspace{21mm} \textrm{for}\quad  T_{\rm rh}/T_{\rm rh}^{\rm ins}=10^{-2}\\
f_\phi/\Mp&\in[ 1.9 \cdot 10^{-2}\, , 2.8 \cdot 10^{-2} ]\hspace{5mm} \textrm{for}\quad  T_{\rm rh}/T_{\rm rh}^{\rm ins}=10^{-6}
\end{aligned} \hspace{1cm} \text{(Maximal estimate)}, \\[2mm]
& \begin{aligned}
f_\phi/\Mp&\in[ 4.1 \cdot 10^{-2}\, ,0.13 ]\hspace{12mm}  \textrm{ for}\quad  T_{\rm rh}/T_{\rm rh}^{\rm ins}=10^{-2}\\
f_\phi/\Mp&\in[ 7.2 \cdot 10^{-4}\, , 1.1 \cdot 10^{-2} ]\hspace{5mm}  \textrm{for}\quad  T_{\rm rh}/T_{\rm rh}^{\rm ins}=10^{-6}
\end{aligned} \hspace{1cm} \text{(Equilibrium estimate)} .
\end{aligned} \nonumber\ee

Let us also mention that the condition (\ref{backreaction-condition}) on the nonbackreaction of the gauge fields on the inflaton, displayed by the black bands, becomes a constraint only at low temperature, $T_{\rm rh}/T_{\rm rh}^{\rm ins} \lesssim 2\cdot 10^{-5} \; (3\cdot 10^{-6})$ for the Schwinger maximal (equilibrium) estimate. Finally, the condition for the nonbackreaction of fermion currents on the gauge fields, which corresponds to $f_\phi\gtrsim 0.19\,\Mp$, is outside the region of validity of the baryogenesis region, which shows that the Schwinger effect can never be neglected in the baryogenesis analysis.

\section{Some phenomenological considerations}
\label{sec:pheno}
In some chaotic inflation models, the mass of the inflaton is constrainted to a high value because of the observational constraint on the scalar perturbations amplitude.
In our model, though, we have two terms in the inflaton potential: while inflation is controlled by the quartic term, dominant at Planckian scales, the quadratic one controls reheating and low energy physics. Thus the value of the inflaton mass is decoupled from the inflationary dynamics.  

In previous sections, we have considered on the one hand the upper value of the inflaton mass as $m\lesssim \mathcal Q_I$, small enough to solve the instability problem of the Higgs potential, and on the other hand we have roughly imposed $m\gtrsim 1$ TeV on phenomenological grounds for the theory to not being excluded by present experimental data. In fact, an inflaton mass at the TeV scale could have implications for low energy physics. 
Therefore, in this section we will make some considerations from the point of view of collider physics and the Standard Model in the presence of the inflaton field with the interactions appearing in the Lagrangian~(\ref{eq:potencial}).

\subsection*{The naturalness problem}
First of all, our theory has two hierarchically separated scales, the inflaton mass $m$ and the Higgs mass $m_h=125.25$~GeV, with $m\gg m_h$. As such, the theory should exhibit a hierarchy problem, which in general implies an unnatural fine-tuning of the parameters. In the absence of any symmetry protecting the EW scale from the high-scale UV physics, one has either to accept the fine-tuning (as it is customary done in the Standard Model) or to lower the value of the mass $m$ as much as possible. More quantitatively, the coupling in the Lagrangian $\mu \phi |\mathcal H|^2=\sqrt{2\delta_\lambda}m|\mathcal H|^2$ generates a contribution to the Higgs mass term $\mu_h^2$ through the one-loop radiative corrections. In the limit $\mu\to 0$ (i.e.~$\delta_\lambda\to 0$), there is an enhanced $\mathbb Z_2$ symmetry $\phi\to -\phi$ indicating that any value of $\mu$, as small as it can be, is natural in the sense of 't Hooft, since in this limit the symmetry is recovered.
Moreover, this coupling induces a correction to the parameter $\mu_h^2$ in the Lagrangian as~\cite{deGouvea:2014xba}
\be
\Delta \mu_h^2\simeq -\frac{\delta_\lambda}{8\pi^2}\,m^2\,\log\frac{m^2}{m_h^2} \,.
\ee 
Naturalness would then require $|\Delta \mu_h^2|\lesssim \mu_h^2=m_h^2/2$, which translates into the bound
\be
m\lesssim  1.2 \textrm{ TeV}\,,
\ee
where we have considered the typical value of the coupling $\delta_\lambda\simeq 0.1$. This leads to the exciting possibility of having an inflaton with an $\mathcal O(\textrm{TeV})$ mass, which does not spoil naturalness, solve the problem of the instability of the EW minimum, and has phenomenological implications for present and future colliders.

\subsection*{The Higgs-inflaton mixing}

Near the vacuum, the potential for the Higgs and $\phi$ fields is given by
\be
V(\phi,\mathcal H)=-\sqrt{2\delta_\lambda}\,m\, \phi |\mathcal H|^2+\frac{1}{2}m^2\phi^2-\mu_h^2|\mathcal H|^2+\lambda_0|\mathcal H|^4\,.
\ee
The vacuum is defined as the solution to the minimum equations $\partial V/\partial \phi=\partial V/\partial h=0$, which provides $\langle h\rangle=v=246$~GeV and $\langle\phi\rangle=v_\phi$, with
\be \mu_h^2=\lambda v^2,\hspace{3cm} v_\phi=\sqrt{\dfrac{\delta_\lambda}{2}}\, \frac{v^2}{m},
\ee
where the parameters $\delta_\lambda$ and $\lambda$ were defined in Eqs.~(\ref{eq:defdelta}) and (\ref{eq:deflambda}), respectively.

In the presence of the parameter $\delta_\lambda$, there is a mixing between the Higgs $h$ and $\phi$ fields given by the squared mass matrix at the minimum 
\be
\mathcal M^2=\begin{pmatrix} 2(\lambda +\delta_\lambda)v^2 & & -\sqrt{2\delta_\lambda}\,mv\\ -\sqrt{2\delta_\lambda}\,mv & & m^2 \,
\label{eq:M2}
\end{pmatrix} .
\ee
This matrix is diagonalized by an orthogonal rotation with angle $\alpha$~\footnote{We are using the notation $c_\alpha\equiv\cos\alpha$, $s_\alpha\equiv\sin\alpha$, $t_\alpha\equiv\tan\alpha$.} as
\be
\begin{pmatrix} c_\alpha & s_\alpha \\ -s_\alpha & c_\alpha \end{pmatrix} \mathcal M^2 \begin{pmatrix} c_\alpha & -s_\alpha \\ s_\alpha & c_\alpha \end{pmatrix}=\begin{pmatrix} m_{\tilde h}^2 & 0 \\ 0 & m_{\tilde\phi}^2 \end{pmatrix}\,,
\ee
such that the mass eigenstates are
\be
\tilde h=c_\alpha\, h+s_\alpha\, \phi,\hspace{2 cm} \tilde \phi=c_\alpha\, \phi-s_\alpha\, h\,,
\label{eq:physicalstates}
\ee
and the mass eigenvalues are
\be
\frac{m^2_{\tilde h,\, \tilde\phi}}{m^2}=\frac{1}{2}+\left(\lambda+\delta_\lambda  \right)\frac{v^2}{m^2}\mp
\sqrt{\frac{1}{4}-\left(\lambda-\delta_\lambda  \right)\frac{v^2}{m^2}+\left(\lambda+\delta_\lambda  \right)^2 \frac{v^4}{m^4}
}.
\label{eq:masas}
\ee

In this way the physical mass eigenstate $\tilde h$ is associated with the Standard Model Higgs, with a mass $m_{\tilde h}=125.25$~GeV, while $\tilde\phi$ is the physical singlet, and both of them are coupled to the SM fields through the mixing angle $\alpha$. 

Hence this theory predicts then the existence of a scalar $\tilde \phi$ that decays mainly into the channel $\tilde\phi\to \tilde h \tilde h$ with a decay rate
\be
\Gamma(\tilde\phi\to \tilde h\tilde h)=\frac{\kappa^2 \,m}{32\pi}\sqrt{1-\frac{4 m_{\tilde h}^2}{m_{\tilde \phi}^2}},\quad \kappa=\sqrt{2\delta_\lambda}c_\alpha(1-3s_\alpha^2)+6 s_\alpha c_\alpha^2(\lambda+\delta_\lambda)\frac{v}{m}.
\ee
which was responsible for the reheating in Sec.~\ref{sec:reheating}.
Contour lines of $\Gamma(\tilde\phi\to \tilde h\tilde h)$ are exhibited in the upper left panel of Fig.~\ref{fig:pheno-all} in the parameter space $(m,\delta_\lambda)$. As we can see, typically the width of the resonance $\tilde \phi$ is around a few GeV. As was already stated in Sec.~\ref{sec:reheating}, there are also subleading decay channels into SM particles ($X\in \textrm{\rm SM}$), as $\tilde\phi\to X\bar X$, induced by the mixing with the Higgs, with very suppressed branching fractions 
\be
\mathcal B(\tilde\phi\to X\bar X)=\mathcal B(\tilde h\to X\bar X) \cdot s^2_\alpha \;\frac{\Gamma_{\tilde h}}{\Gamma_{\tilde \phi}}
 \ee
as $\Gamma_{\tilde h}\simeq 4 c^2_\alpha$ MeV in the SM, $\Gamma_{\tilde \phi}\simeq \Gamma(\tilde\phi\to \tilde h\tilde h)\simeq $ few GeV, so that $s^2_\alpha \Gamma_{\tilde h}/\Gamma_{\tilde \phi}\ll 1$. 
\subsection*{Electroweak precision constraints}
The doublet-singlet mixing can affect the electroweak precision observables (EWPO) through changes in the gauge boson propagators. Explicit expressions for the modified scalar contributions to the $W$ and $Z$ propagators are given in Refs.~\cite{Profumo:2007wc,Barger:2007im}. In particular the contribution to the $S$ and $T$ oblique parameters from the new physics, $\Delta S\equiv S^{\rm NP}-S^{\rm SM}$ and $\Delta T\equiv T^{\rm NP}-T^{\rm SM}$, are found to be given by  
\be
\Delta T\simeq \frac{3}{16\pi}\frac{s_\alpha^2}{s_W^2}\left[\left(
\frac{1}{c_W^2}\frac{m_{\tilde h}^2}{m_{\tilde h}^2-m_Z^2}\log\frac{m_{\tilde h}^2}{m_Z^2}
-\frac{m_{\tilde h}^2}{m_{\tilde h}^2-m_W^2}\log\frac{m_{\tilde h}^2}{m_W^2} \right)-\left( m_{\tilde h}\to m_{\tilde \phi}\right)
\right]
\label{eq:deltaT}
\ee
and
\be
\Delta S=\frac{s_\alpha^2}{12\pi}\left[\frac{\hat m_{\tilde h}^6-9\hat  m_{\tilde h}^4 +3\hat m_{\tilde h}^2+5+12\hat m_{\tilde h}^2\log(\hat m_{\tilde h}^2)}{(\hat m_{\tilde h}^2-1)^3}  -\left(\hat  m_{\tilde h}\to\hat  m_{\tilde \phi} \right)  \right]
\label{eq:deltaS}
\ee
where we are defining masses in units of $m_Z$, i.e.~$\hat m_X\equiv m_X/m_Z$.

The model predictions, Eqs.~(\ref{eq:deltaT}) and (\ref{eq:deltaS}), must be compared with the experimental values, given by~\cite{Zyla:2020zbs}
\be
\Delta T=0.05\pm 0.06,\quad \Delta S=0.0\pm 0.07
\label{eq:Peskin-Takeuchi-constraint}
\ee
and 92\% correlation between the $S$ and $T$ parameters. This gives rise to a $\Delta\chi^2(m,\delta_\lambda)$ distribution, which defines the allowed region in the parameter space $(m,\delta_\lambda)$, exhibited in all panels of Fig.~\ref{fig:pheno-all}. In particular we display, in orange shading, the region in the parameter space $(m,\delta_\lambda)$ for which $\Delta\chi^2(m,\delta_\lambda)<5.99$, that corresponds to the bound at 95\% C.L. As we can see, for large values of the parameter $\delta_\lambda$ the lower bound on $m$ can be near the TeV scale.

\begin{figure}[htb]
\begin{center}
\includegraphics[width = 8.7cm]{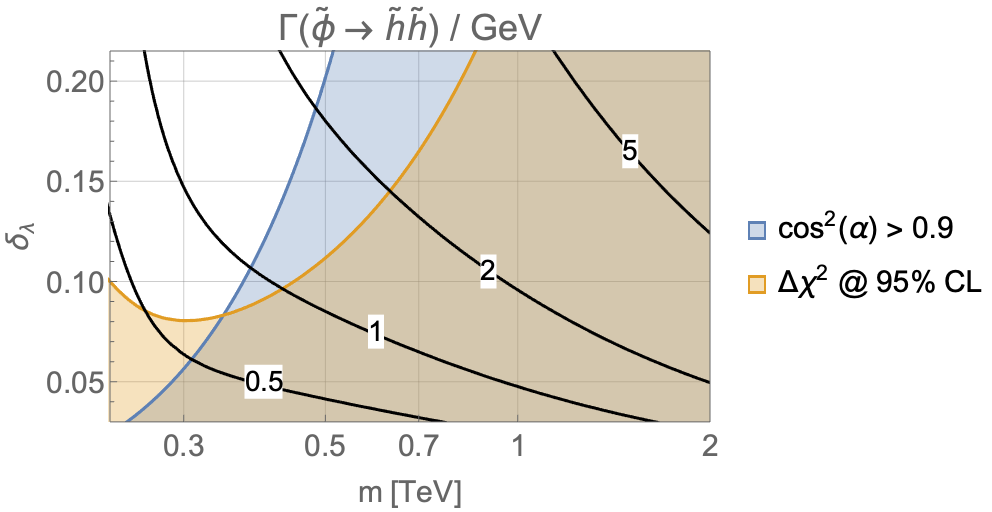}
\includegraphics[width = 6.7cm]{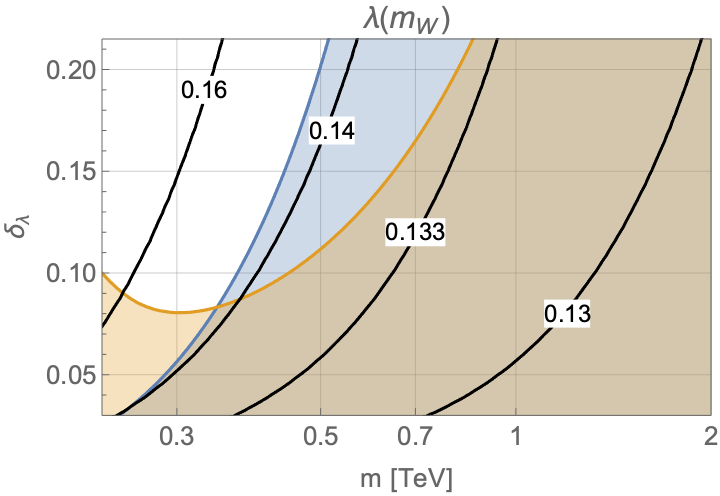}\vspace{2mm}
\includegraphics[width = 6.9cm]{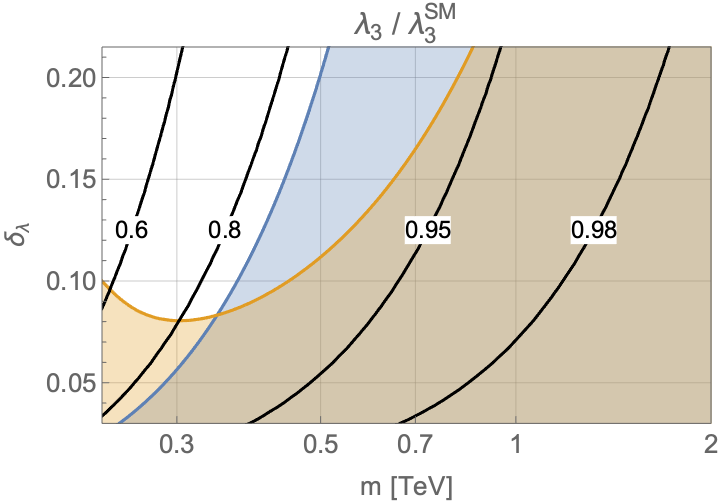}\hspace{16mm}
\includegraphics[width = 6.9cm]{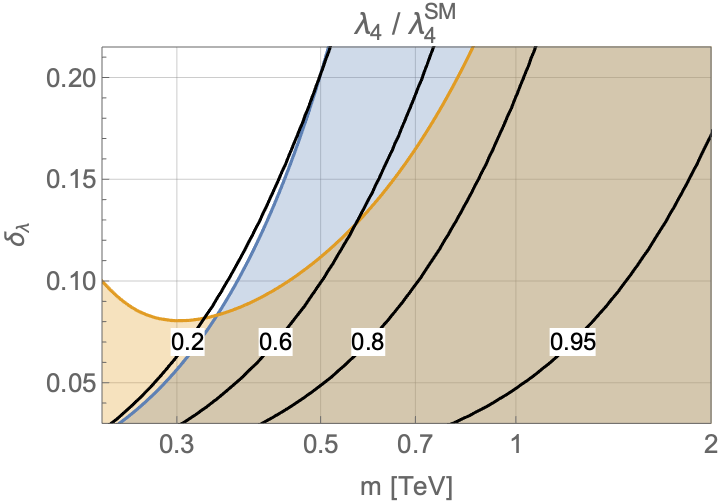}
\caption{\it Contour lines in the plane $(m/\textrm{TeV},\delta_\lambda)$ of the decay rate $\Gamma(\tilde\phi\to \tilde h\tilde h)/$GeV (top left panel), the quartic parameter at the weak scale $\lambda(m_W)$ (top right panel) as well as the Higgs trilinear (bottom left panel) and quartic (bottom right panel) couplings normalized to the SM values with the regions of validity defined by the signal strength modifier (\ref{eq:bound}) (blue) and the constraints from the electroweak parameters (\ref{eq:Peskin-Takeuchi-constraint}) (orange) superimposed. One should read the contour lines in black that pass through the overlapping region, and hence that satisfy both constraints. As discussed in Sec.~\ref{sec:stability}, for $m \sim \mathcal{O}$(TeV), to solve the stability problem and after imposing that the theory remains in the perturbative regime up to the high scale, the parameter $\delta_\lambda$ is constrained to be in the region $0.05\lesssim \delta_\lambda \lesssim 0.2$. 
\label{fig:pheno-all}} 
\end{center}
\end{figure}

\subsection*{LHC constraints}
In this section we will consider several constraints arising from LHC physics where we are led to the exciting possibility to explore the inflaton sector at present and future high energy colliders and, in particular, at the LHC.

\subsubsection*{\it - The Higgs signal strength}
From Eq.~(\ref{eq:physicalstates}) we see that the coupling of the mass eigenstate $\tilde h$ to the SM particles, is suppressed, with respect to the coupling of the SM Higgs $h$, by the factor $c_\alpha$. Given that, the signal strength modifier $r_i^f$ for a specific process $ i\to \tilde h\to f$, is given by
\be
r_i^f=\frac{\sigma_i \, \mathcal B^f}{(\sigma_i)_{\rm SM}\,\mathcal B^f_{\rm SM}}\simeq c^2_\alpha
\ee
where $\sigma_i$ is the production cross section for the initial state into $\tilde h$, and $\mathcal B^f$ its branching fraction on the final state. For the last equality we have considered that the production cross section is suppressed by $c^2_\alpha$ while the branching fraction is approximately equal to the SM one. Experimental data from ATLAS~\cite{ATLAS:2019nkf} and CMS~\cite{CMS:2018uag} provide the global values
\be
r=1.11^{+0.09}_{-0.08}\quad \textrm{(ATLAS)},\hspace{2cm} r=1.17\pm 0.1\quad  \textrm{(CMS)}
\ee
which are consistent with a value of $r=1$ (the SM prediction) with $\sim 10\%$ error, thus providing a lower bound on $c_\alpha$ as
\be
c^2_\alpha\gtrsim 0.9.
\label{eq:bound}
\ee 
For $m\gg v$ the mixing angle is $s_\alpha\simeq \sqrt{2\delta_\lambda}( v/m)\ll 1$ so that the bound (\ref{eq:bound}) is easily satisfied. However for TeV values of $m$ the bound (\ref{eq:bound}) translates into a lower bound on the value of $m$.
We shade in blue, in all panels of Fig.~\ref{fig:pheno-all}, the region in the parameter space $(m,\delta_\lambda)$, where this constraint is satisfied. 
In particular we see that, for $\delta_\lambda=0.1$, the bound (\ref{eq:bound}) is satisfied for $m\gtrsim 0.4$~TeV. For $m\simeq 1$~TeV and $\delta_\lambda=0.1$, the mixing is given by $c^2_\alpha\simeq 0.988$, which is not excluded by the actual LHC data. 

\subsubsection*{\it - Trilinear and quartic Higgs couplings}
As the light state $\tilde h$ is to be identified with the SM Higgs, with mass $m_{\tilde h}=125.25$~GeV, for any fixed value of the parameter $\delta_\lambda$ the experimental value of the Higgs mass fixes the value of the quartic parameter at the weak scale, $\lambda(m_W)$, at a different value than in the SM case. In the upper right panel of Fig.~\ref{fig:pheno-all} we plot contour lines of $\lambda(m_W)$ in the parameter space $(m,\delta_\lambda)$. As we can see $\lambda(m_W)>\lambda_{\rm SM}(m_W)$, and only for values of $m\to\infty$ one recovers the SM value.

Moreover, the mixing of the Higgs with the singlet $\phi$ modifies, in the broken phase, the trilinear $\lambda_3$ and quartic $\lambda_4$ SM couplings. Recent experiments on di-Higgs searches are putting bounds on these two parameters by looking for possible departures with respect to the SM values $\lambda_3^{\rm SM}\equiv v\lambda_{\rm SM}$ and $\lambda_4^{\rm SM}\equiv\lambda_{\rm SM}$. In our theory the $h$-$\phi$ mixing angle $\alpha$ generates such a departure. After going to the broken phase by means of the shifts $\tilde\phi\to\tilde\phi+\tilde v_\phi$, $\tilde h\to \tilde h+\tilde v$ (where $\tilde v\equiv c_\alpha\, v+s_\alpha\, v_\phi$ and  $\tilde v_\phi\equiv c_\alpha\, v_\phi-s_\alpha\, v$), and integrating out the field $\tilde\phi$, which yields the value
\be
\tilde\phi=c_\alpha \left[\sqrt{\frac{\delta_\lambda}{2}}(1-3s_\alpha^2)+3(\lambda+\delta_\lambda)s_\alpha c_\alpha\frac{v}{m}\right]\frac{\tilde h^2}{m}+\cdots\,,
\ee
one gets the Higgs potential, in the broken phase,
\be
V(\tilde h)=\frac{1}{2}m_{\tilde h}^2 \tilde h^2+\lambda_3 \tilde h^3+\frac{1}{4}\lambda_4 \tilde h^4+\cdots
\ee
where the ellipses are higher order terms, giving rise to powers $\tilde h^n$ ($n>4$) in the potential, and
\be\begin{aligned}
\lambda_3=& \; c^3_\alpha\,v\left[ \lambda+\delta_\lambda-t_\alpha \, \sqrt{\frac{\delta_\lambda}{2}} \, \frac{m}{v}\right],\\
\lambda_4=& \;c_\alpha^4\lambda+c_\alpha^2(-c_\alpha^4-4 s_\alpha^4+4 c_\alpha^2 s_\alpha^2+c_\alpha^2)\delta_\lambda\\
-&6\sqrt{2\delta_\lambda}\,c_\alpha^3 s_\alpha(c_\alpha^2-2 s_\alpha^2)(\lambda+\delta_\lambda)\frac{v}{m}-18 s_\alpha^2 c_\alpha^4(\lambda+\delta_\lambda)^2\frac{v^2}{m^2}.
\end{aligned}
\ee

The model can then, in the future, be excluded or confirmed by experimental data on trilinear (and quartic) Higgs couplings data. Notice that in the limit $m\gg m_h$ the mixing angle behaves as $s_\alpha\simeq \sqrt{2\delta_\lambda}\,v/m$ so that $\lambda_3\simeq \lambda_3^{\rm SM}$ and $\lambda_4\simeq \lambda_4^{\rm SM}$~\footnote{Of course, in the limit $m\gg v$, $\lambda\simeq \lambda_{\rm SM}$ as exhibited in the top right panel of Fig.~\ref{fig:pheno-all}.}, and the decoupling is automatic. We plot in the bottom panels of Fig.~\ref{fig:pheno-all} contour lines of the trilinear and quartic couplings, normalized to the corresponding SM values, as functions of the parameters $m$ and $\delta_\lambda$. At present, with 89 fb$^{-1}$ of LHC data, the triple Higgs coupling has been constrained by the ATLAS collaboration to be $\lambda_3/\lambda_3^{\rm SM}=4.0^{+4.3}_{-4.1}$, excluding it outside the interval  $[-3.2\, ,11.9]$ at 95\% C.L.~\cite{ATL-PHYS-PUB-2019-009}, while the CMS collaboration finds $\lambda_3/\lambda_3^{\rm SM}=0.6^{+6.3}_{-1.8}$, excluding it outside the interval $[-3.3\,,8.5]$ at 95\% C.L.~\cite{CMS:2020tkr}. Theoretical studies based on the HE-LHC at $\sqrt{s}=27$~TeV and 15 ab$^{-1}$ luminosity foresee exploring the interval range $\lambda_3/\lambda_3^{\rm SM}\in[0.6\, ,1.46]$ at 68\% C.L.~\cite{Homiller:2018dgu}, while a future 100 TeV hadron collider could achieve the trilinear coupling measurement within better than 5\% accuracy~\cite{Goncalves:2018qas}, thus potentially imposing strong constraints on $m$ from the plots in Fig.~\ref{fig:pheno-all}.

\subsubsection*{\it - Heavy Higgs production}
Finally the state $\tilde \phi$ can be produced at the LHC by the same mechanisms of Higgs production with a cross section given by
\be
\sigma(pp\to\tilde\phi+X)=s^2_\alpha \,\sigma(pp\to H+X)
\ee
where $H$ is a heavy SM-like Higgs with a mass equal to $m$. Using the results of inclusive cross sections for $\sigma(pp\to H)$ for the leading mechanism of gluon-gluon fusion (ggf)~\cite{LHCHiggsCrossSectionWorkingGroup:2011wcg} we plot, in Fig.~\ref{fig:cross-section},
the cross section $\sigma_{ggf}(pp\to \tilde\phi)$ as a function of $m$ for two relevant values of the parameter $\delta_\lambda$ for $m\lesssim 1$~TeV and a center of mass energy $\sqrt{s}=13$~TeV. 
Given that, as we have explained earlier in this section $\mathcal B(\tilde\phi\to \tilde h\tilde h)\simeq 1$, we can compare these cross sections with the SM cross sections for di-Higgs production $\sigma(pp\to hh)$ given by $\sigma_{ggf}^{\rm SM}(hh)\simeq  33.5$ fb~\cite{LHCHiggsCrossSectionWorkingGroup:2016ypw}. 

\begin{figure}[htb]
\begin{center}
\includegraphics[width = 11cm]{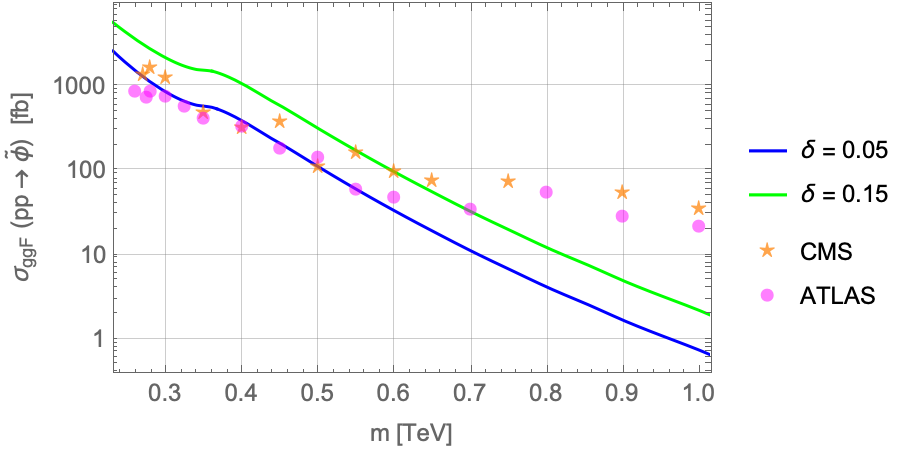}
\caption{\it Plots of cross section $\sigma(pp\to\tilde\phi)$ in fb for relevant values of $\delta_\lambda=$ 0.05, 0.15. 
%The horizontal line shows the SM value for $\sigma_{ggf}^{\rm SM}(pp\to hh)\simeq 33.5$ fb for gluon-gluon fusion for $\sqrt{s}=13$~TeV.
The dots (stars) are the 95\% C.L. upper bound from ATLAS~\cite{ATLAS:2019qdc} (CMS~\cite{CMS:2018ipl}), that bring the approximate constraint $m\gtrsim 0.55$-$0.7$~TeV, depending on the value of $\delta_\lambda$. 
\label{fig:cross-section}} 
\end{center}
\end{figure}
The predicted cross sections in Fig.~\ref{fig:cross-section} are compared with the present experimental upper bounds at 95\% C.L. on the production of a scalar field ($\tilde\phi$) which decays into two Higgs bosons, from ATLAS with luminosities 27.5-36.1 fb$^{-1}$~\cite{ATLAS:2019qdc} and CMS with luminosity 35.9 fb$^{-1}$~\cite{CMS:2018ipl}, at present LHC center of mass energies, $\sqrt{s}=13$~TeV (see Fig.~\ref{fig:cross-section}). We conclude from here that the present lower bounds on the value of $m$ are
\be
m\gtrsim 0.55 \ (0.7) \ \textrm{TeV @ 95\% C.L.,} \quad \textrm{for } \delta_\lambda=0.05\ (0.15) ,
\ee 
while in the future much stronger bounds could be achieved.

\section{Conclusion}
\label{sec:conclusions}

In this paper, we have explored the possibility of modifying the Higgs inflation theory by means of the introduction of an extra scalar field $\phi$, with the Ricci coupling $(g/2)\phi^2R $, and an interaction term $\mu\, \phi h^2 $ to solve the stability problem of the electroweak vacuum.
Both fields, $\phi$ and $h$, participate in the dynamics of inflation through the two-field potential $V(\phi,h)$, which has the shape of a valley in which they are related by simple analytical expressions so that we can express one field in term of the other. 
This allows us to define the true inflaton field $\chi$ as the one following the valley with canonical kinetic term although we kept the description in terms of $\phi$ for mathematical convenience.

A key point is that we have considered for the $\phi$ field a quartic coupling $\lambda_\phi$ and a mass  $m$, such that inflation is driven by the quartic coupling term, while reheating is driven by the mass term. 
The Lagrangian coupling $\delta_\lambda=\mu^2/2m^2$ triggers a positive contribution to the $\beta$ function of the Higgs quartic coupling such that, if the mass scale $m$ is in the range $1 \textrm{ TeV}\lesssim m\lesssim\mathcal Q_I$, where $\mathcal Q_I\simeq 10^{11}$~GeV is the instability scale of the electroweak potential, the instability problem of the electroweak vacuum can be solved just by roughly imposing the mild constraint $\delta_\lambda\lesssim \mathcal O(1)$.

We find that the beginning of inflation $\phi=\phi_\ast$ ($N_\ast = 60$) is mainly driven by the scalar field $\phi$, and since the amplitude of density perturbations is fixed by the $\phi$ quartic coupling (and not by the Higgs quartic coupling), the main problem of Higgs inflation is easily solved with $g\lesssim 1$. On the other hand, the end of inflation ($N\simeq 0$), where the hypermagnetic helicity will be produced, is equally driven by both the scalar $\phi$ and the Higgs $h$ quartic terms, so that the role played by the Higgs field is relevant. Both regimes are separated, for $g\simeq 0.01$ (0.05), by a critical value of the field $\phi_c/\Mp\simeq 10$ (4), which corresponds to the critical number of $e$-folds $N_c\simeq 12$ (2).
After imposing the Planck and BICEP/Keck conditions on the slow roll parameters and the unitarity condition $\phi_*\lesssim \Mp/g$ (see however footnote~\ref{foot} for a nuance) we obtain the allowed interval on the parameter $g$, $0.01\lesssim g\lesssim 0.05$, which translates into the 
prediction for the cosmological observables in agreement with observations, and with a Hubble parameter
almost saturating the Planck upper bound $H_*^{\rm obs}<6\cdot 10^{13}$ GeV:
%
%\begin{equation}\nonumber \small \boxed{ \begin{aligned}
%0.965\lesssim n_s\lesssim 0.967,\quad 0.047 \gtrsim r \gtrsim \; 0.012 ,\quad 5.5\cdot 10^{13}\textrm{ GeV}\gtrsim H(\phi_*)\gtrsim 2.8\cdot 10^{13} \textrm{ GeV} \end{aligned}}
%\end{equation}
%
\begin{center}\begin{tcolorbox}[enhanced,width=14.3cm,center upper,fontupper=\small,drop shadow southwest] 
$0.965\lesssim n_s\lesssim 0.967,\hspace{1cm} 0.047 \gtrsim r \gtrsim \; 0.012 ,\hspace{1cm} 5.5\cdot 10^{13}$ GeV $\gtrsim H(\phi_*)\gtrsim 2.8\cdot 10^{13} $ GeV \end{tcolorbox}\end{center}

During the last $e$-folds of inflation we generate maximally helical magnetic fields that will source the BAU via the $(B+L)$ anomaly of the SM during the EWPT~\cite{Anber:2006xt,Bamba:2006km,Bamba:2007hf,Anber:2009ua,Anber:2015yca,Cado:2016kdp,Sfakianakis:2018lzf,Kamada:2016eeb,Kamada:2016cnb,Jimenez:2017cdr,Domcke:2019mnd,Cado:2021bia}. This results from the introduction of a coupling of the Chern-Simons term of the hypercharge gauge group with the inflaton, as $\phi Y_{\mu\nu}\widetilde Y^{\mu\nu}$, with coupling strength $\Mp/f_\phi$, that breaks the $\mathcal{CP}$ symmetry. This effective $\mathcal{CP}$ breaking operator can be easily obtained from a UV completion with a $\mathcal{CP}$-violating Yukawa coupling of $\phi$ to a hypercharged vector like heavy fermion, as it is shown in App.~\ref{sec:UV}. This source of $\mathcal{CP}$-violation is needed by the Sakharov conditions~\cite{Sakharov:1967dj}, the two other conditions being provided by the chiral anomaly of the SM, which violates the baryon number, and the helical magnetic fields conversion to baryon asymmetry at the EWPT, which happens during EW sphalerons freeze out, when they go out of thermal equilibrium.

%The inflaton coupling to gauge fields in the effective $\mathcal {CP}$ breaking operator can generate electromagnetic gauge fields nonperturbatively with a net magnetic helicity.
%At reheating this helicity is converted into hypermagnetic helicity which, under certain conditions, can survive until the EWPT.
%The produced (hyper)helicity and (hyper) electric/magnetic energy densities are exponentially sensitive to the coupling strength $\Mp/f_\phi$, and a window for the right amount of the BAU is obtained.
%During the electroweak crossover, the hypermagnetic helicity is converted back into ordinary magnetic helicity. The former enters the anomaly equation of the current for baryon plus lepton number, $(B+L)$, while the latter does not. The conversion of the helicity during the EWPT therefore leads to the production of a compensating $(B + L)$-asymmetry. This is partly washed out by the EW sphalerons, but since sphalerons freeze-out while the phase transition is in progress, a sizable net $(B + L)$-asymmetry can remain, in a rate which depends on the reheating temperature of the inflaton.

We have undertaken both backreation processes, namely the one of the gauge fields on the inflaton, and the one of the thermally produced chiral fermions on the gauge fields, known as the Schwinger effect. The latter significantly reduces the amount of electromagnetic energy and helicity generated at the end of inflation as, for $f_\phi/\Mp \lesssim 0.19$, we have to trade their exponential behavior with two polynomial cases: the maximal and the equilibrium estimates. %Hence larger values of the coupling strength, i.e.~$f_\phi/\Mp\gtrsim 0.02$, are required to achieve the BAU, while in the absence of the Schwinger effect, $f_\phi/ \Mp \gtrsim 0.11$. 
This raises the effective coupling of the $\mathcal{CP}$-violating term $\Mp/f_\phi$, thus compensating its lowest overall value since electromagnetic fields are simultaneously weakened.

As for the former issue, we have found a critical value of the coupling strength of $\phi$ to gauge bosons in the $\mathcal{CP}$-violating operator, $f_\phi^c$, such that for $f_\phi \gtrsim f^c_\phi$, the backreaction of the gauge field on the inflaton can be neglected. In particular we find $f_\phi^c/\Mp\simeq 0.02 \; (7 \cdot 10^{-4})$ for the Schwinger maximal (equilibrium) estimate. On the contrary, for $f_\phi\lesssim f_\phi^c$, the field $\phi$ is \textit{strongly} coupled to the gauge fields, the backreaction of the latter on the inflaton equations of motion cannot be neglected, and the preheating of the Universe proceeds by the nonperturbative production of gauge fields.
In this paper we have concentrated in the case $f_\phi\gtrsim f_\phi^c$, where the field $\phi$ is \textit{weakly} coupled to the gauge fields, the backreaction of gauge fields on the inflaton dynamics can be neglected and the Universe reheating proceeds by the perturbative decay of the inflaton into SM particles. %Therefore, by imposing no backreaction of the gauge fields on the inflaton dynamics, the baryon asymmetry mechanism using the Schwinger maximal (equilibrium) estimate imposes the lower bound on the parameter $f_\phi  \gtrsim 0.02 \; (7 \cdot 10^{-4}) \, \Mp$.
Besides, we have considered the constraints from non-Gaussianity of primordial fluctuations, and baryon isocurvature perturbations, and find that they have no influence on our model. 

Concerning the value of the baryon asymmetry of the Universe
generated at the EW crossover, $\eta_B$ depends on the value of the reheat temperature $T_{\rm rh}$, and in particular on its ratio with respect to the reference instant reheat temperature $T_{\rm rh}/T_{\rm rh}^{\rm ins}$
(in our model $T_{\rm rh}^{\rm ins}\simeq 2\cdot 10^{15}$ GeV). As we are imposing no backreaction of gauge fields on the inflaton dynamics, and reheating should proceed by perturbative inflaton decays, the value of $T_{\rm rh}/T_{\rm rh}^{\rm ins}$ depends on the inflaton decay width $\Gamma_\chi$. In our model the inflaton mainly decays through the channel $\chi\to hh$, with a width which increases with the value of the inflaton mass $m$ and ranges in the interval $1 \textrm{ GeV}\lesssim\Gamma_\chi\lesssim 10^9 \textrm{ GeV}$, which corresponds to  $10^{-6}\lesssim T_{\rm rh}/T_{\rm rh}^{\rm ins}\lesssim 10^{-2}$, for $1 \textrm{ TeV}\lesssim m\lesssim 5\cdot 10^{10} \textrm{ GeV}$.

Moreover, as helicity is converted into baryon asymmetry at the EWPT, while it was produced at the end of inflation, it has to survive MHD processes between these two times. By imposing that the magnetic Reynolds number is bigger than unity, and that the chiral plasma instability effect does not washout the produced gauge fields, the available parameter window is reduced by an amount which depends on the value of the reheating temperature. We have shown that all these constraints are satisfied for a large range of the parameters $f_\phi$ and $T_{\rm rh}$, very insensitive to the value of the  parameter $g$, inside its allowed range from cosmological observables:
%for the Schwinger maximal estimate, we find the allowed range $0.14\lesssim f_\phi/\Mp\lesssim 0.17$, for 
%$r_{\rm rh}\simeq 10^{-2}$,
% %which corresponds to $m\simeq 5\cdot 10^{10}$~GeV,
%and $0.02\lesssim f_\phi/\Mp\lesssim 0.03$, for $r_{\rm rh}\simeq 10^{-6}$.
%%, for $m\simeq 10^{3}$~GeV. 
%For the equilibrium estimate the allowed ranges become $0.04\lesssim f_\phi/\Mp\lesssim 0.13$ for 
%$r_{\rm rh}\simeq 10^{-2}$, and $0.0007\lesssim f_\phi/\Mp\lesssim 0.01$ for $r_{\rm rh}\simeq 10^{-6}$.
%
%In particular, 
%\be \small \boxed{ \begin{aligned} & \begin{aligned}
%f_\phi/\Mp&\in[ 0.14\, ,0.17 ]\hspace{20mm} \textrm{for}\quad  T_{\rm rh}/T_{\rm rh}^{\rm ins}=10^{-2}\\
%f_\phi/\Mp&\in[ 1.9 \cdot 10^{-2}\, , 2.8 \cdot 10^{-2} ]\ \textrm{ for}\quad  T_{\rm rh}/T_{\rm rh}^{\rm ins}=10^{-6}
%\end{aligned} \hspace{1cm} \text{(Maximal estimate)} \\[2mm]
%& \begin{aligned}
%f_\phi/\Mp&\in[ 4.1 \cdot 10^{-2}\, ,0.13 ]\hspace{10mm}  \textrm{ for}\quad  T_{\rm rh}/T_{\rm rh}^{\rm ins}=10^{-2}\\
%f_\phi/\Mp&\in[ 7.2 \cdot 10^{-4}\, , 1.1 \cdot 10^{-2} ]\ \textrm{ for}\quad  T_{\rm rh}/T_{\rm rh}^{\rm ins}=10^{-6}
%\end{aligned} \hspace{1cm} \text{(Equilibrium estimate)} 
%\end{aligned} }\nonumber\ee
%
\begin{center}\begin{tcolorbox}[enhanced,width=13.2cm,center upper,fontupper=\small,drop shadow southwest] 
$\begin{aligned} & \begin{aligned}
f_\phi/\Mp&\in[ 0.14\, ,0.17 ]\hspace{17mm} \textrm{for}\quad  T_{\rm rh}/T_{\rm rh}^{\rm ins}=10^{-2}\\
f_\phi/\Mp&\in[ 1.9 \cdot 10^{-2}\, , 2.8 \cdot 10^{-2} ]\ \textrm{ for}\quad  T_{\rm rh}/T_{\rm rh}^{\rm ins}=10^{-6}
\end{aligned} \hspace{1cm} \text{(Maximal estimate)} \\[2mm]
& \begin{aligned}
f_\phi/\Mp&\in[ 4.1 \cdot 10^{-2}\, ,0.13 ]\hspace{8.5mm}  \textrm{ for}\quad  T_{\rm rh}/T_{\rm rh}^{\rm ins}=10^{-2}\\
f_\phi/\Mp&\in[ 7.2 \cdot 10^{-4}\, , 1.1 \cdot 10^{-2} ]\ \textrm{ for}\quad  T_{\rm rh}/T_{\rm rh}^{\rm ins}=10^{-6}
\end{aligned} \hspace{1cm} \text{(Equilibrium estimate)} 
\end{aligned} $
\end{tcolorbox}\end{center}
The complete available parameter region is summarized in Fig.~\ref{fig:parameter-space}.

Notice that the fact that the inflaton potential has both quadratic and quartic terms allows to decouple the mass $m$ from the actual value of the amplitude of density perturbations, which in the absence of a quartic term would fix its value to $m\simeq 10^{12}$~GeV (or smaller at the price of the introduction of a curvaton scalar), allowing any value $m<\mathcal Q_I$ in order to stabilize the electroweak vacuum. This is achieved by the contribution, to the Higgs quartic coupling $\beta$ function, provided by the coupling $\delta_\lambda$ in the Lagrangian term, $\sqrt{2\delta_\lambda}\,m\,\phi\,|\mathcal H|^2$. We have found for the parameter $\delta_\lambda$ the absolute bounds, $\delta_\lambda\gtrsim 0.05$ in order to solve the stability problem, and $\delta_\lambda\lesssim 0.35$ to not spoil the perturbativity of the theory, although its particular range depends on the actual value of $m$. Nevertheless, values $m\gg m_h$ create a naturalness/fine-tuning problem, essentially given by the fact that there appears a loop correction to the Higgs squared mass term $\mu_h^2$. It translates into a fine-tuning of the order of $4\pi^2/(\delta_\lambda \rho^2 \log\rho^2)$ where $\rho\equiv m/m_h$. While for $\delta_\lambda\simeq 0.1$ and $m=10^{10}$~GeV the fine tuning is $\sim 10^{-14}$ (similar to the SM fine-tuning), and for $m=10$~TeV it is $\sim 0.01$, there is essentially no fine-tuning for values $m\lesssim 1$ TeV. This leads to the exciting possibility of a light inflaton which could possibly be detected by direct measurements at LHC and/or future colliders.

The key point here was that the Lagrangian term $\sqrt{2\delta_\lambda}\,m\,\phi\,|\mathcal H|^2$ creates a $\phi$-$h$ mixing, sizable for low values of the mass $m$, leading to an interesting phenomenology for high energy colliders. In fact all the collider phenomenology is triggered by the mixing angle $\alpha$. 
%\begin{itemize}
%\item
The mass eigenstates $(\tilde\phi,\tilde h)$, where $\tilde h$ should be identified with the experimentally detected Higgs with a mass equal to 125.25 GeV, are related to the weak states $(\phi,h)$ by a rotation with angle $\alpha$. This fact triggers that $\lambda(m_W)$ be different from $\lambda_{\rm SM}(m_W)\simeq 0.13$, which leads to predictions on the ratios $\lambda_3/\lambda_3^{\rm SM}$ and $\lambda_4/\lambda_4^{\rm SM}$ which could be probed by future experiments, as HE-LHC and/or a 100 TeV collider.
%\item
The mixing is already bounded by present ATLAS and CMS results on the SM Higgs signal strengths, which provides the bound $m\gtrsim 0.3 \ (0.45)$~TeV for $\delta_\lambda=0.05\ (0.15)$. 
%\item
It also generates a contribution to the oblique electroweak observables, and yields for e.g.~$\delta_\lambda=0.15$ the lower bound $m\gtrsim 0.5$~TeV. 
%\item
Finally, the mixing is responsible for the inflaton production and decay. In particular $\tilde\phi\to \tilde h\tilde h$, triggered by the coupling $\delta_\lambda$, is the main decay channel, while other decay channels into the SM particles, via the mixing $s_\alpha$, are subleading. The inflaton $\tilde \phi$ can also be produced mainly by the gluon-gluon fusion mechanism through its Higgs mixing. Present data from ATLAS and CMS translate into lower bounds $m\gtrsim 0.55 \ (0.7)$~TeV at 95\% C.L. for $\delta_\lambda=0.05\ (0.15)$.
%\end{itemize}

There are a number of research lines which could be safely explored in the future. First of all, we have considered models of inflation based on the Ricci coupling $\phi^2R$, and a $\phi$ dependent potential dominated, for large values of $\phi$, by the quartic coupling. This kind of theories, when considered in the Einstein frame, give naturally rise, for large values of $\phi$, to flat potentials, appropriate for inflation, without invoking any particular symmetry. It is clear that similar results could be obtained for theories with a Ricci coupling as $F(\phi)R$, and a Jordan frame potential behaving, for large values of $\phi$, as $U(\phi)\simeq F^2(\phi)$. In particular it would be interesting to see what kind of theories would produce enough baryon asymmetry in the presence of a period of preheating, by the nonperturbative production of gauge fields. A very recent work~\cite{Kodama:2021yrm} has 
already explored a general class of inflationary potentials and shown consistency with cosmological observables. In particular our model, labeled therein by $(n,p)=(2,4)$, give results for the cosmological observables, which are in good agreement with this paper. These general theories are therefore good candidates to generate also the observed value of the BAU, provided they contain the inflaton coupling to the Chern-Simons term. 
In addition, as far as we are aware of, there are no in-depth studies in the literature of preheating mechanisms taking into account the Schwinger effect, which has led us to make some shortcuts in this article. Therefore we leave for future work a more rigorous study of nonperturbative production of gauge fields at preheating, leading to the BAU, that takes this effect into account.
Lastly, at the level of particle physics it remains as an exciting playground the possibility of detecting the inflaton at present or future colliders, or that future experimental results on the production of heavy scalars, coupled to the SM fields, or on the measurements of the trilinear and quartic Higgs couplings, by di-Higgs production, could start cornering the present theory and put stronger bounds on the mass of the inflaton and its mixing with the SM Higgs.

\vspace{0.5cm}
\section*{Acknowledgments}
%%%%%%%%%
This work is supported by the Departament d’Empresa i Coneixement, Generalitat de Catalunya Grant No.~2017SGR1069, by the Ministerio de Economía y Competitividad Grant No.~FPA2017-88915-P. IFAE is partially funded by Centres de Recerca de Catalunya. Y. C. is supported by the European Union’s Horizon 2020 research and innovation programme under the Marie Skłodowska-Curie Actions No.~754558.

\appendix

\section{UV completion for $\mathcal{CP}$-violation}
\label{sec:UV}
$\mathcal{CP}$-violation in our model is driven by the effective dimension-five operator
\be
S_{\cancel{\mathcal {CP}}}=- \int d^4x \; \frac{\phi}{4 \tilde f_\phi}Y_{\mu\nu}\tilde Y^{\mu\nu}
\ee
where $Y^{\mu\nu}$ is the hypercharge field strength. 

A simple UV completion generating such effective operator can be a massive (with mass $M$) hypercharged vectorlike fermion $\psi$ with a $\mathcal{CP}$-violating Yukawa coupling to $\phi$ as
\be\mathcal L=-\bar\psi_L(M+|\lambda_\psi|e^{i\theta_\lambda}\phi)\psi_R+\text{h.c.}=-|\lambda_\psi|\phi\left[\cos\theta_\lambda\bar\psi\psi+\sin\theta_\lambda\bar\psi i\gamma_5\psi  \right]\label{lag}
\ee
where $\mathcal{CP}$-violation is induced by the angle $\theta_\lambda$.
The $\mathcal{CP}$-even $\phi Y_{\mu\nu}Y^{\mu\nu}$, and  $\mathcal{CP}$-odd $\phi Y_{\mu\nu}\tilde Y^{\mu\nu}$, couplings are generated by loop diagrams where the fermion $\psi$ propagates in the loop and emits two gauge bosons $Y_\mu$, via the $\cos\theta_\lambda$ and $\sin\theta_\lambda$ couplings in Eq.~(\ref{lag}), respectively. The corresponding Feynman diagrams are finite and thus one gets $\tilde f_\phi\propto M$.
For maximal $\mathcal{CP}$-violation, i.e.~$\theta_\lambda=\pm\pi/2$, only the coupling $\phi Y_{\mu\nu}\tilde Y^{\mu\nu}$ is generated such that
\be M\simeq \frac{|\lambda_\psi|g_Y^2}{4\pi^2}\tilde f_\phi\simeq 8\cdot 10^{15} \textrm{ GeV}\,|\lambda_\psi|\,(\tilde f_\phi/\Mp)  \,. \ee

\subsection*{\it - Stability of the inflationary potential}

The UV completion here proposed could affect the stability of the inflationary potential through radiative corrections in the high energy theory.
In fact, the coupling in Eq.~(\ref{lag}) provides a correction to the $\beta$ function of the coupling $\lambda_\phi$, similar to the correction to the $\beta$ function of the Higgs quartic coupling coupling $\beta_\lambda$ from the top quark Yukawa coupling. This contribution comes from the box diagram with four $\phi$ external legs, where the fermion $\psi$ is exchanged, and the resulting contribution to $\beta_{\lambda_\phi}$ is given by
\be \Delta\beta_{\lambda_\phi}=-\dfrac{2|\lambda_\psi|^4}{16\pi^2}\,\theta(t-t_M),\quad t-t_M=\log \mathcal (Q/M).
\label{eq:Deltabetalambdaphi}
\ee
Notice that the correction given by Eq.~(\ref{eq:Deltabetalambdaphi}) is negative, as it arises from a fermion loop, which can lead the coupling $\lambda_\phi$ to negative values and thus destabilize the whole inflationary scenario, a process similar to the destabilization of the EW vacuum by the loop corrections induced by the top quark. It is then required to prevent such destabilization.
A sufficient condition to not destabilize the quartic inflaton coupling, without any tuning of parameters, is to impose $|\lambda_\psi|\lesssim \lambda_\phi^{1/4}$ which translates, using the typical value, from Fig.~\ref{fig:lambdaphi},  $\lambda_\phi\simeq 10^{-12}$, into $|\lambda_\psi|\lesssim 10^{-3}$, and so into an upper value of the $\psi$-mass as 
\be M\lesssim 10^{13}\;\textrm{GeV}(\tilde f_\phi/\Mp). \ee
Notice that in the limit $\lambda_\psi\to 0$ the UV Lagrangian has the enhanced $\mathbb Z_2$ symmetry, $\phi\to -\phi$, and thus any small value of $\lambda_\psi$ is natural in the sense of 't Hooft. For instance, values of $\lambda_\psi \sim 10^{-12}$ would lead to values of $M\simeq \mathcal O(\textrm{TeV})$. 

\subsection*{\it - Naturalness problem}

The UV completion brings a new naturalness problem as there is the hierarchy of masses $M\gg m_h$.
In fact, the presence of the vectorlike fermion $\psi$ coupled to the field $\phi$ through the coupling (\ref{lag}), along with the $\phi$-$h$ mixing generates the Lagrangian
\be
\mathcal L=|\lambda_\psi| s_\alpha\, \tilde h \,\bar\psi i\gamma_5\psi+|\lambda_\psi|  c_\alpha\,\tilde\phi\,\bar\psi i\gamma_5\psi
\label{interaccionA}
\ee
whose first term provides at one-loop (for scales $\mathcal Q\gtrsim M$) a contribution to the mass parameter $\mu_h^2$ as
\be
\Delta \mu_h^2\simeq \frac{1}{4\pi^2}\, s^2_\alpha\, |\lambda_\psi|^2\, M^2\log\frac{M^2}{m_h^2}
\ee
which would require, for large values of $M$, a fine-tuning. In particular, the naturalness condition $\Delta\mu_h^2 \lesssim m_h^2/2$ implies, for
$m\simeq 1$~TeV, the upper bounds on $M$ and $|\lambda_\psi|$ given by
\be
M\lesssim (7.6,\,2.5,\,0.8)\cdot 10^8 \ \textrm{GeV},\quad |\lambda_\psi| \lesssim (1,\,3,\,10)\cdot 10^{-6}\,,
\label{eq:valuesM}
\ee
where the values in parenthesis correspond to $\tilde f_\phi/\Mp=(0.1,0.01,0.001)$, respectively, and where we have used $\delta_\lambda=0.15$.

Of course the second term of (\ref{interaccionA}) can create a second naturalness problem, as $M\gg m$ by radiative corrections providing a one-loop contribution to $m_{\tilde\phi}^2$ as 
\be
\Delta m_{\tilde\phi}^2\simeq \frac{1}{4\pi^2}\, c^2_\alpha\, |\lambda_\psi|^2\, M^2\log\frac{M^2}{m^2}
\ee
However, once we have solved the naturalness problem between $M$ and $m_h$, as $m^2\gg m_h^2$, the second naturalness problem between $M$ and $m$ is automatically solved as, for all values in Eq.~(\ref{eq:valuesM}), it turns out that $\Delta m_{\tilde\phi}^2/m^2\simeq 0.4$.

\subsection*{\it - Cosmological problems}
The Lagrangian (\ref{lag}) has the ($\psi$ number) discrete $\mathbb Z_2$ symmetry $\psi\to-\psi$ making the fermion $\psi$ cosmologically stable, inconsistent with direct Dark Matter detection, and possibly overclosing the Universe. A simple way out is explicitly breaking the $\mathbb Z_2$ symmetry. For instance we can identify $\psi\equiv E$ with a heavy vectorlike, $SU(2)$ singlet, lepton $E=(E_L,E_R)^T$, with hypercharge -1, as the SM right-handed leptons $e_{R_i}$. We can then generate a tiny mixing of e.g.~the third generation leptons with $E$ by means of the Yukawa coupling $Y'_3$
\be\mathcal L_E=-M\bar E_L E_R-Y_3\bar\ell_{L_3} H\tau_{R}-Y'_3\bar\ell_{L_3} HE_R+h.c.\label{mixing}\ee

The mixing in (\ref{mixing}) generates a mass matrix as
\be\begin{pmatrix} \bar\tau_L & \bar E_L\end{pmatrix}\mathcal M \begin{pmatrix} \tau_R\\ E_R\end{pmatrix},\quad \mathcal M=\begin{pmatrix}
m_3 & m'_3\\0 & M \end{pmatrix}\ee
where $m_3=Y_3 v/\sqrt{2}$ is the $\tau$-lepton mass in the absence of the mixing with the heavy fermion, and $m'_3\equiv Y'_3 v/\sqrt{2}$. One can diagonalize the mass matrix $\mathcal M$ with left and right unitary transformations, with angles $\theta_L$ and $\theta_R$, respectively, as
\be\mathcal M_d=U_L^\dagger \mathcal M U_R, \qquad U_{L/R} = \begin{pmatrix}
\sin\theta_{L/R} & \cos\theta_{L/R}\\-\cos\theta_{L/R} & \sin\theta_L \end{pmatrix} . \ee
In the limit $M\gg m_3,m'_3$ we get
\be\sin\theta_L\simeq \frac{m'_3}{M}\left[ 1+\frac{m_3^2}{M^2}+\cdots\right],\quad \sin\theta_R\simeq \frac{m_3m'_3}{M^2}\left[ 1+\frac{m_3^2-m_3^{\prime 2}}{M^2}+\cdots\right] . \ee

As a consequence of the mixing the mass eigenfunctions are shifted as
\bse
\tau_R\to & \tau_R+\dfrac{m_3m'_3}{M^2}E_R,\quad E_R&\to E_R-\dfrac{m_3m'_3}{M^2}\tau_R.\\
\tau_L\to & \tau_L+\dfrac{m'_3}{M}E_L,\quad  E_L&\to E_L-\dfrac{m'_3}{M}\tau_L.
\ese
and the mass eigenvalues as
\be m_3\to m_{\tau}=m_3\left[1-\frac{m'^2_3}{2M^2}+\cdots  \right],\quad M\to M\left[1+\frac{m'^2_3}{2M^2}+\cdots    \right] \ee
by which the fermion $E$ decays as $E\to H \tau$, as well as to leptons and gauge bosons through the mixing with $\tau_L$ and $\tau_R$, as
$E\to W\nu_\tau$ or $E\to \tau Z,\tau\gamma$. These decays prevent the heavy fermion from overclosing the Universe.

\section{Baryon isocurvature perturbations}
\label{sec:isocurvature}

Baryon isocurvature perturbations can be generated by the presence of strong gauge fields~\cite{Kamada:2020bmb}. To be conservative, in this section we will consider the case where the generated gauge fields are as strong as possible: where one neglects the backreaction from the fermionic Schwinger currents (the backreactionless case).
Borrowing the notation from~\cite{Kamada:2020bmb}, we have in our case for the symmetric and antisymmetric combinations, $S(k)=(|A_+(k)|^2+|A_-(k)|^2)/2$ and $A(k)=(|A_+(k)|^2-|A_-(k)|^2)/2$. For the case of maximally helical gauge fields one obtains
\be S(k)\simeq A(k)\simeq \frac{|A_+|^2}{2}\simeq\frac{1}{4k_\lambda} \left(\frac{k}{k_\lambda} \right)^{-\frac{1}{2}} \; \frac{e^{2\pi\xi}}{\xi}\;\exp\left(-4\sqrt{\frac{k}{k_\lambda}}\right), \ee
where we are choosing e.g.~$A_+(k)$ as the amplified mode, and (\ref{Amplified-mode}) was used together with the definition $k_\lambda=a_{\rm E} H_E / 2\xi \simeq 10^{12}$~GeV, which corresponds to the spectrum peak of $A_+$.
Writing $B^2 \simeq 2 \rho_B$ in term of $k_\lambda$, and using (\ref{magnetic-energy}), we obtain a relation for the spectrum given by~\cite{Kamada:2020bmb} %(sec. 4.2)
\be A(k)\simeq \frac{1024\,\pi^2}{315} \frac{B^2}{k_\lambda^5} \left(\frac{k}{k_\lambda} \right)^{-\frac{1}{2}} \exp{\left(-4\sqrt{\frac{k}{k_\lambda}}\right)}. \ee
It may be interesting to note that in these terms, the magnetic field and the helicity are written as
\be B^2\simeq \frac{1}{4\pi^2} \frac{315}{1024} \frac{e^{2\pi\xi}}{2\xi}\; k_\lambda^4,\hspace{1.5cm} \mathscr H  \simeq\frac{2}{7} \frac{B^2}{k_\lambda}. \ee

From this we can estimate the baryon isocurvature perturbation at the BBN as
\be \overline{\mathcal S^2}_{\rm B,BBN}\simeq \frac{7\sqrt{\pi}}{20\sqrt{3}}\left(\frac{k_d}{k_\lambda} \right)^{3}\left(\frac{k_\lambda}{k_\lambda^{\rm EWPT}}\right)^3+\mathcal O\left(\frac{k_d}{k_\lambda}  \right)^{5} , \label{SBBN-constraint}\ee
where $k_d$ is the comoving neutron diffusion scale at the BBN, $k_d^{-1}\simeq 0.0025$~pc. From the expansion ratio $k_d / k_\lambda \sim 10^{-42}$, we can see that Eq.~(\ref{SBBN-constraint}) is suppressed provided that $k_\lambda / k_\lambda^{\rm EWPT}$ is not too big, which we will next demonstrate.

Eq.~(\ref{SBBN-constraint}) should be evaluated at the time of baryon asymmetry production at $T_{\rm EWPT}\simeq 135$~GeV, hence the rescaling for the wave number $k_\lambda$. At first glance, this rescaling could appear to be exactly one since $k_\lambda$ is comoving, but because of the peculiar dynamic of the plasma decribed by the MHD equations, comoving quantities do scale with the expansion of the Universe after reheating, as already stated in section~\ref{sec:evolution-after-reheating}.

We shall now study how the comoving coherence length scales until the EWPT. Every plasma quantity (field amplitude, correlation length, wave number) evolves adiabatically from reheating until the eddy turnover temperature $T_t \simeq v T_{\rm rh}$ where $v$ is the typical bulk velocity of the plasma.
For $T<T_t$ the scaling regime depends on the value of the electric Reynolds number at the end of inflation. 
The velocity of the plasma is
\bse  \mathcal R_e < 1 \quad \Rightarrow\quad v &  \approx & 2.9\cdot 10^{-10}\; \frac{\ell_{B_Y} \rho_{B_Y}}{H_E^3} \left( \frac{H_E}{10^{13} \, \text{GeV}} \right)^{\frac{3}{2}} \left(\frac{T_{\rm rh}}{T_{\rm rh}^{\rm ins}} \right) \\\mathcal R_e > 1 \quad \Rightarrow\quad  v &  \approx & 5.3 \cdot 10^{-6}\; \frac{\sqrt{\rho_{B_Y}}}{H_E^2}\left( \frac{H_E}{10^{13} \, \text{GeV}} \right) \left(\frac{T_{\rm rh}}{T_{\rm rh}^{\rm ins}} \right)^{\frac{2}{3}} \ese
For $\mathcal R_e^{\rm rh}<1$, as $\mathcal R_e$ grows with time, we eventually reach the point where it becomes one, at temperature~\cite{Domcke:2019mnd}
\be T_1\equiv T(\mathcal R_e=1)= \mathcal R_e^{\rm rh} \; T_t.\ee
Once $\mathcal R_e>1$, the scaling regimes for comoving quantities become (\ref{eq:ScalingTurbulence}) until recombination.

In summary, the magnetic energy and correlation length scale adiabatically until the eddy turnover temperature $T_t$, then they scale according to (\ref{eq:ScalingViscous}) until $\mathcal R_e=1$, where the regime changes to (\ref{eq:ScalingTurbulence}) until recombination. However we compute the scaling only until $T_{\rm EWPT}=135$~GeV since the comparison with the neutron diffusion scale must be done at the EWPT temperature~\cite{Kamada:2020bmb}.
This yields a total dilution factor for comoving quantities as
\bse \frac{B_Y^{\rm EWPT}}{B_Y^{\rm rh}} &=& \left(\frac{T_{\rm EWPT}}{T_1}\right)^{\frac{1}{3}}\left(\frac{T_1}{T_t}\right)^{\frac{1}{2}}, \\
\frac{\ell_{B_Y}^{\rm EWPT}}{\ell_{B_Y}^{\rm rh}} &=&\left(\frac{T_{\rm EWPT}}{T_1}\right)^{-\frac{2}{3}}\left(\frac{T_1}{T_t}\right)^{-1}.\ese

We stress that $T_t$ and $T_1$ depend on $v$, which in turn depends on $\ell_{B_Y}$ and $\rho_B$. For values of the parameters space yielding the correct BAU, e.g.~for $g=0.01$, $ f_\phi=0.15\Mp$ and $T_{\rm rh}\simeq 10^{-2}\,T_{\rm rh}^{\rm ins}$ (blue region in Fig.~\ref{fig:baryoRe}) we find that $T_t \simeq 2\cdot 10^{8}$~GeV and $T_1 \simeq 3\cdot 10^{7}$~GeV.
Then we get that the comoving quantities $B_Y$ and $\ell_{B_Y}$ get scaled between the reheating and EWPT temperatures as
\be  \frac{B_Y^{\rm EWPT}}{B_Y^{\rm rh}} \sim 10^{-2},\quad
\frac{\ell^{\rm EWPT}_{B_Y}}{\ell_{B_Y}^{\rm rh}} \sim10^{4}.\ee
Going back to the baryon isocurvature perturbation (\ref{SBBN-constraint}), we hence have
\be \left(\frac{k_\lambda}{k_\lambda^{\rm EWPT}}\right)^3\propto\left( \frac{T_{\rm rh}}{T_{\rm rh}^{\rm ins}} \right)^{-4},\quad
 \left(\frac{k_\lambda}{k_\lambda^{\rm EWPT}}\right)^3
 \sim  10^{-13},\ee
which therefore get for the observable $\overline{\mathcal S^2}_{\rm B,BBN}$ an exceedingly small value. A similar result is obtained for all allowed values of the parameters $(g,f_\phi,T_{\rm rh})$, so for our model the prediction is  $\overline{\mathcal S^2}_{\rm B,BBN} \simeq 0$. This result also holds for the case where the Schwinger effect is considered, as in this case gauge fields are much weaker than in the backreactionless case studied above, and so their contribution to $\overline{\mathcal S^2}_{\rm B,BBN}$ is expected to be much smaller.

\bibliographystyle{JHEP}
\bibliography{refs}

\providecommand{\href}[2]{#2}\begingroup\raggedright\begin{thebibliography}{10}

\bibitem{Sakharov:1967dj}
A.~D. Sakharov, \emph{{Violation of $\mathcal{CP}$ Invariance, $\mathcal{C}$
  asymmetry, and baryon asymmetry of the universe}},
  \href{http://dx.doi.org/10.1070/PU1991v034n05ABEH002497}{\emph{Pisma Zh.
  Eksp. Teor. Fiz.} {\bfseries 5} (1967) 32--35}.

\bibitem{Cohen:1993nk}
A.~G. Cohen, D.~B. Kaplan and A.~E. Nelson, \emph{{Progress in electroweak
  baryogenesis}},
  \href{http://dx.doi.org/10.1146/annurev.ns.43.120193.000331}{\emph{Ann. Rev.
  Nucl. Part. Sci.} {\bfseries 43} (1993) 27--70},
  [\href{https://arxiv.org/abs/hep-ph/9302210}{{\ttfamily hep-ph/9302210}}].

\bibitem{Quiros:1994dr}
M.~Quiros, \emph{{Field theory at finite temperature and phase transitions}},
  {\emph{Helv. Phys. Acta} {\bfseries 67} (1994) 451--583}.

\bibitem{Rubakov:1996vz}
V.~A. Rubakov and M.~E. Shaposhnikov, \emph{{Electroweak baryon number
  nonconservation in the early universe and in high-energy collisions}},
  \href{http://dx.doi.org/10.1070/PU1996v039n05ABEH000145}{\emph{Usp. Fiz.
  Nauk} {\bfseries 166} (1996) 493--537},
  [\href{https://arxiv.org/abs/hep-ph/9603208}{{\ttfamily hep-ph/9603208}}].

\bibitem{Carena:1997ys}
M.~Carena and C.~E.~M. Wagner, \emph{{Electroweak baryogenesis and Higgs
  physics}}, \href{http://dx.doi.org/10.1142/9789812819505_0010}{\emph{Adv.
  Ser. Direct. High Energy Phys.} {\bfseries 17} (1997) 320--358},
  [\href{https://arxiv.org/abs/hep-ph/9704347}{{\ttfamily hep-ph/9704347}}].

\bibitem{Quiros:1999jp}
M.~Quiros, \emph{{Finite temperature field theory and phase transitions}},  in
  \emph{{ICTP Summer School in High-Energy Physics and Cosmology}},
  pp.~187--259, 1, 1999.
\newblock \href{https://arxiv.org/abs/hep-ph/9901312}{{\ttfamily
  hep-ph/9901312}}.

\bibitem{Morrissey:2012db}
D.~E. Morrissey and M.~J. Ramsey-Musolf, \emph{{Electroweak baryogenesis}},
  \href{http://dx.doi.org/10.1088/1367-2630/14/12/125003}{\emph{New J. Phys.}
  {\bfseries 14} (2012) 125003},
  [\href{https://arxiv.org/abs/1206.2942}{{\ttfamily 1206.2942}}].

\bibitem{DOnofrio:2015gop}
M.~D'Onofrio and K.~Rummukainen, \emph{{Standard model cross-over on the
  lattice}}, \href{http://dx.doi.org/10.1103/PhysRevD.93.025003}{\emph{Phys.
  Rev. D} {\bfseries 93} (2016) 025003},
  [\href{https://arxiv.org/abs/1508.07161}{{\ttfamily 1508.07161}}].

\bibitem{Kajantie:1996qd}
K.~Kajantie, M.~Laine, K.~Rummukainen and M.~E. Shaposhnikov, \emph{{A
  Nonperturbative analysis of the finite T phase transition in SU(2) x U(1)
  electroweak theory}},
  \href{http://dx.doi.org/10.1016/S0550-3213(97)00164-8}{\emph{Nucl. Phys. B}
  {\bfseries 493} (1997) 413--438},
  [\href{https://arxiv.org/abs/hep-lat/9612006}{{\ttfamily hep-lat/9612006}}].

\bibitem{Anber:2006xt}
M.~M. Anber and L.~Sorbo, \emph{{N-flationary magnetic fields}},
  \href{http://dx.doi.org/10.1088/1475-7516/2006/10/018}{\emph{JCAP} {\bfseries
  10} (2006) 018}, [\href{https://arxiv.org/abs/astro-ph/0606534}{{\ttfamily
  astro-ph/0606534}}].

\bibitem{Bamba:2006km}
K.~Bamba, \emph{{Baryon asymmetry from hypermagnetic helicity in dilaton
  hypercharge electromagnetism}},
  \href{http://dx.doi.org/10.1103/PhysRevD.74.123504}{\emph{Phys. Rev. D}
  {\bfseries 74} (2006) 123504},
  [\href{https://arxiv.org/abs/hep-ph/0611152}{{\ttfamily hep-ph/0611152}}].

\bibitem{Bamba:2007hf}
K.~Bamba, C.~Q. Geng and S.~H. Ho, \emph{{Hypermagnetic Baryogenesis}},
  \href{http://dx.doi.org/10.1016/j.physletb.2008.05.027}{\emph{Phys. Lett. B}
  {\bfseries 664} (2008) 154--156},
  [\href{https://arxiv.org/abs/0712.1523}{{\ttfamily 0712.1523}}].

\bibitem{Anber:2009ua}
M.~M. Anber and L.~Sorbo, \emph{{Naturally inflating on steep potentials
  through electromagnetic dissipation}},
  \href{http://dx.doi.org/10.1103/PhysRevD.81.043534}{\emph{Phys. Rev. D}
  {\bfseries 81} (2010) 043534},
  [\href{https://arxiv.org/abs/0908.4089}{{\ttfamily 0908.4089}}].

\bibitem{Anber:2015yca}
M.~M. Anber and E.~Sabancilar, \emph{{Hypermagnetic Fields and Baryon Asymmetry
  from Pseudoscalar Inflation}},
  \href{http://dx.doi.org/10.1103/PhysRevD.92.101501}{\emph{Phys. Rev. D}
  {\bfseries 92} (2015) 101501},
  [\href{https://arxiv.org/abs/1507.00744}{{\ttfamily 1507.00744}}].

\bibitem{Cado:2016kdp}
Y.~Cado and E.~Sabancilar, \emph{{Asymmetric Dark Matter and Baryogenesis from
  Pseudoscalar Inflation}},
  \href{http://dx.doi.org/10.1088/1475-7516/2017/04/047}{\emph{JCAP} {\bfseries
  04} (2017) 047}, [\href{https://arxiv.org/abs/1611.02293}{{\ttfamily
  1611.02293}}].

\bibitem{Sfakianakis:2018lzf}
E.~I. Sfakianakis and J.~van~de Vis, \emph{{Preheating after Higgs Inflation:
  Self-Resonance and Gauge boson production}},
  \href{http://dx.doi.org/10.1103/PhysRevD.99.083519}{\emph{Phys. Rev. D}
  {\bfseries 99} (2019) 083519},
  [\href{https://arxiv.org/abs/1810.01304}{{\ttfamily 1810.01304}}].

\bibitem{Kamada:2016eeb}
K.~Kamada and A.~J. Long, \emph{{Baryogenesis from decaying magnetic
  helicity}}, \href{http://dx.doi.org/10.1103/PhysRevD.94.063501}{\emph{Phys.
  Rev. D} {\bfseries 94} (2016) 063501},
  [\href{https://arxiv.org/abs/1606.08891}{{\ttfamily 1606.08891}}].

\bibitem{Kamada:2016cnb}
K.~Kamada and A.~J. Long, \emph{{Evolution of the Baryon Asymmetry through the
  Electroweak Crossover in the Presence of a Helical Magnetic Field}},
  \href{http://dx.doi.org/10.1103/PhysRevD.94.123509}{\emph{Phys. Rev. D}
  {\bfseries 94} (2016) 123509},
  [\href{https://arxiv.org/abs/1610.03074}{{\ttfamily 1610.03074}}].

\bibitem{Jimenez:2017cdr}
D.~Jim\'enez, K.~Kamada, K.~Schmitz and X.-J. Xu, \emph{{Baryon asymmetry and
  gravitational waves from pseudoscalar inflation}},
  \href{http://dx.doi.org/10.1088/1475-7516/2017/12/011}{\emph{JCAP} {\bfseries
  12} (2017) 011}, [\href{https://arxiv.org/abs/1707.07943}{{\ttfamily
  1707.07943}}].

\bibitem{Domcke:2019mnd}
V.~Domcke, B.~von Harling, E.~Morgante and K.~Mukaida, \emph{{Baryogenesis from
  axion inflation}},
  \href{http://dx.doi.org/10.1088/1475-7516/2019/10/032}{\emph{JCAP} {\bfseries
  10} (2019) 032}, [\href{https://arxiv.org/abs/1905.13318}{{\ttfamily
  1905.13318}}].

\bibitem{Cado:2021bia}
Y.~Cado, B.~von Harling, E.~Mass\'o and M.~Quir\'os, \emph{{Baryogenesis via
  gauge field production from a relaxing Higgs}},
  \href{http://dx.doi.org/10.1088/1475-7516/2021/07/049}{\emph{JCAP} {\bfseries
  07} (2021) 049}, [\href{https://arxiv.org/abs/2102.13650}{{\ttfamily
  2102.13650}}].

\bibitem{Bezrukov:2007ep}
F.~L. Bezrukov and M.~Shaposhnikov, \emph{{The Standard Model Higgs boson as
  the inflaton}},
  \href{http://dx.doi.org/10.1016/j.physletb.2007.11.072}{\emph{Phys. Lett. B}
  {\bfseries 659} (2008) 703--706},
  [\href{https://arxiv.org/abs/0710.3755}{{\ttfamily 0710.3755}}].

\bibitem{Bezrukov:2008ej}
F.~L. Bezrukov, A.~Magnin and M.~Shaposhnikov, \emph{{Standard Model Higgs
  boson mass from inflation}},
  \href{http://dx.doi.org/10.1016/j.physletb.2009.03.035}{\emph{Phys. Lett. B}
  {\bfseries 675} (2009) 88--92},
  [\href{https://arxiv.org/abs/0812.4950}{{\ttfamily 0812.4950}}].

\bibitem{Bezrukov:2010jz}
F.~Bezrukov, A.~Magnin, M.~Shaposhnikov and S.~Sibiryakov, \emph{{Higgs
  inflation: consistency and generalisations}},
  \href{http://dx.doi.org/10.1007/JHEP01(2011)016}{\emph{JHEP} {\bfseries 01}
  (2011) 016}, [\href{https://arxiv.org/abs/1008.5157}{{\ttfamily 1008.5157}}].

\bibitem{Rubio:2018ogq}
J.~Rubio, \emph{{Higgs inflation}},
  \href{http://dx.doi.org/10.3389/fspas.2018.00050}{\emph{Front. Astron. Space
  Sci.} {\bfseries 5} (2019) 50},
  [\href{https://arxiv.org/abs/1807.02376}{{\ttfamily 1807.02376}}].

\bibitem{Han:2004wt}
T.~Han and S.~Willenbrock, \emph{{Scale of quantum gravity}},
  \href{http://dx.doi.org/10.1016/j.physletb.2005.04.040}{\emph{Phys. Lett. B}
  {\bfseries 616} (2005) 215--220},
  [\href{https://arxiv.org/abs/hep-ph/0404182}{{\ttfamily hep-ph/0404182}}].

\bibitem{Burgess:2009ea}
C.~P. Burgess, H.~M. Lee and M.~Trott, \emph{{Power-counting and the Validity
  of the Classical Approximation During Inflation}},
  \href{http://dx.doi.org/10.1088/1126-6708/2009/09/103}{\emph{JHEP} {\bfseries
  09} (2009) 103}, [\href{https://arxiv.org/abs/0902.4465}{{\ttfamily
  0902.4465}}].

\bibitem{Barbon:2009ya}
J.~L.~F. Barbon and J.~R. Espinosa, \emph{{On the Naturalness of Higgs
  Inflation}}, \href{http://dx.doi.org/10.1103/PhysRevD.79.081302}{\emph{Phys.
  Rev. D} {\bfseries 79} (2009) 081302},
  [\href{https://arxiv.org/abs/0903.0355}{{\ttfamily 0903.0355}}].

\bibitem{Lerner:2009na}
R.~N. Lerner and J.~McDonald, \emph{{Higgs Inflation and Naturalness}},
  \href{http://dx.doi.org/10.1088/1475-7516/2010/04/015}{\emph{JCAP} {\bfseries
  04} (2010) 015}, [\href{https://arxiv.org/abs/0912.5463}{{\ttfamily
  0912.5463}}].

\bibitem{Burgess:2010zq}
C.~P. Burgess, H.~M. Lee and M.~Trott, \emph{{Comment on Higgs Inflation and
  Naturalness}}, \href{http://dx.doi.org/10.1007/JHEP07(2010)007}{\emph{JHEP}
  {\bfseries 07} (2010) 007},
  [\href{https://arxiv.org/abs/1002.2730}{{\ttfamily 1002.2730}}].

\bibitem{Hertzberg:2010dc}
M.~P. Hertzberg, \emph{{On Inflation with Non-minimal Coupling}},
  \href{http://dx.doi.org/10.1007/JHEP11(2010)023}{\emph{JHEP} {\bfseries 11}
  (2010) 023}, [\href{https://arxiv.org/abs/1002.2995}{{\ttfamily 1002.2995}}].

\bibitem{Antoniadis:2021axu}
I.~Antoniadis, A.~Guillen and K.~Tamvakis, \emph{{Ultraviolet behaviour of
  Higgs inflation models}},
  \href{http://dx.doi.org/10.1007/JHEP05(2022)074}{\emph{JHEP} {\bfseries 08}
  (2021) 018}, [\href{https://arxiv.org/abs/2106.09390}{{\ttfamily
  2106.09390}}].

\bibitem{Ito:2021ssc}
A.~Ito, W.~Khater and S.~Rasanen, \emph{{Tree-level unitarity in Higgs
  inflation in the metric and the Palatini formulation}},
  \href{https://arxiv.org/abs/2111.05621}{{\ttfamily 2111.05621}}.

\bibitem{Karananas:2022byw}
G.~K. Karananas, M.~Shaposhnikov and S.~Zell, \emph{{Field redefinitions,
  perturbative unitarity and Higgs inflation}},
  \href{http://dx.doi.org/10.1007/JHEP06(2022)132}{\emph{JHEP} {\bfseries 06}
  (2022) 132}, [\href{https://arxiv.org/abs/2203.09534}{{\ttfamily
  2203.09534}}].

\bibitem{Bezrukov:2014ipa}
F.~Bezrukov, J.~Rubio and M.~Shaposhnikov, \emph{{Living beyond the edge: Higgs
  inflation and vacuum metastability}},
  \href{http://dx.doi.org/10.1103/PhysRevD.92.083512}{\emph{Phys. Rev. D}
  {\bfseries 92} (2015) 083512},
  [\href{https://arxiv.org/abs/1412.3811}{{\ttfamily 1412.3811}}].

\bibitem{Giudice:2010ka}
G.~F. Giudice and H.~M. Lee, \emph{{Unitarizing Higgs Inflation}},
  \href{http://dx.doi.org/10.1016/j.physletb.2010.10.035}{\emph{Phys. Lett. B}
  {\bfseries 694} (2011) 294--300},
  [\href{https://arxiv.org/abs/1010.1417}{{\ttfamily 1010.1417}}].

\bibitem{Barbon:2015fla}
J.~L.~F. Barbon, J.~A. Casas, J.~Elias-Miro and J.~R. Espinosa, \emph{{Higgs
  Inflation as a Mirage}},
  \href{http://dx.doi.org/10.1007/JHEP09(2015)027}{\emph{JHEP} {\bfseries 09}
  (2015) 027}, [\href{https://arxiv.org/abs/1501.02231}{{\ttfamily
  1501.02231}}].

\bibitem{Dine:1990fj}
M.~Dine, P.~Huet, R.~L. Singleton, Jr and L.~Susskind, \emph{{Creating the
  baryon asymmetry at the electroweak phase transition}},
  \href{http://dx.doi.org/10.1016/0370-2693(91)91905-B}{\emph{Phys. Lett. B}
  {\bfseries 257} (1991) 351--356}.

\bibitem{Carena:2018cjh}
M.~Carena, M.~Quir\'os and Y.~Zhang, \emph{{Electroweak Baryogenesis from
  Dark-Sector CP Violation}},
  \href{http://dx.doi.org/10.1103/PhysRevLett.122.201802}{\emph{Phys. Rev.
  Lett.} {\bfseries 122} (2019) 201802},
  [\href{https://arxiv.org/abs/1811.09719}{{\ttfamily 1811.09719}}].

\bibitem{Carena:2019xrr}
M.~Carena, M.~Quir\'os and Y.~Zhang, \emph{{Dark CP violation and gauged lepton
  or baryon number for electroweak baryogenesis}},
  \href{http://dx.doi.org/10.1103/PhysRevD.101.055014}{\emph{Phys. Rev. D}
  {\bfseries 101} (2020) 055014},
  [\href{https://arxiv.org/abs/1908.04818}{{\ttfamily 1908.04818}}].

\bibitem{Domcke:2018eki}
V.~Domcke and K.~Mukaida, \emph{{Gauge Field and Fermion Production during
  Axion Inflation}},
  \href{http://dx.doi.org/10.1088/1475-7516/2018/11/020}{\emph{JCAP} {\bfseries
  11} (2018) 020}, [\href{https://arxiv.org/abs/1806.08769}{{\ttfamily
  1806.08769}}].

\bibitem{Gorbar:2021zlr}
E.~V. Gorbar, K.~Schmitz, O.~O. Sobol and S.~I. Vilchinskii,
  \emph{{Hypermagnetogenesis from axion inflation: Model-independent
  estimates}}, \href{http://dx.doi.org/10.1103/PhysRevD.105.043530}{\emph{Phys.
  Rev. D} {\bfseries 105} (2022) 043530},
  [\href{https://arxiv.org/abs/2111.04712}{{\ttfamily 2111.04712}}].

\bibitem{Espinosa:2015qea}
J.~R. Espinosa, G.~F. Giudice, E.~Morgante, A.~Riotto, L.~Senatore, A.~Strumia
  et~al., \emph{{The cosmological Higgstory of the vacuum instability}},
  \href{http://dx.doi.org/10.1007/JHEP09(2015)174}{\emph{JHEP} {\bfseries 09}
  (2015) 174}, [\href{https://arxiv.org/abs/1505.04825}{{\ttfamily
  1505.04825}}].

\bibitem{Khoury:2021zao}
J.~Khoury and T.~Steingasser, \emph{{Gauge hierarchy from electroweak vacuum
  metastability}},
  \href{http://dx.doi.org/10.1103/PhysRevD.105.055031}{\emph{Phys. Rev. D}
  {\bfseries 105} (2022) 055031},
  [\href{https://arxiv.org/abs/2108.09315}{{\ttfamily 2108.09315}}].

\bibitem{DeSimone:2008ei}
A.~De~Simone, M.~P. Hertzberg and F.~Wilczek, \emph{{Running Inflation in the
  Standard Model}},
  \href{http://dx.doi.org/10.1016/j.physletb.2009.05.054}{\emph{Phys. Lett. B}
  {\bfseries 678} (2009) 1--8},
  [\href{https://arxiv.org/abs/0812.4946}{{\ttfamily 0812.4946}}].

\bibitem{Planck:2018jri}
{\scshape Planck} collaboration, Y.~Akrami et~al., \emph{{Planck 2018 results.
  X. Constraints on inflation}},
  \href{http://dx.doi.org/10.1051/0004-6361/201833887}{\emph{Astron.
  Astrophys.} {\bfseries 641} (2020) A10},
  [\href{https://arxiv.org/abs/1807.06211}{{\ttfamily 1807.06211}}].

\bibitem{Lyth:1996im}
D.~H. Lyth, \emph{{What would we learn by detecting a gravitational wave signal
  in the cosmic microwave background anisotropy?}},
  \href{http://dx.doi.org/10.1103/PhysRevLett.78.1861}{\emph{Phys. Rev. Lett.}
  {\bfseries 78} (1997) 1861--1863},
  [\href{https://arxiv.org/abs/hep-ph/9606387}{{\ttfamily hep-ph/9606387}}].

\bibitem{Linde:1990flp}
A.~D. Linde, \emph{{Particle physics and inflationary cosmology}},
  {\emph{Contemp. Concepts Phys.} {\bfseries 5} (1990) 1--362},
  [\href{https://arxiv.org/abs/hep-th/0503203}{{\ttfamily hep-th/0503203}}].

\bibitem{Herranen:2014cua}
M.~Herranen, T.~Markkanen, S.~Nurmi and A.~Rajantie, \emph{{Spacetime curvature
  and the Higgs stability during inflation}},
  \href{http://dx.doi.org/10.1103/PhysRevLett.113.211102}{\emph{Phys. Rev.
  Lett.} {\bfseries 113} (2014) 211102},
  [\href{https://arxiv.org/abs/1407.3141}{{\ttfamily 1407.3141}}].

\bibitem{BICEPKeck:2021gln}
{\scshape BICEP/Keck} collaboration, P.~A.~R. Ade et~al., \emph{{Improved
  Constraints on Primordial Gravitational Waves using Planck, WMAP, and
  BICEP/Keck Observations through the 2018 Observing Season}},
  \href{http://dx.doi.org/10.1103/PhysRevLett.127.151301}{\emph{Phys. Rev.
  Lett.} {\bfseries 127} (2021) 151301},
  [\href{https://arxiv.org/abs/2110.00483}{{\ttfamily 2110.00483}}].

\bibitem{Kodama:2021yrm}
T.~Kodama and T.~Takahashi, \emph{{Relaxing inflation models with nonminimal
  coupling: A general study}},
  \href{http://dx.doi.org/10.1103/PhysRevD.105.063542}{\emph{Phys. Rev. D}
  {\bfseries 105} (2022) 063542},
  [\href{https://arxiv.org/abs/2112.05283}{{\ttfamily 2112.05283}}].

\bibitem{book:Abramowitz}
M.~Abramowitz and I.~A. Stegun, \emph{Handbook of Mathematical Functions: With
  Formulas, Graphs, and Mathematical Tables}.
\newblock Dover Publications, 1965.

\bibitem{Durrer:2013pga}
R.~Durrer and A.~Neronov, \emph{{Cosmological Magnetic Fields: Their
  Generation, Evolution and Observation}},
  \href{http://dx.doi.org/10.1007/s00159-013-0062-7}{\emph{Astron. Astrophys.
  Rev.} {\bfseries 21} (2013) 62},
  [\href{https://arxiv.org/abs/1303.7121}{{\ttfamily 1303.7121}}].

\bibitem{Gorbar:2021rlt}
E.~V. Gorbar, K.~Schmitz, O.~O. Sobol and S.~I. Vilchinskii, \emph{{Gauge-field
  production during axion inflation in the gradient expansion formalism}},
  \href{http://dx.doi.org/10.1103/PhysRevD.104.123504}{\emph{Phys. Rev. D}
  {\bfseries 104} (2021) 123504},
  [\href{https://arxiv.org/abs/2109.01651}{{\ttfamily 2109.01651}}].

\bibitem{Kofman:1997yn}
L.~Kofman, A.~D. Linde and A.~A. Starobinsky, \emph{{Towards the theory of
  reheating after inflation}},
  \href{http://dx.doi.org/10.1103/PhysRevD.56.3258}{\emph{Phys. Rev. D}
  {\bfseries 56} (1997) 3258--3295},
  [\href{https://arxiv.org/abs/hep-ph/9704452}{{\ttfamily hep-ph/9704452}}].

\bibitem{Adshead:2015pva}
P.~Adshead, J.~T. Giblin, T.~R. Scully and E.~I. Sfakianakis,
  \emph{{Gauge-preheating and the end of axion inflation}},
  \href{http://dx.doi.org/10.1088/1475-7516/2015/12/034}{\emph{JCAP} {\bfseries
  12} (2015) 034}, [\href{https://arxiv.org/abs/1502.06506}{{\ttfamily
  1502.06506}}].

\bibitem{Cuissa:2018oiw}
J.~R.~C. Cuissa and D.~G. Figueroa, \emph{{Lattice formulation of axion
  inflation. Application to preheating}},
  \href{http://dx.doi.org/10.1088/1475-7516/2019/06/002}{\emph{JCAP} {\bfseries
  06} (2019) 002}, [\href{https://arxiv.org/abs/1812.03132}{{\ttfamily
  1812.03132}}].

\bibitem{Chung:1998rq}
D.~J. Chung, E.~W. Kolb and A.~Riotto, \emph{{Production of massive particles
  during reheating}},
  \href{http://dx.doi.org/10.1103/PhysRevD.60.063504}{\emph{Phys.\ Rev.\ D}
  {\bfseries 60} (1999) 063504},
  [\href{https://arxiv.org/abs/hep-ph/9809453}{{\ttfamily hep-ph/9809453}}].

\bibitem{Giudice:2000ex}
G.~F. Giudice, E.~W. Kolb and A.~Riotto, \emph{{Largest temperature of the
  radiation era and its cosmological implications}},
  \href{http://dx.doi.org/10.1103/PhysRevD.64.023508}{\emph{Phys.\ Rev.\ D}
  {\bfseries 64} (2001) 023508},
  [\href{https://arxiv.org/abs/hep-ph/0005123}{{\ttfamily hep-ph/0005123}}].

\bibitem{Kofman:1994rk}
L.~Kofman, A.~D. Linde and A.~A. Starobinsky, \emph{{Reheating after
  inflation}}, \href{http://dx.doi.org/10.1103/PhysRevLett.73.3195}{\emph{Phys.
  Rev. Lett.} {\bfseries 73} (1994) 3195--3198},
  [\href{https://arxiv.org/abs/hep-th/9405187}{{\ttfamily hep-th/9405187}}].

\bibitem{Cosme:2022htl}
C.~Cosme, D.~G. Figueroa and N.~Loayza, \emph{{Gravitational wave production
  from preheating with trilinear interactions}},
  \href{https://arxiv.org/abs/2206.14721}{{\ttfamily 2206.14721}}.

\bibitem{Zyla:2020zbs}
{\scshape Particle Data Group} collaboration, P.~Zyla et~al., \emph{{Review of
  Particle Physics}}, \href{http://dx.doi.org/10.1093/ptep/ptaa104}{\emph{PTEP}
  {\bfseries 2020} (2020) 083C01}.

\bibitem{Giovannini:1997eg}
M.~Giovannini and M.~E. Shaposhnikov, \emph{{Primordial hypermagnetic fields
  and triangle anomaly}},
  \href{http://dx.doi.org/10.1103/PhysRevD.57.2186}{\emph{Phys. Rev. D}
  {\bfseries 57} (1998) 2186--2206},
  [\href{https://arxiv.org/abs/hep-ph/9710234}{{\ttfamily hep-ph/9710234}}].

\bibitem{Vachaspati:2020blt}
T.~Vachaspati, \emph{{Progress on cosmological magnetic fields}},
  \href{http://dx.doi.org/10.1088/1361-6633/ac03a9}{\emph{Rept. Prog. Phys.}
  {\bfseries 84} (2021) 074901},
  [\href{https://arxiv.org/abs/2010.10525}{{\ttfamily 2010.10525}}].

\bibitem{Banerjee:2004df}
R.~Banerjee and K.~Jedamzik, \emph{{The Evolution of cosmic magnetic fields:
  From the very early universe, to recombination, to the present}},
  \href{http://dx.doi.org/10.1103/PhysRevD.70.123003}{\emph{Phys. Rev. D}
  {\bfseries 70} (2004) 123003},
  [\href{https://arxiv.org/abs/astro-ph/0410032}{{\ttfamily
  astro-ph/0410032}}].

\bibitem{Kamada:2018tcs}
K.~Kamada, \emph{{Return of grand unified theory baryogenesis: Source of
  helical hypermagnetic fields for the baryon asymmetry of the universe}},
  \href{http://dx.doi.org/10.1103/PhysRevD.97.103506}{\emph{Phys. Rev. D}
  {\bfseries 97} (2018) 103506},
  [\href{https://arxiv.org/abs/1802.03055}{{\ttfamily 1802.03055}}].

\bibitem{Joyce:1997uy}
M.~Joyce and M.~E. Shaposhnikov, \emph{{Primordial magnetic fields,
  right-handed electrons, and the Abelian anomaly}},
  \href{http://dx.doi.org/10.1103/PhysRevLett.79.1193}{\emph{Phys. Rev. Lett.}
  {\bfseries 79} (1997) 1193--1196},
  [\href{https://arxiv.org/abs/astro-ph/9703005}{{\ttfamily
  astro-ph/9703005}}].

\bibitem{Boyarsky:2011uy}
A.~Boyarsky, J.~Frohlich and O.~Ruchayskiy, \emph{{Self-consistent evolution of
  magnetic fields and chiral asymmetry in the early Universe}},
  \href{http://dx.doi.org/10.1103/PhysRevLett.108.031301}{\emph{Phys. Rev.
  Lett.} {\bfseries 108} (2012) 031301},
  [\href{https://arxiv.org/abs/1109.3350}{{\ttfamily 1109.3350}}].

\bibitem{Akamatsu:2013pjd}
Y.~Akamatsu and N.~Yamamoto, \emph{{Chiral Plasma Instabilities}},
  \href{http://dx.doi.org/10.1103/PhysRevLett.111.052002}{\emph{Phys. Rev.
  Lett.} {\bfseries 111} (2013) 052002},
  [\href{https://arxiv.org/abs/1302.2125}{{\ttfamily 1302.2125}}].

\bibitem{Hirono:2015rla}
Y.~Hirono, D.~Kharzeev and Y.~Yin, \emph{{Self-similar inverse cascade of
  magnetic helicity driven by the chiral anomaly}},
  \href{http://dx.doi.org/10.1103/PhysRevD.92.125031}{\emph{Phys. Rev. D}
  {\bfseries 92} (2015) 125031},
  [\href{https://arxiv.org/abs/1509.07790}{{\ttfamily 1509.07790}}].

\bibitem{Yamamoto:2016xtu}
N.~Yamamoto, \emph{{Scaling laws in chiral hydrodynamic turbulence}},
  \href{http://dx.doi.org/10.1103/PhysRevD.93.125016}{\emph{Phys. Rev. D}
  {\bfseries 93} (2016) 125016},
  [\href{https://arxiv.org/abs/1603.08864}{{\ttfamily 1603.08864}}].

\bibitem{Rogachevskii:2017uyc}
I.~Rogachevskii, O.~Ruchayskiy, A.~Boyarsky, J.~Fr\"ohlich, N.~Kleeorin,
  A.~Brandenburg et~al., \emph{{Laminar and turbulent dynamos in chiral
  magnetohydrodynamics-I: Theory}},
  \href{http://dx.doi.org/10.3847/1538-4357/aa886b}{\emph{Astrophys. J.}
  {\bfseries 846} (2017) 153},
  [\href{https://arxiv.org/abs/1705.00378}{{\ttfamily 1705.00378}}].

\bibitem{Komatsu:2001ysk}
E.~Komatsu, \emph{{The pursuit of non-gaussian fluctuations in the cosmic
  microwave background}}.
\newblock PhD thesis, Tohoku U., 2001.
\newblock \href{https://arxiv.org/abs/astro-ph/0206039}{{\ttfamily
  astro-ph/0206039}}.

\bibitem{WMAP:2010qai}
{\scshape WMAP} collaboration, E.~Komatsu et~al., \emph{{Seven-Year Wilkinson
  Microwave Anisotropy Probe (WMAP) Observations: Cosmological
  Interpretation}},
  \href{http://dx.doi.org/10.1088/0067-0049/192/2/18}{\emph{Astrophys. J.
  Suppl.} {\bfseries 192} (2011) 18},
  [\href{https://arxiv.org/abs/1001.4538}{{\ttfamily 1001.4538}}].

\bibitem{Barnaby:2010vf}
N.~Barnaby and M.~Peloso, \emph{{Large Nongaussianity in Axion Inflation}},
  \href{http://dx.doi.org/10.1103/PhysRevLett.106.181301}{\emph{Phys. Rev.
  Lett.} {\bfseries 106} (2011) 181301},
  [\href{https://arxiv.org/abs/1011.1500}{{\ttfamily 1011.1500}}].

\bibitem{Barnaby:2011qe}
N.~Barnaby, E.~Pajer and M.~Peloso, \emph{{Gauge Field Production in Axion
  Inflation: Consequences for Monodromy, non-Gaussianity in the CMB, and
  Gravitational Waves at Interferometers}},
  \href{http://dx.doi.org/10.1103/PhysRevD.85.023525}{\emph{Phys. Rev. D}
  {\bfseries 85} (2012) 023525},
  [\href{https://arxiv.org/abs/1110.3327}{{\ttfamily 1110.3327}}].

\bibitem{Planck:2019kim}
{\scshape Planck} collaboration, Y.~Akrami et~al., \emph{{Planck 2018 results.
  IX. Constraints on primordial non-Gaussianity}},
  \href{http://dx.doi.org/10.1051/0004-6361/201935891}{\emph{Astron.
  Astrophys.} {\bfseries 641} (2020) A9},
  [\href{https://arxiv.org/abs/1905.05697}{{\ttfamily 1905.05697}}].

\bibitem{Neronov:2010gir}
A.~Neronov and I.~Vovk, \emph{{Evidence for strong extragalactic magnetic
  fields from Fermi observations of TeV blazars}},
  \href{http://dx.doi.org/10.1126/science.1184192}{\emph{Science} {\bfseries
  328} (2010) 73--75}, [\href{https://arxiv.org/abs/1006.3504}{{\ttfamily
  1006.3504}}].

\bibitem{Tavecchio:2010mk}
F.~Tavecchio, G.~Ghisellini, L.~Foschini, G.~Bonnoli, G.~Ghirlanda and
  P.~Coppi, \emph{{The intergalactic magnetic field constrained by Fermi/LAT
  observations of the TeV blazar 1ES 0229+200}},
  \href{http://dx.doi.org/10.1111/j.1745-3933.2010.00884.x}{\emph{Mon. Not.
  Roy. Astron. Soc.} {\bfseries 406} (2010) L70--L74},
  [\href{https://arxiv.org/abs/1004.1329}{{\ttfamily 1004.1329}}].

\bibitem{Ando:2010rb}
S.~Ando and A.~Kusenko, \emph{{Evidence for Gamma-Ray Halos Around Active
  Galactic Nuclei and the First Measurement of Intergalactic Magnetic Fields}},
  \href{http://dx.doi.org/10.1088/2041-8205/722/1/L39}{\emph{Astrophys. J.
  Lett.} {\bfseries 722} (2010) L39},
  [\href{https://arxiv.org/abs/1005.1924}{{\ttfamily 1005.1924}}].

\bibitem{Kamada:2020bmb}
K.~Kamada, F.~Uchida and J.~Yokoyama, \emph{{Baryon isocurvature constraints on
  the primordial hypermagnetic fields}},
  \href{http://dx.doi.org/10.1088/1475-7516/2021/04/034}{\emph{JCAP} {\bfseries
  04} (2021) 034}, [\href{https://arxiv.org/abs/2012.14435}{{\ttfamily
  2012.14435}}].

\bibitem{Inomata:2018htm}
K.~Inomata, M.~Kawasaki, A.~Kusenko and L.~Yang, \emph{{Big Bang
  Nucleosynthesis Constraint on Baryonic Isocurvature Perturbations}},
  \href{http://dx.doi.org/10.1088/1475-7516/2018/12/003}{\emph{JCAP} {\bfseries
  12} (2018) 003}, [\href{https://arxiv.org/abs/1806.00123}{{\ttfamily
  1806.00123}}].

\bibitem{deGouvea:2014xba}
A.~de~Gouvea, D.~Hernandez and T.~M.~P. Tait, \emph{{Criteria for Natural
  Hierarchies}},
  \href{http://dx.doi.org/10.1103/PhysRevD.89.115005}{\emph{Phys. Rev. D}
  {\bfseries 89} (2014) 115005},
  [\href{https://arxiv.org/abs/1402.2658}{{\ttfamily 1402.2658}}].

\bibitem{Profumo:2007wc}
S.~Profumo, M.~J. Ramsey-Musolf and G.~Shaughnessy, \emph{{Singlet Higgs
  phenomenology and the electroweak phase transition}},
  \href{http://dx.doi.org/10.1088/1126-6708/2007/08/010}{\emph{JHEP} {\bfseries
  08} (2007) 010}, [\href{https://arxiv.org/abs/0705.2425}{{\ttfamily
  0705.2425}}].

\bibitem{Barger:2007im}
V.~Barger, P.~Langacker, M.~McCaskey, M.~J. Ramsey-Musolf and G.~Shaughnessy,
  \emph{{LHC Phenomenology of an Extended Standard Model with a Real Scalar
  Singlet}}, \href{http://dx.doi.org/10.1103/PhysRevD.77.035005}{\emph{Phys.
  Rev. D} {\bfseries 77} (2008) 035005},
  [\href{https://arxiv.org/abs/0706.4311}{{\ttfamily 0706.4311}}].

\bibitem{ATLAS:2019nkf}
{\scshape ATLAS} collaboration, G.~Aad et~al., \emph{{Combined measurements of
  Higgs boson production and decay using up to $80$ fb$^{-1}$ of proton-proton
  collision data at $\sqrt{s}=$ 13 TeV collected with the ATLAS experiment}},
  \href{http://dx.doi.org/10.1103/PhysRevD.101.012002}{\emph{Phys. Rev. D}
  {\bfseries 101} (2020) 012002},
  [\href{https://arxiv.org/abs/1909.02845}{{\ttfamily 1909.02845}}].

\bibitem{CMS:2018uag}
{\scshape CMS} collaboration, A.~M. Sirunyan et~al., \emph{{Combined
  measurements of Higgs boson couplings in proton\textendash{}proton collisions
  at $\sqrt{s}=13\,\text {Te}\text {V} $}},
  \href{http://dx.doi.org/10.1140/epjc/s10052-019-6909-y}{\emph{Eur. Phys. J.
  C} {\bfseries 79} (2019) 421},
  [\href{https://arxiv.org/abs/1809.10733}{{\ttfamily 1809.10733}}].

\bibitem{ATL-PHYS-PUB-2019-009}
{\scshape ATLAS} collaboration, \emph{{Constraint of the Higgs boson
  self-coupling from Higgs boson differential production and decay
  measurements}},  tech. rep., CERN, Geneva, Mar, 2019.

\bibitem{CMS:2020tkr}
{\scshape CMS} collaboration, A.~M. Sirunyan et~al., \emph{{Search for
  nonresonant Higgs boson pair production in final states with two bottom
  quarks and two photons in proton-proton collisions at $ \sqrt{s} $ = 13
  TeV}}, \href{http://dx.doi.org/10.1007/JHEP03(2021)257}{\emph{JHEP}
  {\bfseries 03} (2021) 257},
  [\href{https://arxiv.org/abs/2011.12373}{{\ttfamily 2011.12373}}].

\bibitem{Homiller:2018dgu}
S.~Homiller and P.~Meade, \emph{{Measurement of the Triple Higgs Coupling at a
  HE-LHC}}, \href{http://dx.doi.org/10.1007/JHEP03(2019)055}{\emph{JHEP}
  {\bfseries 03} (2019) 055},
  [\href{https://arxiv.org/abs/1811.02572}{{\ttfamily 1811.02572}}].

\bibitem{Goncalves:2018qas}
D.~Gon\c{c}alves, T.~Han, F.~Kling, T.~Plehn and M.~Takeuchi, \emph{{Higgs
  boson pair production at future hadron colliders: From kinematics to
  dynamics}}, \href{http://dx.doi.org/10.1103/PhysRevD.97.113004}{\emph{Phys.
  Rev. D} {\bfseries 97} (2018) 113004},
  [\href{https://arxiv.org/abs/1802.04319}{{\ttfamily 1802.04319}}].

\bibitem{LHCHiggsCrossSectionWorkingGroup:2011wcg}
{\scshape LHC Higgs Cross Section Working Group} collaboration, S.~Dittmaier
  et~al., \emph{{Handbook of LHC Higgs Cross Sections: 1. Inclusive
  Observables}},  \href{https://arxiv.org/abs/1101.0593}{{\ttfamily
  1101.0593}}.

\bibitem{LHCHiggsCrossSectionWorkingGroup:2016ypw}
{\scshape LHC Higgs Cross Section Working Group} collaboration, D.~de~Florian
  et~al., \emph{{Handbook of LHC Higgs Cross Sections: 4. Deciphering the
  Nature of the Higgs Sector}},
  \href{https://arxiv.org/abs/1610.07922}{{\ttfamily 1610.07922}}.

\bibitem{ATLAS:2019qdc}
{\scshape ATLAS} collaboration, G.~Aad et~al., \emph{{Combination of searches
  for Higgs boson pairs in $pp$ collisions at $\sqrt{s} = $13 TeV with the
  ATLAS detector}},
  \href{http://dx.doi.org/10.1016/j.physletb.2019.135103}{\emph{Phys. Lett. B}
  {\bfseries 800} (2020) 135103},
  [\href{https://arxiv.org/abs/1906.02025}{{\ttfamily 1906.02025}}].

\bibitem{CMS:2018ipl}
{\scshape CMS} collaboration, A.~M. Sirunyan et~al., \emph{{Combination of
  searches for Higgs boson pair production in proton-proton collisions at
  $\sqrt{s} = $ 13 TeV}},
  \href{http://dx.doi.org/10.1103/PhysRevLett.122.121803}{\emph{Phys. Rev.
  Lett.} {\bfseries 122} (2019) 121803},
  [\href{https://arxiv.org/abs/1811.09689}{{\ttfamily 1811.09689}}].

\end{thebibliography}\endgroup

\end{document}